\documentclass[preprint,prdlaps,amsmath,amssymb,superscriptaddress,floatfix]{revtex4-1}
\pdfoutput=1 
\usepackage{graphicx}
\usepackage{dcolumn}
\usepackage{bm}
\usepackage{subfigure}
\usepackage{epsfig}
\usepackage{float}
\usepackage{enumerate}
\usepackage{tabularx} 

\usepackage{color}

\usepackage{txfonts}
\usepackage{morefloats}

\newcommand{\beq}{\begin{equation}}
\newcommand{\eeq}{\end{equation}}
\newcommand{\bea}{\begin{eqnarray}}
\newcommand{\eea}{\end{eqnarray}}

\newcommand*\chem[1]{\ensuremath{\mathrm{#1}}}

\usepackage{psfrag}
\usepackage{url}
\usepackage{fancybox}
\usepackage{graphicx}
\usepackage{afterpage}
\usepackage{color}
\usepackage{subfigure}
\usepackage{epsfig}
\usepackage{pifont}
\usepackage{exscale}
\usepackage{relsize}
\newcommand{\tickYes}{\checkmark}
\newcommand{\tickNo}{\hspace{1pt}\ding{55}}

\begin{document}

\rightline{UdeM-GPP-TH-16-251}
\rightline{YGHP-16-04}

\vskip 10pt

\title{\large{Head butting sheep: Kink Collisions in the Presence of False Vacua}}

\vskip 10pt
                                                                                                                                                                                                     
\author{Jennifer Ashcroft}%
 \email{J.E.Ashcroft@kent.ac.uk}
\affiliation{%
School of Mathematics, Statistics and Actuarial Science,\\ University of Kent, Canterbury CT2 7NF, U.K.\\
}%

\author{Minoru Eto}%
\email{meto@sci.kj.yamagata-u.ac.jp}

\affiliation{%
Department of Physics, Yamagata University, Yamagata 990-8560, Japan
}%

\author{Mareike Haberichter}%
 \email{mkh@phys-h.keio.ac.jp}

\affiliation{%
School of Mathematics, Statistics and Actuarial Science,\\ University of Kent, Canterbury CT2 7NF, U.K.\\
}%

\affiliation{%
Department of Physics, and Research and Education Center for Natural Sciences, Keio University, Hiyoshi 4-1-1, Yokohama, Kanagawa 223-8521, Japan
}%

\author{Muneto Nitta}%
\email{nitta@phys-h.keio.ac.jp}

\affiliation{%
Department of Physics, and Research and Education Center for Natural Sciences, Keio University, Hiyoshi 4-1-1, Yokohama, Kanagawa 223-8521, Japan
}%

\author{M.~B.~Paranjape}
\email{paranj@lps.umontreal.ca}

\affiliation{Groupe de Physique des Particules, D\'epartement de physique, Universit\'{e} de Montr\'{e}al, C.~P.~6128, Succursale Centre-ville, Montreal, Qu\'{e}bec, Canada, H3C 3J7}

\date{\today}

\vskip 10pt

\begin{abstract}
 
We investigate numerically kink collisions in a $1+1$ dimensional scalar field theory with multiple vacua. The domain wall model we are interested in involves two scalar fields and a potential term built from an asymmetric double well and (double) sine-Gordon potential together with an interaction term. Depending on the initial kink setup and impact velocities, the model allows for a wide range of scattering behaviours. Kinks can repel each other, annihilate, form true or false domain walls and reflect off each other.

\end{abstract}

\pacs{Valid PACS appear here}
\maketitle

\section{Introduction}

Scattering of kink solutions in  $1+1$ dimensional field theories revealed a rich phenomenology. Depending on the impact velocity, very different scattering outcomes can be observed in kink-antikink collisions in $\phi^4$ \cite{Moshir:1981ja,Campbell:1983xu,Belova:1985fg,Anninos:1991un,Halavanau:2012dv} and $\phi^6$ \cite{Dorey:2011yw,Weigel:2013kwa} models. There exist ``windows" or velocity ranges for which kink and antikink collide, reflect, recede to a finite separation, then return to collide again before they escape to infinity. Such a velocity regime is called a two-bounce window, labeled by the number of reflections -- the bounce number. Windows of higher bounce numbers have been studied in the past. There also exist velocities for which the kink and antikink pair capture each other and form a long-lived, oscillatory bound state (an oscillon)  which can persist for thousands of oscillations before decaying to the vacuum. Numerical and analytical studies \cite{Campbell:1983xu,Belova:1985fg} of kink-antikink collisions in $\phi^4$ models showed that reflection and annhilation alternate resulting in a nested structure of  ``resonance windows". Furthermore, novel types of interactions have been found when studying multi-kink scattering in the presence potential wells and barriers \cite{Goatham:2010dg} or recently in the presence of boundaries \cite{Dorey:2015sha,Arthur:2015mva}. 

In this article, we are particularly interested in the dynamics of kink solutions in a model with two scalar fields \cite{Haberichter:2015xga}. The $1+1$ dimensional model recently introduced in Ref.~\cite{Haberichter:2015xga} admits solitons which interpolate between discrete vacua of different energy. The minima of lowest energy are called \emph{true vacua} and the minima of higher energy are identified with \emph{false vacua}. Domain walls connecting different vacua, true or false, are known to be important in cosmology, where they have been discussed as a candidate to explain the dark energy of the Universe without recourse to a non-vanishing cosmological constant \cite{Battye:1999eq,Battye:2006pf,Battye:2010dk,Battye:2011ff}. For a detailed discussion of the cosmological consequences of domain walls and other topological defects we refer the interested reader to the review article \cite{Vilenkin:1984ib}.

When several such domain walls are present, domain wall collisions may occur. Domain wall collisions in a single-field scalar theory with an asymmetric double well potential and hence with vacua of different energies have for example been explored in Refs.~\cite{Braden:2014cra,Braden:2015vza,Bond:2015zfa}. When kink and antikink are evolved in this model, they attract and undergo multiple collisions \cite{Braden:2014cra} radiating away their energy. Ultimately, the kink-antikink pair decays to the true vacuum by radiation. Note that this scattering behaviour differs from the one observed in a symmetric double well. In this case, for suitably chosen initial velocities, the kinks are able to escape to infinity after colliding with each other. 

The domain wall model \cite{Haberichter:2015xga} we are interested in is more complicated and allows for a wider range of scattering behaviours since it involves two scalar fields and a potential term built from an asymmetric double well and sine-Gordon potential together with an interaction term. The inclusion of a sine-Gordon potential is particularly interesting because sine-Gordon kinks can be experimentally observed in a wide range of condensed matter systems. Examples include magnetic flux quanta in Josephson junctions of two superconductors \cite{Ustinov1998315,Fiore2015}, domain walls in two-component Bose-Einstein condensates \cite{Son:2001td} and in two-band superconductors \cite{PhysRevLett.88.017002,PhysRevLett.90.047004}, magnetic bubbles in ferromagnets \cite{Malozemoff:101883}, defects in superfluid Helium-3 \cite{He_Volovik}, soliton excitations within DNA chains \cite{Englander,PhysRevA.44.5292,PhysRevE.66.016614,Yakushevich:2004,Cuenda2006214}, ideal switches for weak signal detection  \cite{PhysRevE.71.056620} and thermodynamic excitations in classical statistical mechanics \cite{PhysRevLett.56.2233}. Domain wall formation and their decay have also been studied during chiral phase transition in QCD-like theories \cite{Eto:2013bxa}. For a detailed discussion of topological solitons in dense QCD and their phenomenological implications, we refer the interested reader to the review article \cite{Eto:2013hoa} and references therein.

Here, double-sine-Gordon systems \cite{Popov2005309,Riazi:1998dw} -- sine-Gordon models generalised by adding a harmonic term  to the ordinary sine-Gordon potential --  are of particular interest in physical applications. For example, soliton creation in the $A$ phase of superfluid Helium-3 can be modelled as domain wall formation in the sine-Gordon model, whereas the dynamics in the $B$ phase is governed by a double-sine-Gordon equation \cite{1402-4896-20-3-4-019,PhysRevB.14.118}.  Double sine-Gordon equations also emerge in the study of spin excitations in  ferromagnetic chains  \cite{PhysRevB.27.2877} and   can help to explain self-induced transparency in dielectrics \cite{DUCKWORTH197619}. Double sine-Gordon models have been used to simulate phase transitions in \chem{K_2SeO_4} and \chem{(NH_4)_2BeF_4} crystals \cite{0022-3719-16-14-009,0022-3719-16-14-010}, to describe domain wall formation in ferroelectric crystals \cite{PhysRevB.29.7082,PhysRevB.30.5306,PhysRevB.31.4633} and to model fluxon dynamics in Josephson junctions \cite{Hatakenaka2000563}. Moreover, diffraction patterns of crystal surfaces have been successfully reproduced \cite{PhysRevLett.58.2762} by double-sine-Gordon soliton-like atomic arrangements. Double-sine-Gordon models have also been proposed  \cite{Uchiyama:1976cz} to describe the quark-confinement mechanism.  

Finally, another motivation for including a sine-Gordon type potential comes from the fact that unlike kinks and antikinks in the double-well potential sine-Gordon kinks pass through each other when colliding. Hence, the addition of a sine-Gordon potential term may not only result in the fusion or fission of solitons but also give rise to elastic soliton scattering. Elastic properties are relevant when for example modelling nuclear collisions by soliton scattering \cite{Manton:2011mi,Foster:2015cpa}.

 In the recent publication  \cite{Haberichter:2015xga}, it was shown that combining an asymmetric double well potential and double-sine-Gordon potential together with a nonlinear interaction potential which prevents double well kinks and sine-Gordon kinks from passing through each other  (but whose specific form is not crucial and other choices are possible \cite{Dupuis:2015fza}) can result in the formation of novel soliton configurations, esp. \emph{false} domain walls. False domain walls interpolate between distinct false vacua, with true vacuum trapped in the core of the domain wall. Here, the double-sine Gordon field acts as a \emph{shepherd} field herding the solitons of the double well potential (the \emph{sheep}) so that they are bunched together.The shepherd field is unstable to quantum tunnelling to its true vacuum; once this occurs, the sheep are released and will spread out to infinity. In Ref.~\cite{Haberichter:2015xga}, the amplitude for such a decay has been calculated. In this article, we want to extend the analysis of Ref.~\cite{Haberichter:2015xga} and explore numerically the collision phenomenology of soliton solutions in this model in detail. The model investigated here may be artificial and not particularly realistic, but it does allow us to study the effect of solitons on vacuum stability and the confinement of solitons in a simple 1+1 dimensional setting.

The interaction properties of double-sine-Gordon solitons have been investigated e.g. in Refs.~\cite{Campbell:1986nu,Riazi:1998dw,1402-4896-20-3-4-019,Gani:1998jb,Peyravi:2009,Popov:2014sma}. As for the $\phi^4$ potential case, there is a system of resonance windows in the range of initial kink velocities. Note that collisions between sine-Gordon kinks have been recently studied \cite{:/content/aip/journal/jmp/56/9/10.1063/1.4928927} in a very similar setting \cite{133778} to ours. In particular, the authors of Ref.~\cite{:/content/aip/journal/jmp/56/9/10.1063/1.4928927}  carried out a systematic analysis of the scattering dynamics of kinks connecting true and false vacuum minima in a system of three coupled long Josephson junctions. However, in their model kinks interpolating between true and false vacuum states arise as solutions of a set of two coupled sine-Gordon equations. Interactions of chiral domain walls with properties resembling those found in our model have been analyzed in Refs.~\cite{PhysRevE.75.026604,PhysRevE.75.026605} in the nonlinear Schr\"odinger (NLS) equation using variational methods and direct numerical simulations. The authors constructed ``solitonic bubbles'', i.e. stationary tightly bound states formed by  stable and unstable domain walls. In particular, they found a (unstable) tightly bound soliton complex (compare Fig.~13 in Ref.~\cite{PhysRevE.75.026604}) that bears some ressemblance to the domain walls considered in this article. Finally, there exists a vast literature on kink interactions in multi-component NLS equations. For example pulse propagation in nonlinear optical fibers can be described by a system of coupled NLS equations \cite{Crosignani:81,Crosignani:82} and scattering of bright solitons in Bose-Einstein condensates can be modelled by a system of two coupled 2D NLS equations \cite{PhysRevA.76.013606}.

In this article, our objective is to consider the dynamics of kinks connecting local potential minima of different depths in a model built from an asymmetric double well and (double) sine-Gordon potential together with an interaction term.

This article is structured as follows. In section~\ref{Sec1}, we briefly review some of the features of the domain wall model introduced recently in Ref.~\cite{Haberichter:2015xga}. In section~\ref{Sec_Scat}, we present our numerical results of various kink collisions and analyze in detail the observed scattering behaviour. Finally, we summarize our conclusions in section~\ref{Sec_Con}. In the appendix, we discuss a  point-particle approximation which can be used to mimic some of the features of the complicated scattering dynamics in this model. Note that a point particle approximation very much similar to ours has been previously discussed in Salmi and Sutcliffe's work \cite{Salmi:2015wvi} on skyrmion solutions in the lightly bound baby Skyrme model. Their motivation for studying the dynamics of baby Skyrmions \cite{Piette:1994ug,Foster:2009vk,Ashcroft:2015jwa} is that it is a toy model for Skyrmions \cite{Skyrme:1961vq,Skyrme:1962vh}, a soliton model for nuclear physics. Although our kink model is completely unrelated to the Skyrme model, we find that similar particle approximations work and hence our model reproduces features of the lightly bound \cite{Gillard:2015eia,Gudnason:2016mms,Gudnason:2016cdo,Salmi:2014hsa} and conventional Skyrme models. This gives us another motivation to study kink dynamics in this $1+1$ dimensional field theory.

\section{The Model}\label{Sec1}
We consider the model \cite{Haberichter:2015xga}  of two real scalar fields $\phi$ and $\psi$ defined by the Lagrangian density
\beq\label{Lag_false}
{\cal L}= {\frac{1}{2}}{\left(\partial_\mu\psi\partial^\mu\psi +\partial_\mu\phi\partial^\mu\phi\right)} -V(\psi,\phi)\,,
\eeq
where $V(\psi,\phi)$ is a scalar potential given by
\beq
V(\psi,\phi)=V_\psi(\psi)+V_\phi(\phi)+V_{\psi\phi}(\psi,\phi)-V_0\,,
\label{Pot_tot}
\eeq
with the individual potential terms
\begin{align}
 V_\phi(\phi)&=\alpha\Big(\sin^2(\pi\phi)+\epsilon_\phi\sin^2(\pi\phi/2)\Big)\label{Vphi2}\,,\\
V_\psi(\psi)&=\beta (\psi+a)^2\Big((\psi-a)^2+\epsilon_\psi^2\Big)\label{Vpsi}\,,
\end{align}
and the interaction potential
\begin{align}\label{Vint}
V_{\psi\phi}(\psi,\phi)&=\lambda\frac{(\psi-a)^2\Big((\psi+a)^2+\epsilon_\psi^2\Big)}{\Big(V_\phi(\phi)-V_\phi(1/2)\Big)^2+\gamma^2}\,.
\end{align}

The model (\ref{Lag_false}) admits kink-type soliton solutions. For vanishing asymmetry parameters ($\epsilon_\phi=\epsilon_\psi=0$) and vanishing coupling constant $\lambda$, there are essentially two different types of solitons: $\phi$ is a sine-Gordon field and the $\psi$ field describes kink solutions connecting the two different vacua $\psi_{\text{vac}}=\pm a$ of the double-well potential. Switching on the asymmetry parameters $\epsilon_\phi$ and $\epsilon_\psi$, gives rise to a set of discrete vacua of different energy. 

In Ref.~\cite{Haberichter:2015xga}, we considered the special case of domain walls interpolating between two distinct false vacua, with true vacuum in the core of the domain wall. Such ``false" domain walls can be constructed by trapping the solitons of the $\phi$ field, which we call sheep, within the solitons of the $\psi$ field, the shepherd. The sheep field $\phi$ is in its false vacuum outside the domain wall and in its true vacuum inside the domain wall. These sheep are prevented from spreading out to infinity by the shepherd field $\psi$. The $\psi$ field is in its true vacuum outside the domain wall, but in its false vacuum inside the domain wall. For non-zero coupling constant $\lambda$, the interaction potential $V_{\psi\phi}(\psi,\phi)$ provides an energy barrier which prevents $\phi$ solitons from passing through the $\psi$ solitons. For numerical calculations, we introduce in (\ref{Vint}) a small, non-zero parameter $\gamma$ to ensure that the contribution to the energy of the interaction term remains finite. We refer the interested reader to Ref.~\cite{Haberichter:2015xga} for a detailed discussion of ``false'' domain walls in model (\ref{Lag_false}) and their decay via quantum tunnelling.

In this article, we want to focus on the dynamics of domain walls interpolating between the distinct discrete vacua of model (\ref{Lag_false}). We choose the following set of parameters:
\begin{align}\label{Para}
\alpha=0.5\,,\quad \beta=0.5\,,\quad \gamma=0.01\,,\quad a=1\,,\quad \epsilon_\psi=1\,,\quad \epsilon_\phi =0.01\, ,\quad \lambda=0.1\,.
\end{align}

\begin{figure}[!htb]
\subfigure[\, ]{\includegraphics[totalheight=6.0cm]{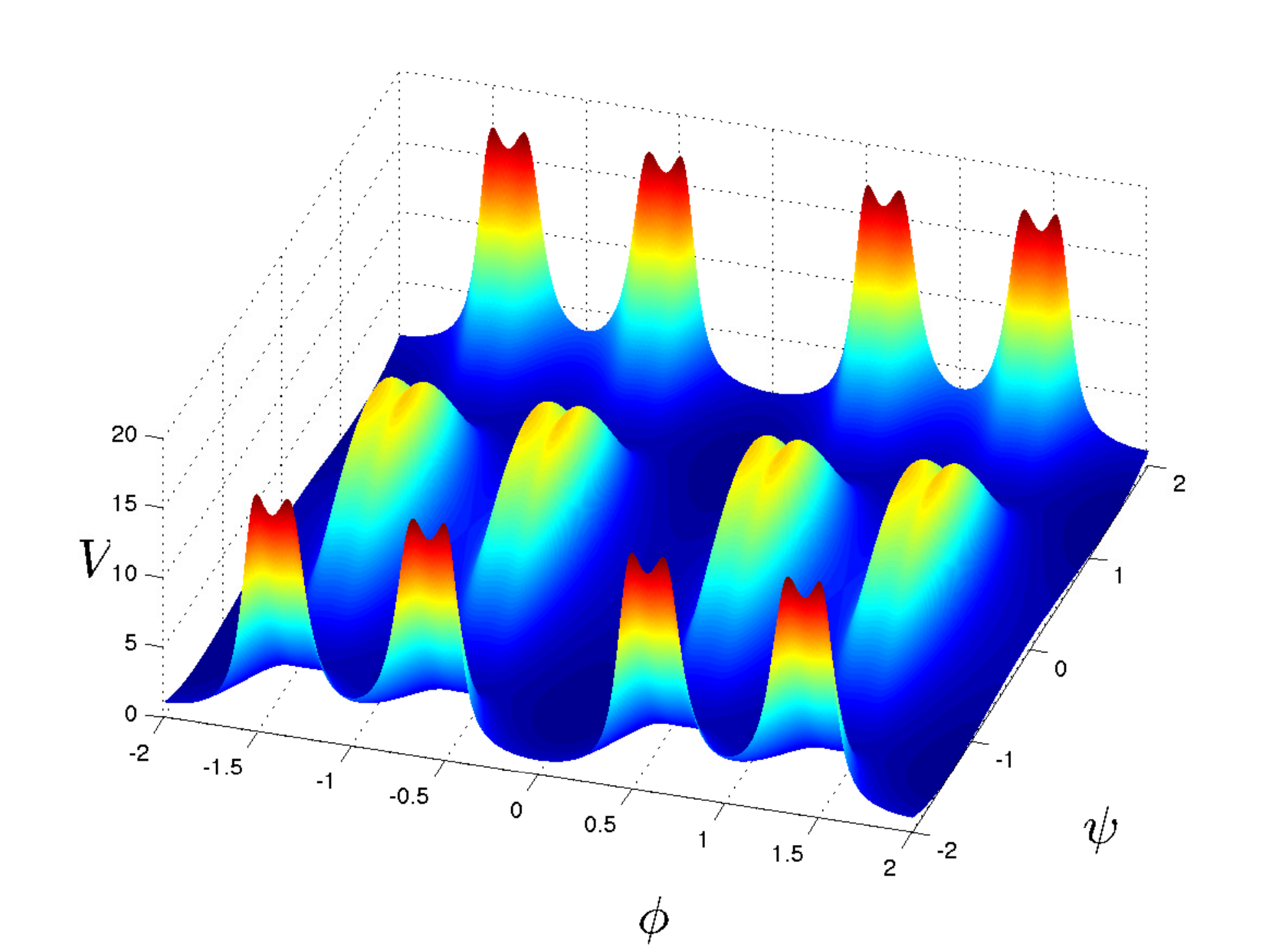}}
\subfigure[\, ]{\includegraphics[totalheight=6.0cm]{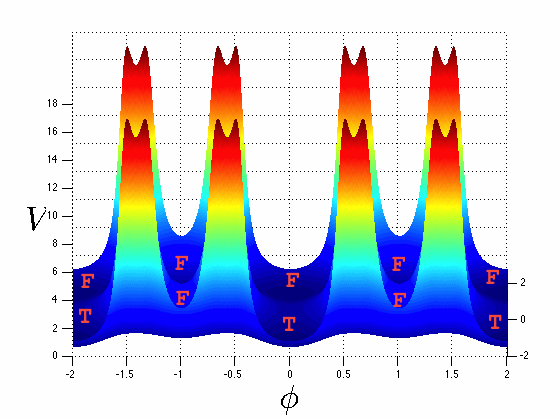}}
\subfigure[\, ]{\includegraphics[totalheight=8.0cm]{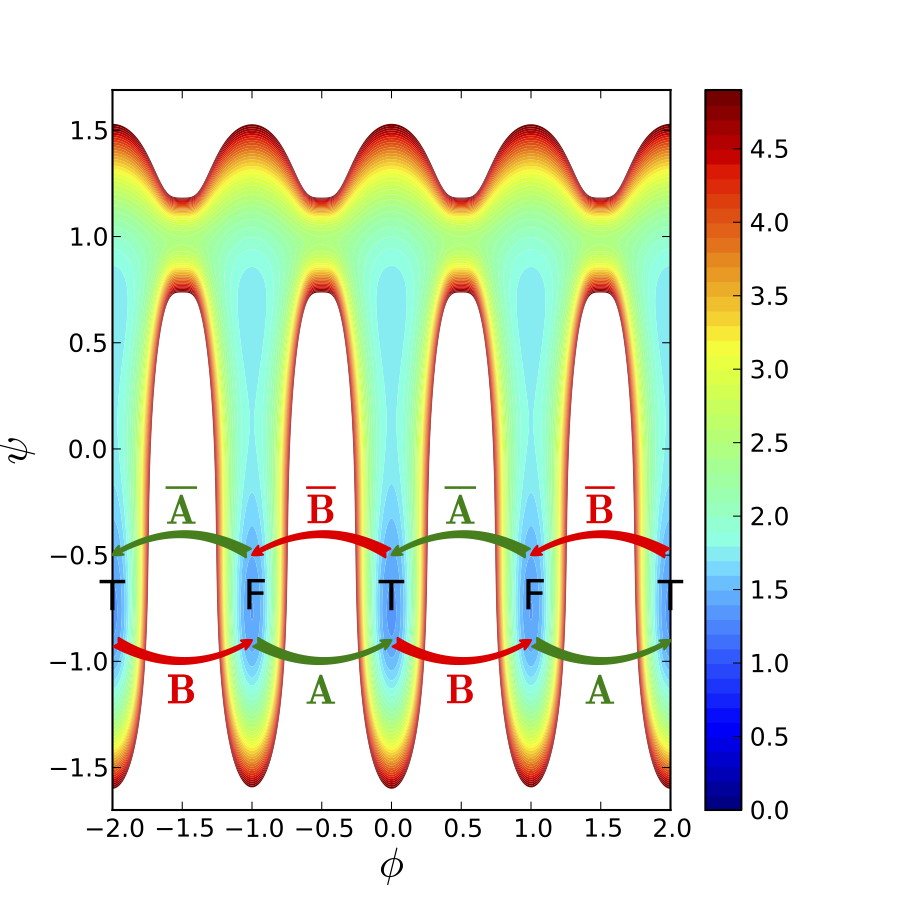}}
\caption{(a), (b) Surface plots for two different viewing angles and (c) contour plot of the full scalar potential~(\ref{Pot_tot}) for the parameter choice (\ref{Para}), but with $\gamma$ set to $0.1$ for illustrative purposes. Here, the true (T) vacua correspond to $\{\psi=-0.7593,\phi=2k\}$, while false vacua (F) occur at $\{\psi=-0.7552,\phi=2k+1\}$. The white regions in the contour plot indicate the insurmountable energy barriers separating true and false vacua. The green and red arrows illustrate the kinks we call of type $A$ and $B$, respectively. Note that the vacua at $\{\psi=0.6396, \phi=2k \}$ and $\{\psi=0.6463,~\phi=2k+1 \}$ (see (b)) are all raised in energy and hence they are false vacua.}
\label{Pot_dens}
\end{figure}

 For the parameter choice (\ref{Para}), the density plots of the full scalar potential (\ref{Pot_tot}) in Fig.~\ref{Pot_dens} visualize the complicated vacuum structure of the model. The true minima of $V\left(\psi,\phi\right)$ occur at even integers $\phi=2k$ for any integer $k$, while the false vacua correspond to odd integers $\phi=2k+1$. Note that due to the non-zero coupling $\lambda$ the $\psi$ field takes the vacuum value $\psi=-0.7593$ at even integers $\phi=2k$ and $\psi=-0.7552$ at odd integers $\phi=2k+1$. True and false vacua are separated by insurmountable potential barriers indicated by the white coloured regions in Fig.~\ref{Pot_dens} (b). Note that there also occur vacua at $\{\psi=0.6396, \phi=2k \}$ and $\{\psi=0.6463,~\phi=2k+1 \}$. However all these vacua are raised in energy and hence they are false vacua.

\begin{figure}[!htb]
\subfigure[\, ]{\includegraphics[totalheight=8.0cm]{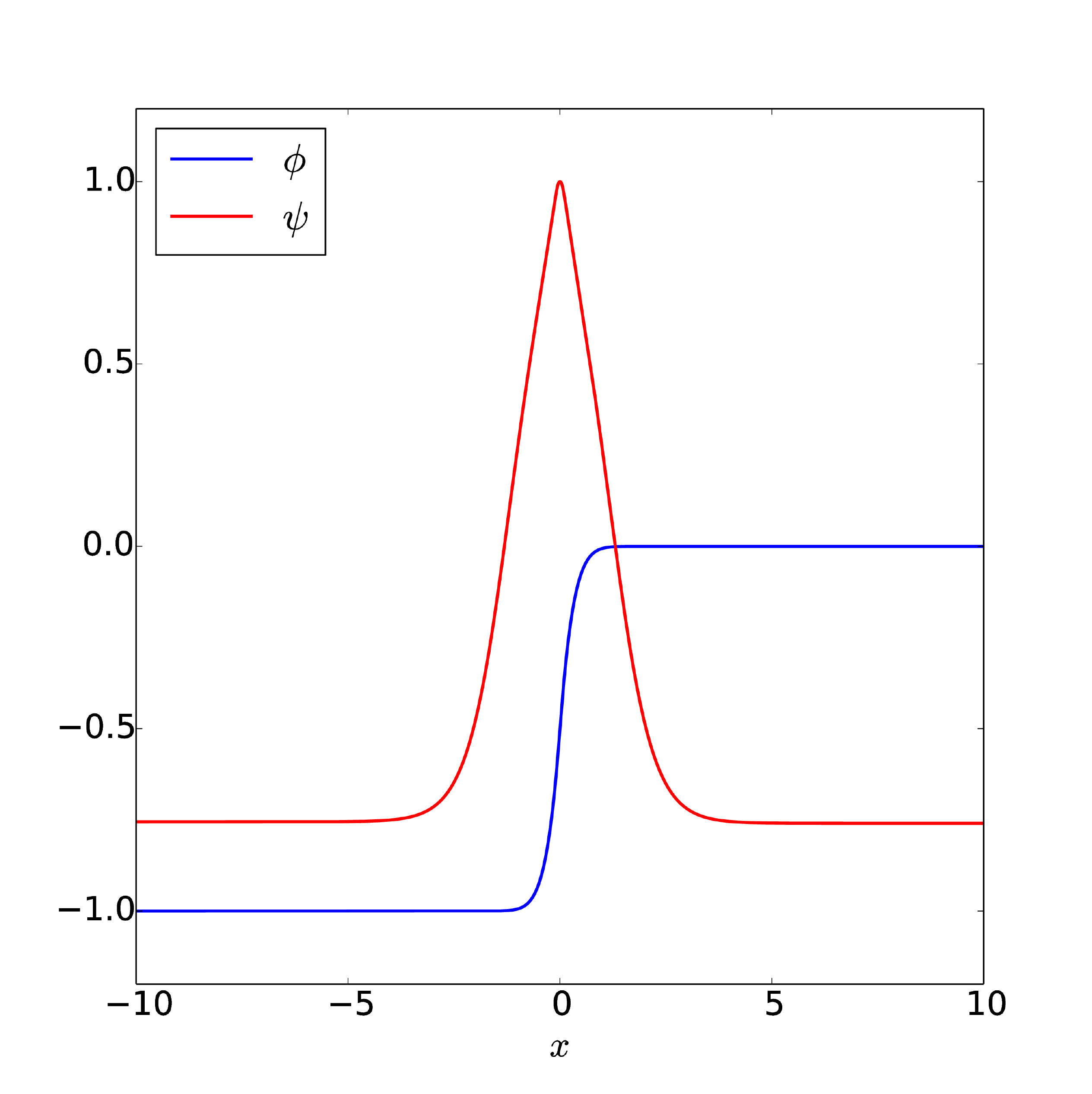}}
\subfigure[\, ]{\includegraphics[totalheight=8.0cm]{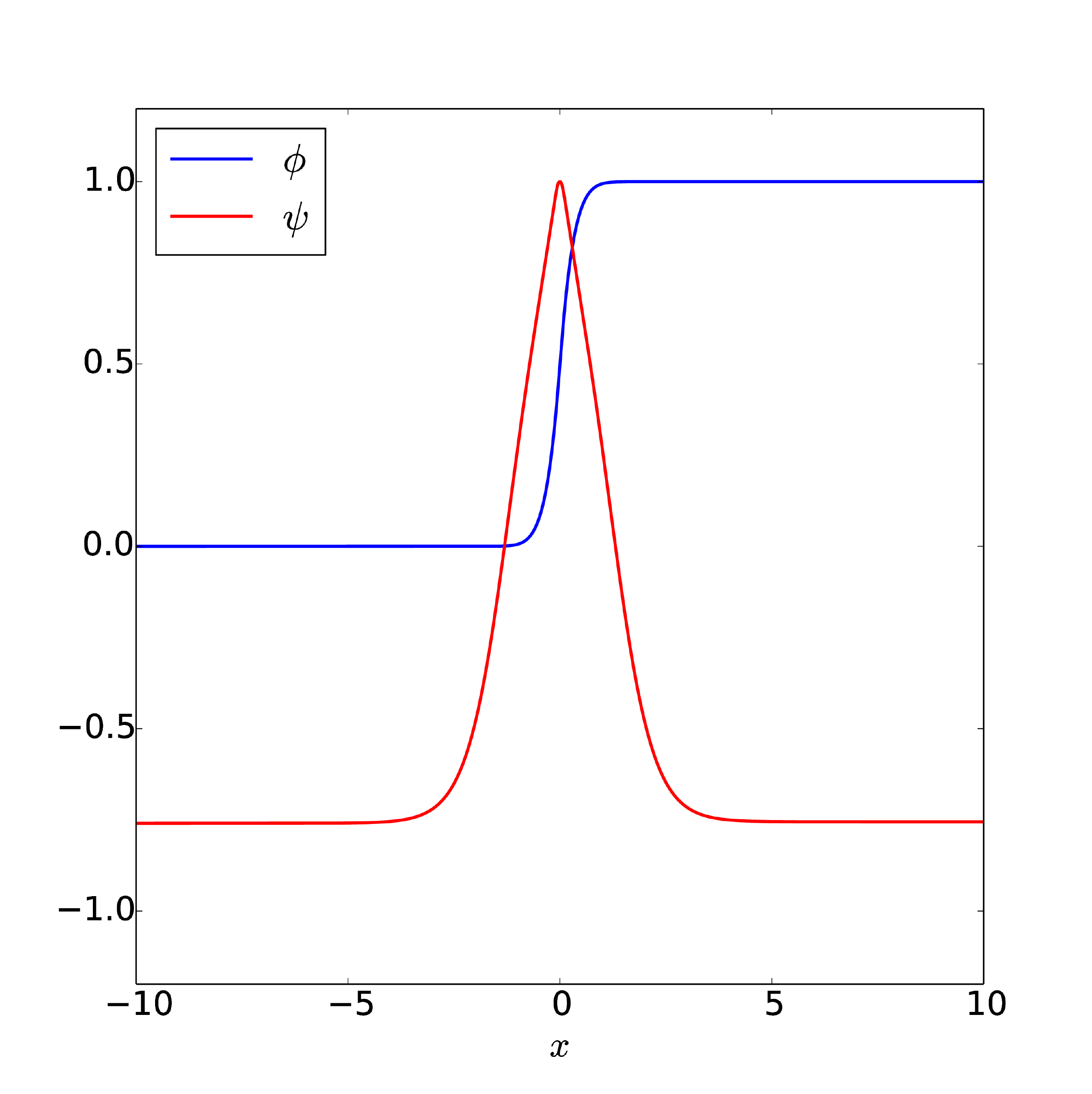}}
\caption{Examples of kink configurations we denote by (a) A and (b) $B$, respectively. Here, the graph displayed in red represents the shepherd field $\psi$ and the blue graph is called the sheep field $\phi$. Recall that in our convention a kink of type $A$ interpolates from false vacuum  $\{\psi=-0.7552,\phi=2k+1\}$ to true vacuum $\{\psi=-0.7593,\phi=2k\}$ as $x$ goes from minus to plus infinity, see Fig.~(a). Kinks passing from true to false vacuum as $x$ increases are denoted by $B$, see Fig.~(b).}
\label{SheepA_B}
\end{figure}

Throughout this paper, kinks interpolating from false vacuum to true vacuum, e.g. $\phi=-1\rightarrow\phi=0$, as $x$ goes from minus to plus infinity are labelled by $A$ and kinks passing from true vacuum to false vacuum, e.g. $\phi=0\rightarrow\phi=+1$, are denoted by $B$. Kinks in the reverse direction are called $\overline{A}$ and $\overline{B}$, respectively. That is, $\overline{A}$ interpolates from false vacuum  to true vacuum $\phi=+1\rightarrow\phi=0$, whereas $\overline{B}$ passes from true vacuum to false vacuum $\phi=0\rightarrow\phi=-1$. Hence in our notation, $A$ and $\overline{A}$ connect true to false vacua and $B$ and $\overline{B}$ link false to true vacua, see Fig.~\ref{Pot_dens} (b). We display in Fig.~\ref{SheepA_B} examples of  kinks labelled by $A$ and $B$, respectively. To obtain solutions such as those shown in Fig.~\ref{SheepA_B}, we first apply a numerical relaxation method. We construct minimal energy solutions in the model (\ref{Lag_false}) by solving the gradient flow equations with a crude initial guess (a straight line approximation for the sheep field $\phi$ and an approximation in terms of hyperbolic tangents for the shepherd field $\psi$). The field equations have been given explicitly in Ref.~\cite{Haberichter:2015xga}. The boundary conditions for kinks of type $A$ and $B$ are chosen as stated above. Due to the imbalance in vacuum energies, the $A$ or $B$ kink will move away from the origin during the relaxation process in order to extend the region of true vacuum. We restore the kink's position to the centre by locating the maximum of the shepherd field $\psi$ and shifting the kink so that this is located at the origin, compare Fig.~\ref{SheepA_B}.

 Due to the different asymptotic vacuum energy values as $x\rightarrow\pm\infty$, it is not immediately obvious how to calculate the energy of an $A$ or $B$ kink. Here, to allow the reader to reproduce our results, we briefly explain our approach. For $A$ and $B$ kinks on a spatial grid with 4001 points and spacing ${\text{d}}x=0.01$ we subtract the true vacuum energy at the one side and the false vacuum energy at the other side using a Heaviside step function. This results in the kink mass $3.116$. We checked that the kink mass is not significantly affected by different grid sizes and spacings.

We create suitable initial conditions for our scattering simulations of soliton solutions in model (\ref{Lag_false}) by sticking together kinks of two different types positioned at $x=-20$ and $x=+20$ respectively. In the following, we label each kink combination by $XY$, where $X$ stands for the chosen kink type at the left hand side (negative direction of the $x$-axis) and $Y$ specifies the kink type at the right hand side (positive direction of the $x$-axis). Note that due to the structure of false and true vacua not all combinations of kinks of types $A,~B,~\overline{A}$ and $\overline{B}$ are allowed. We summarize all of the possible combinations in Table~\ref{Tab_repel_attract}. We use cross symbols whenever the structure of false and true vacua excludes a given combination.

\begin{table}[htb]
\caption{For each combination of kinks, we state whether the solitons repel or attract when they are well-separated and initially at rest. Here, the first column specifies the kink type at the left hand side (negative direction of the $x$-axis) and the first row gives the kink type at the right hand side (positive direction of the $x$-axis). The symbol ``\color{red}\tickNo\color{black}\,'' is used whenever a particular combination is excluded. \\}
\begin{tabular}{c|cccc}
\hline\hline
Left {\textbackslash} Right & $A$ & $B$ & $\overline{B}$ & $\overline{A}$\\\hline
$A$ &\color{red}\tickNo\color{black}&\, repulsive&\, repulsive&\color{red}\tickNo\color{black}\\
$B$ &\, attractive&\color{red}\tickNo\color{black}&\color{red}\tickNo\color{black}&\, attractive\\
$\overline{B}$ &\, attractive&\color{red}\tickNo\color{black}&\color{red}\tickNo\color{black}&\, attractive\\
$\overline{A}$ &\color{red}\tickNo\color{black}&\, repulsive&\, repulsive&\color{red}\tickNo\color{black}\\\hline\hline
\end{tabular}
\label{Tab_repel_attract}
\end{table}

We set up our initial conditions by linking either two true or two false vacua. For example, an $AB$ configuration can be created by sticking together a kink of type $A$, e.g. $\phi=-1\rightarrow\phi=0$, with one of type $B$, namely $\phi=0\rightarrow\phi=+1$. A configuration of type $BA$ can be constructed by linking a soliton of type $B$, e.g. $\phi=0\rightarrow\phi=+1$ with one of type $A$, given by $\phi=+1\rightarrow\phi=+2$. 
For each combination, we include in Table~\ref{Tab_repel_attract} whether the solitons repel or attract when they are well-separated and initially at rest. Solitons with regions of true vacuum in between feel a repulsive force and will accelerate away from each other leaving behind them true vacuum. Solitons with false vacuum in between them attract one another.

\section{Scattering}\label{Sec_Scat}

In this section we present and discuss the results of our numerical simulations into the scattering behaviour of the solitons introduced above. We consider combinations of the type $AB$, $BA$, $B\overline{A}$, and $A\overline{B}$. Other combinations, e.g. $\overline{A}\overline{B}$, are omitted from our discussions because their scattering behaviour is equivalent to those already in our list, and so would not provide any new information. Videos of our kink simulations have been added as supplementary material to this article and we urge the reader to view them. In the appendix, we discuss a point-particle approximation which reproduces some of the scattering outcomes observed in this section. 

Our numerical simulations use a finite difference leapfrog method. We choose the timestep $\Delta t=0.002$ and the grid spacing  $\Delta x=0.01$ throughout, and typically work with grids of 20,001 or 40,001 points. Near the boundary, we apply a damping method in each timestep by updating $\dot{\phi}$ and $\dot{\psi}$ to approach zero exponentially. This reduces radiation and reflection from the boundary. Initial conditions are created by boosting two solitons towards each other, by applying a Lorentz transformation. 
We give one kink the initial velocity $v$ and calculate the initial velocity of the other kink accordingly to remain in the centre of mass frame.

\subsection{$AB$ scattering}

\begin{figure}[!htb]
\subfigure[\, ]{\includegraphics[totalheight=4.0cm]{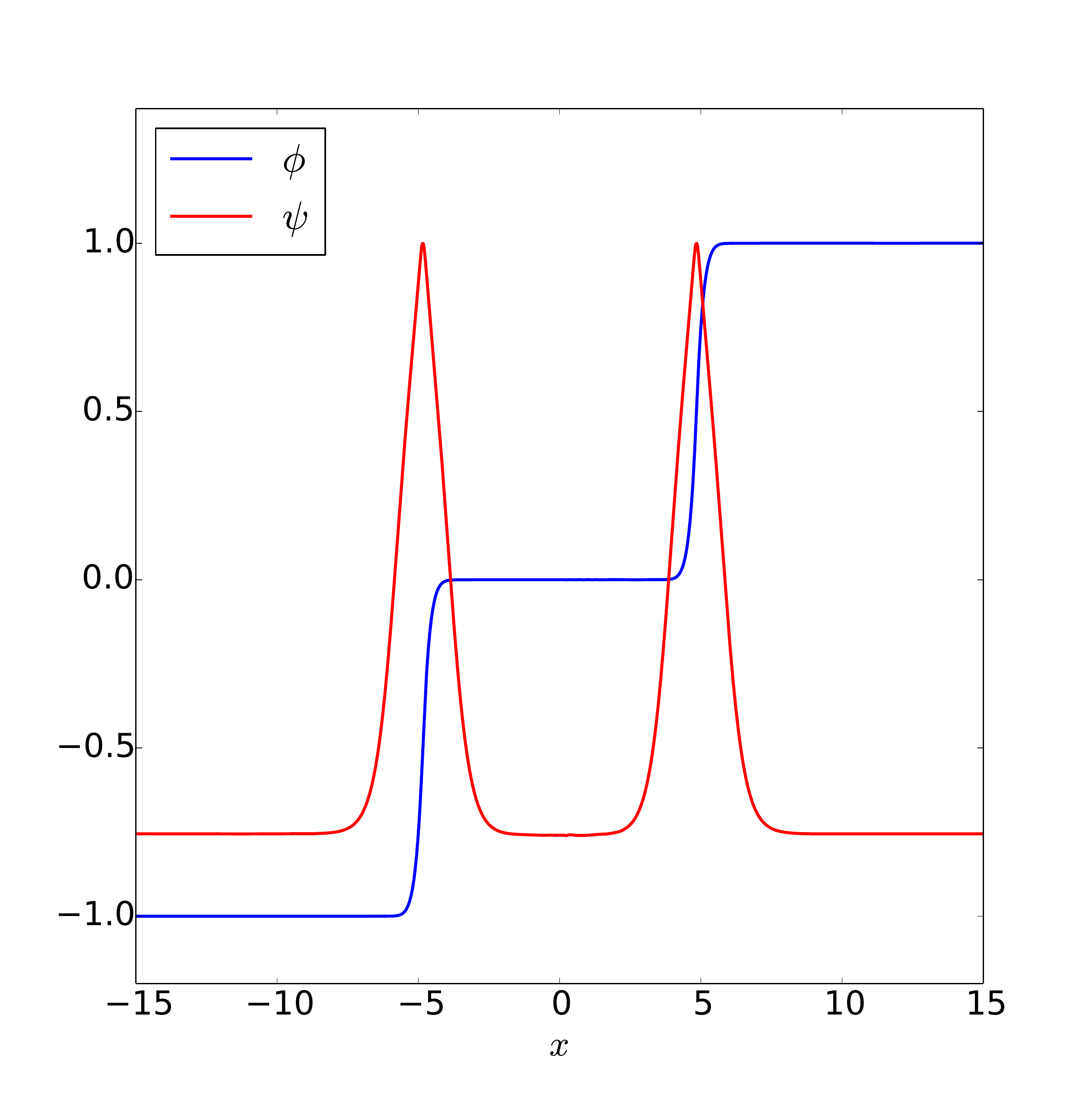}}
\subfigure[\, ]{\includegraphics[totalheight=4.0cm]{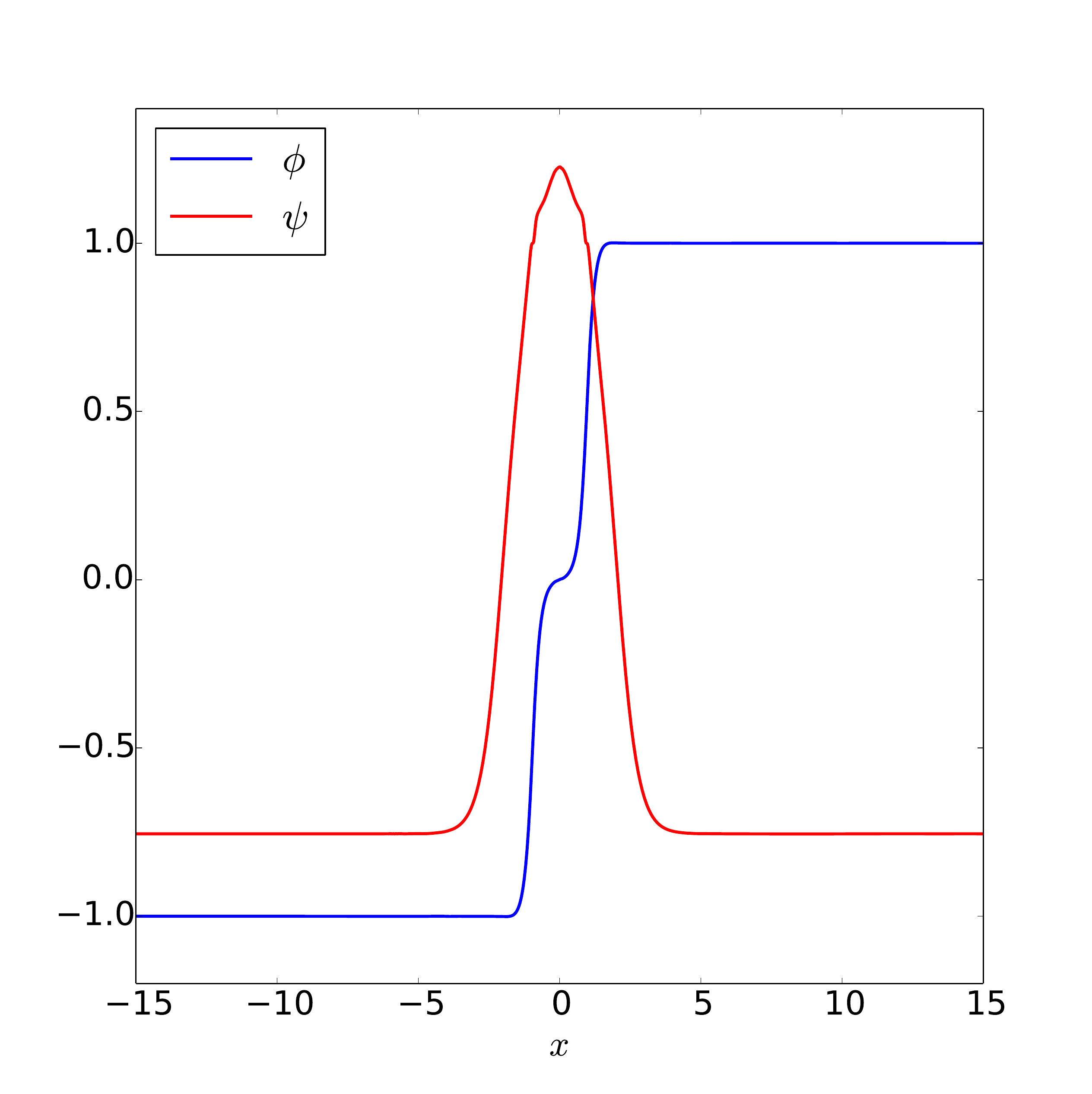}}
\subfigure[\, ]{\includegraphics[totalheight=4.0cm]{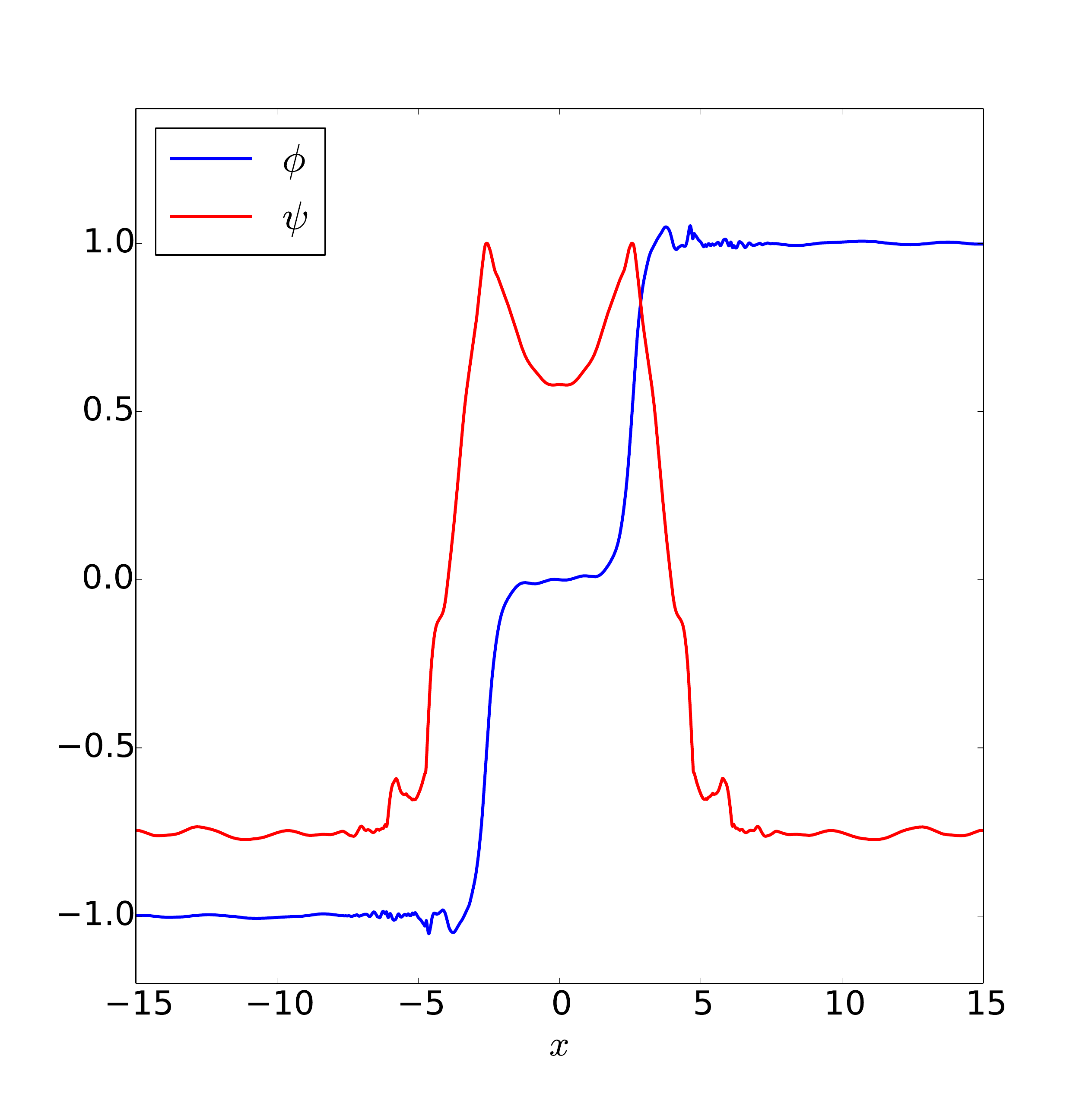}}
\subfigure[\, ]{\includegraphics[totalheight=4.0cm]{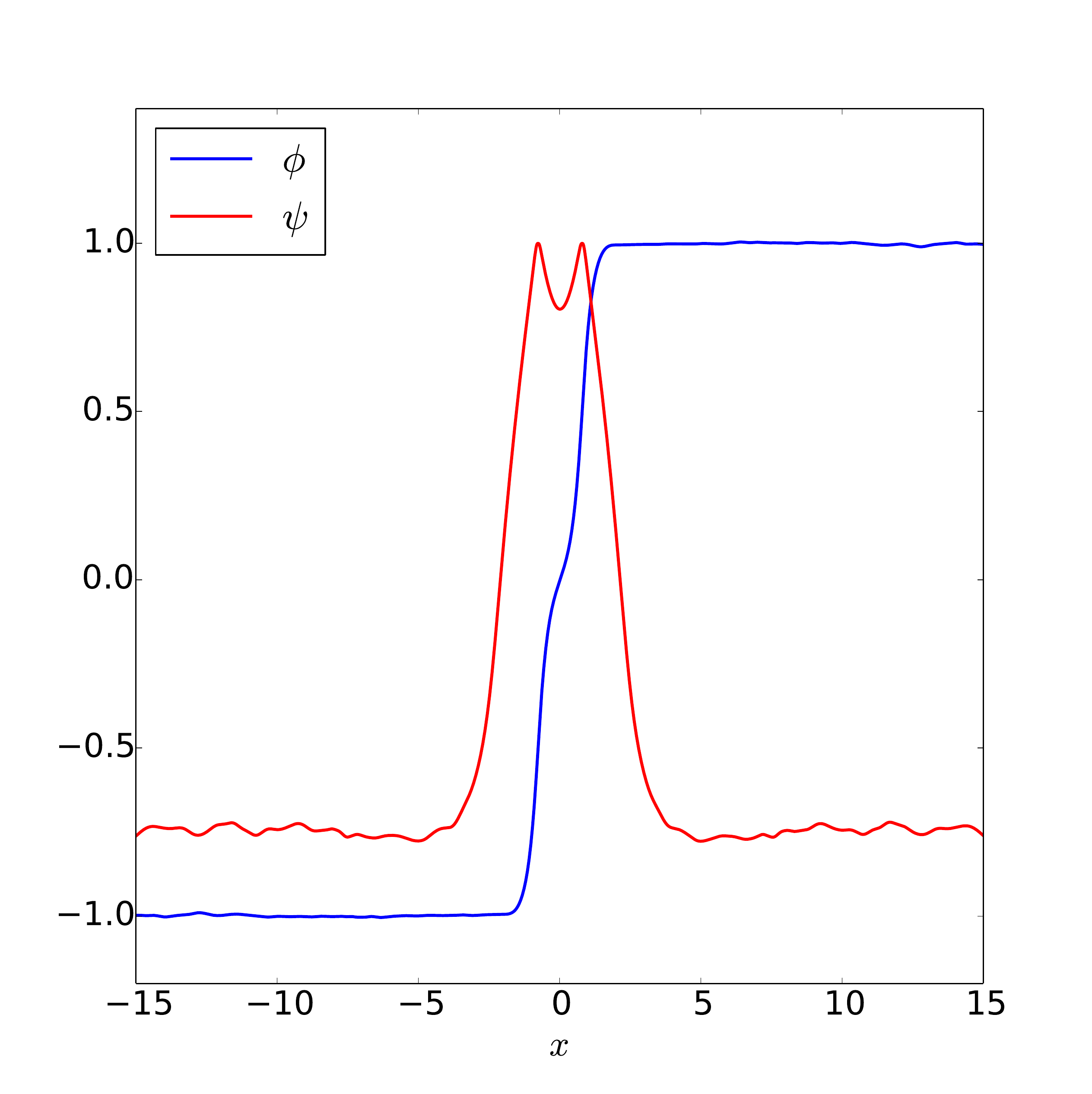}}\\
\subfigure[\, ]{\includegraphics[totalheight=4.cm]{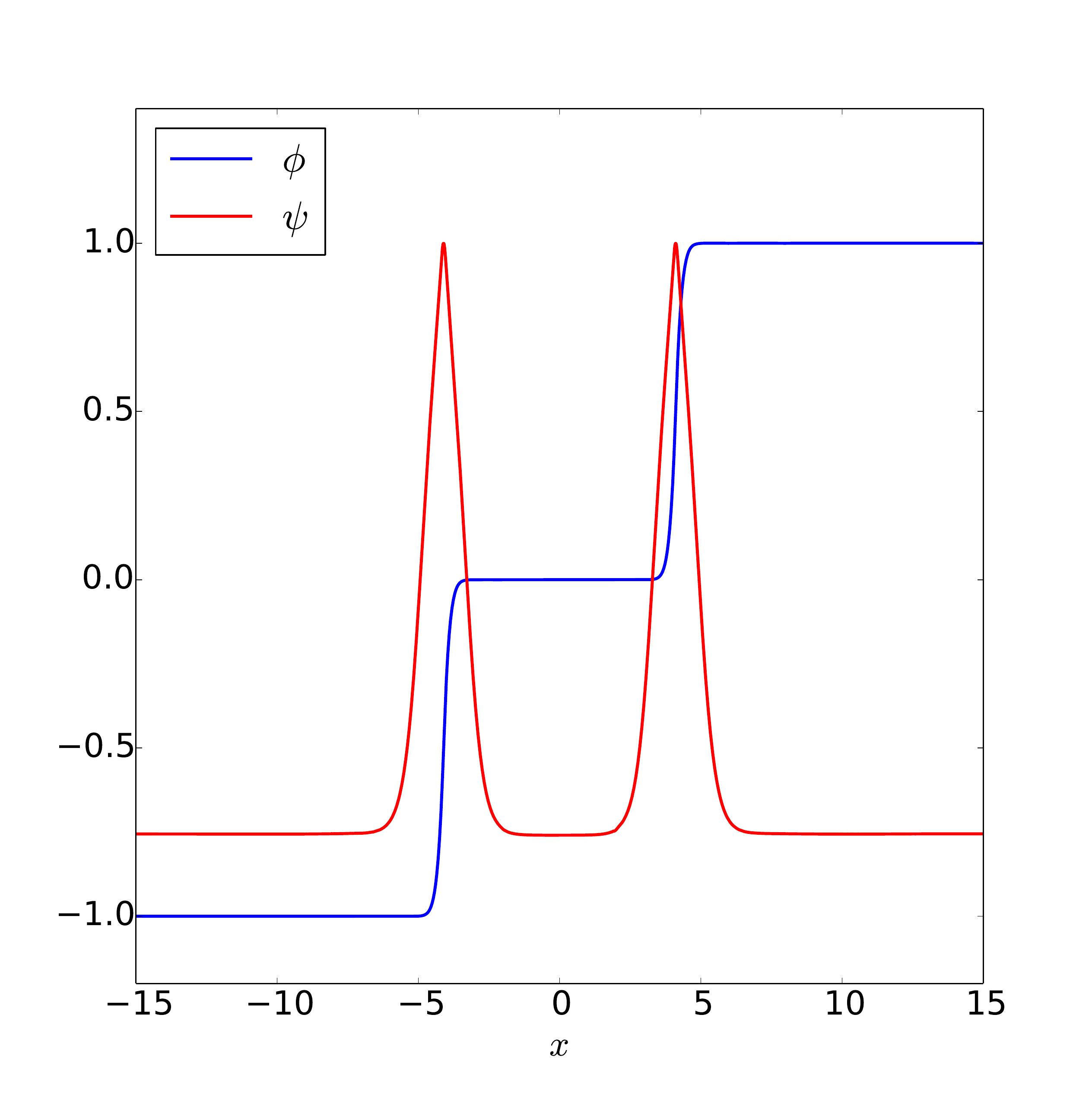}}
\subfigure[\, ]{\includegraphics[totalheight=4.cm]{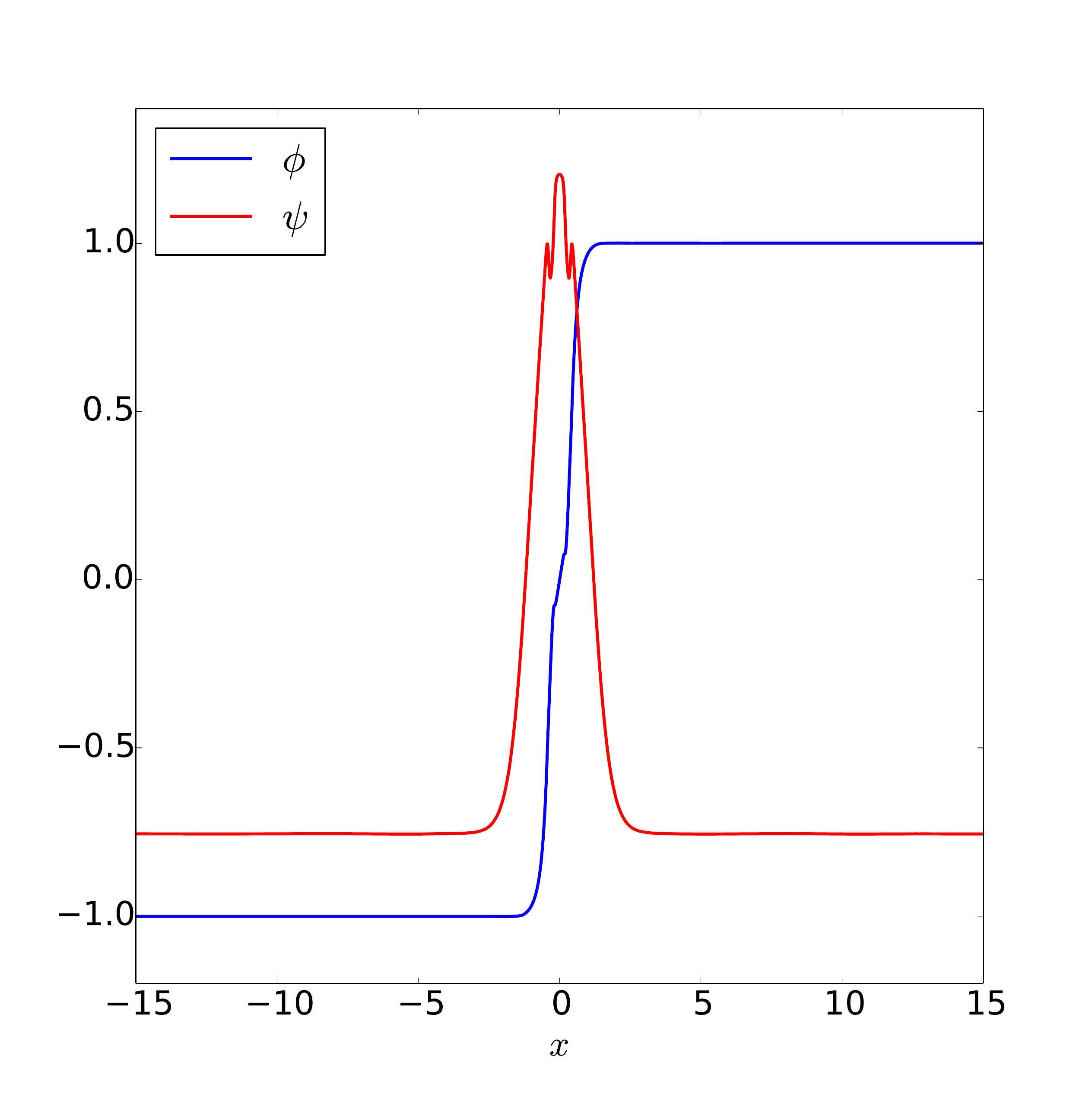}}
\subfigure[\, ]{\includegraphics[totalheight=4.cm]{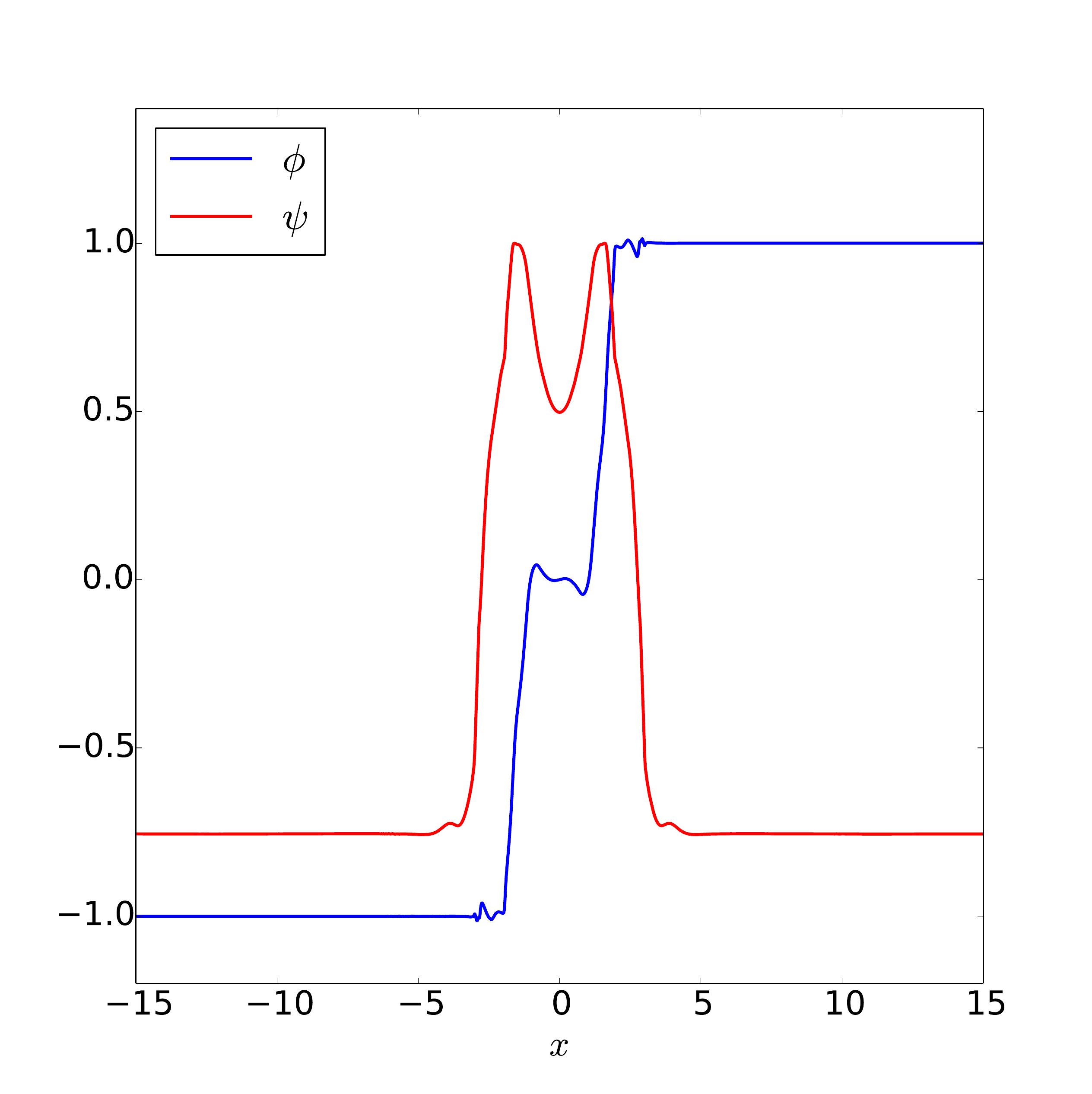}}
\subfigure[\, ]{\includegraphics[totalheight=4.cm]{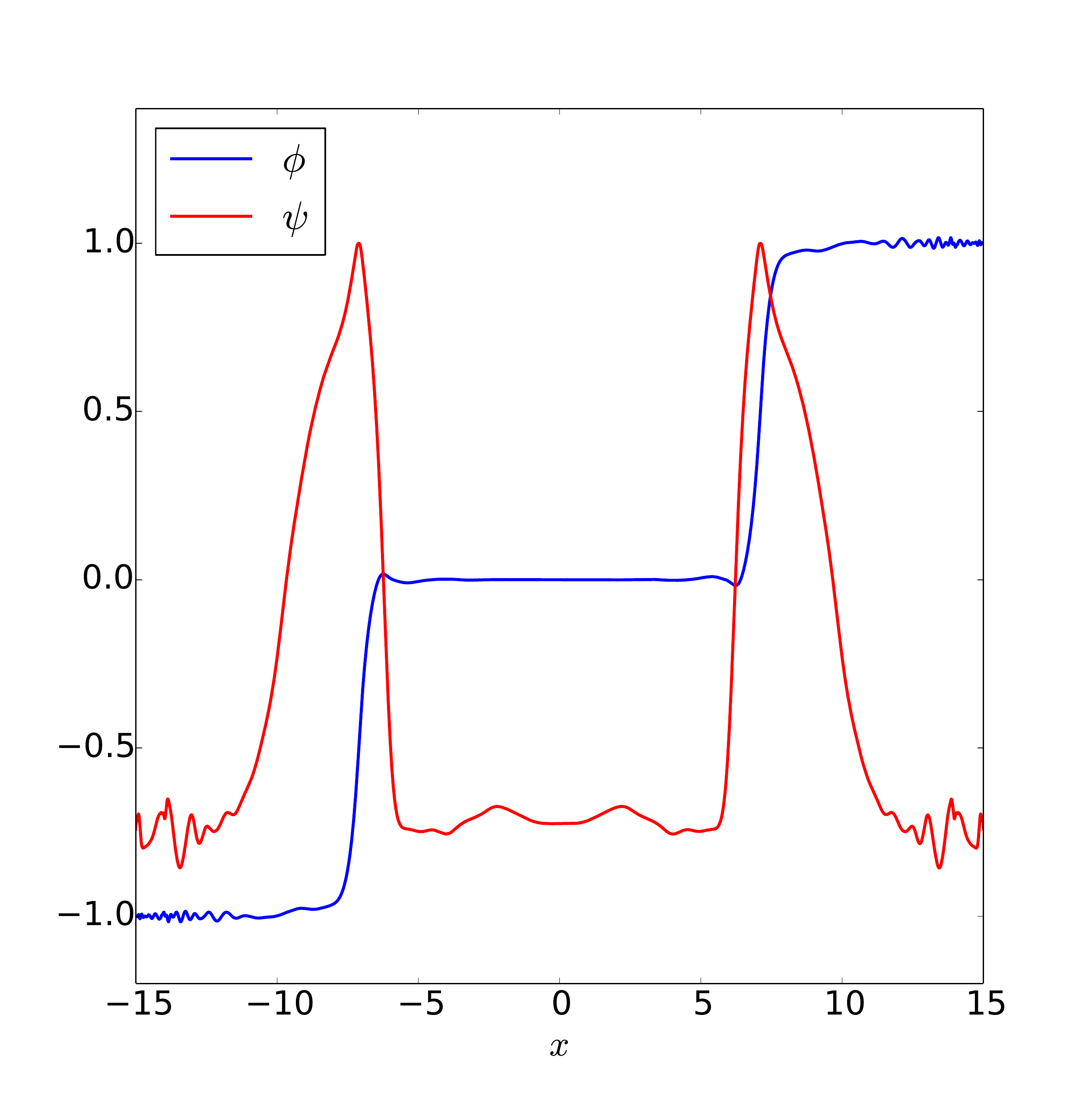}}
\caption{For kink collisions of type $AB$, we display snapshots of the field configuration for two choices of initial velocity:  (a)-(d) For $v=0.7$, the $A$ and $B$ kinks form a false domain wall \cite{Haberichter:2015xga}, that is  the $\phi$ field interpolates between distinct false vacua, with true vacuum in the core of the domain wall. (e)-(h) For $v=0.8$, kinks $A$ and $B$ escape to infinity after reflecting off each other with three bounces in the shepherd field. The corresponding contour plots are given in Fig.~\ref{AB_contour}. }
\label{AB_profile}
\end{figure}

As a first example of kink scattering in model (\ref{Lag_false}), we investigate numerically the collision of a kink of type $A$ with one of type $B$. Suitable initial conditions for this scattering process can be created by  attaching a domain wall of type $B$, e.g. $\phi=0\rightarrow\phi=+1$,  to one of type $A$, namely $\phi=-1\rightarrow\phi=0$. At the beginning of our simulation, solitons $A$ and $B$ are widely separated and are moving towards each other with an initial velocity $v$; see Fig.~\ref{AB_profile}(a) for a snapshot of the field configuration before the collision takes place. 

\begin{figure}[!htb]
\subfigure[\, ]{\includegraphics[totalheight=7.8cm]{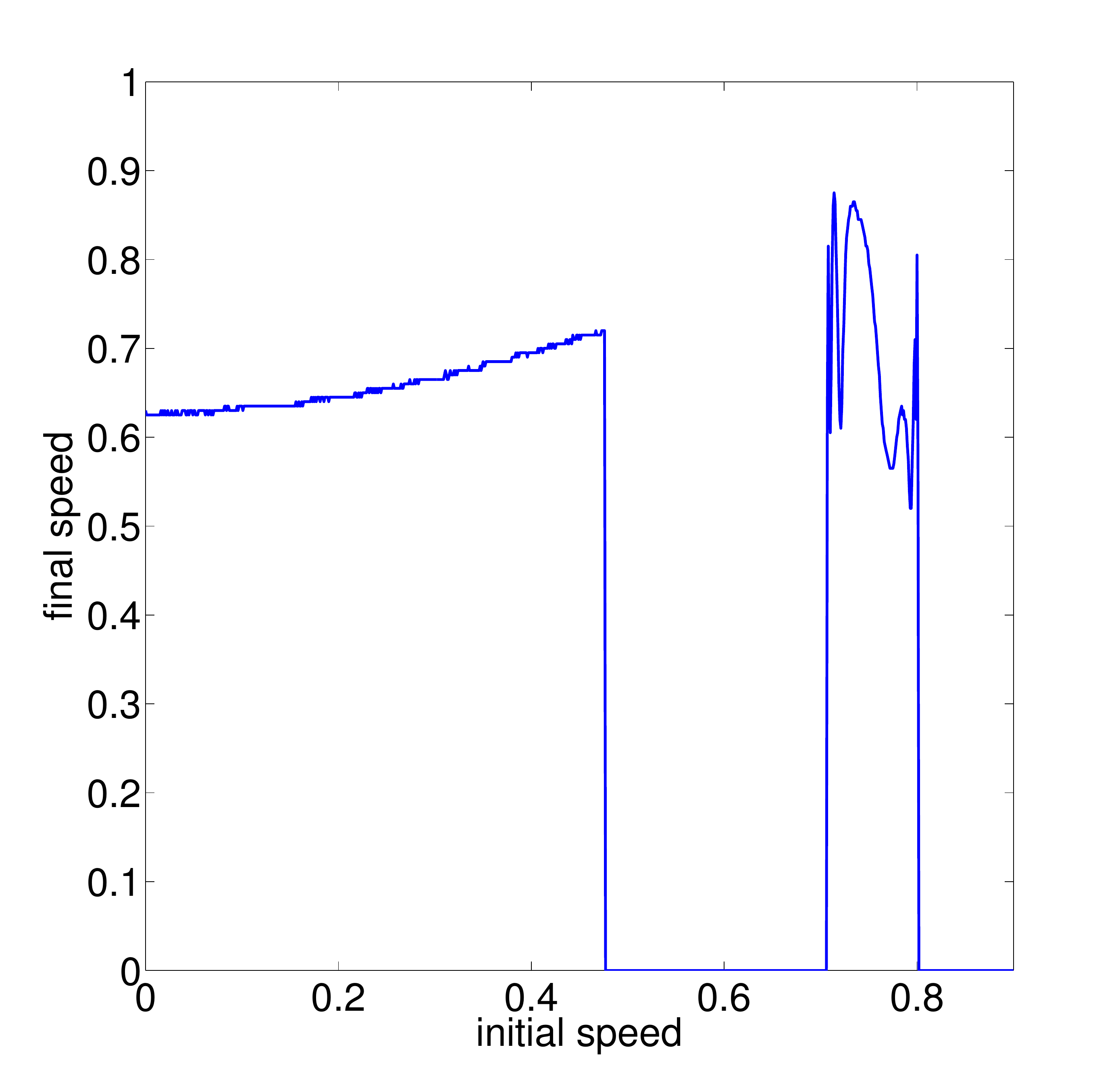}}
\subfigure[\, ]{\includegraphics[totalheight=7.8cm]{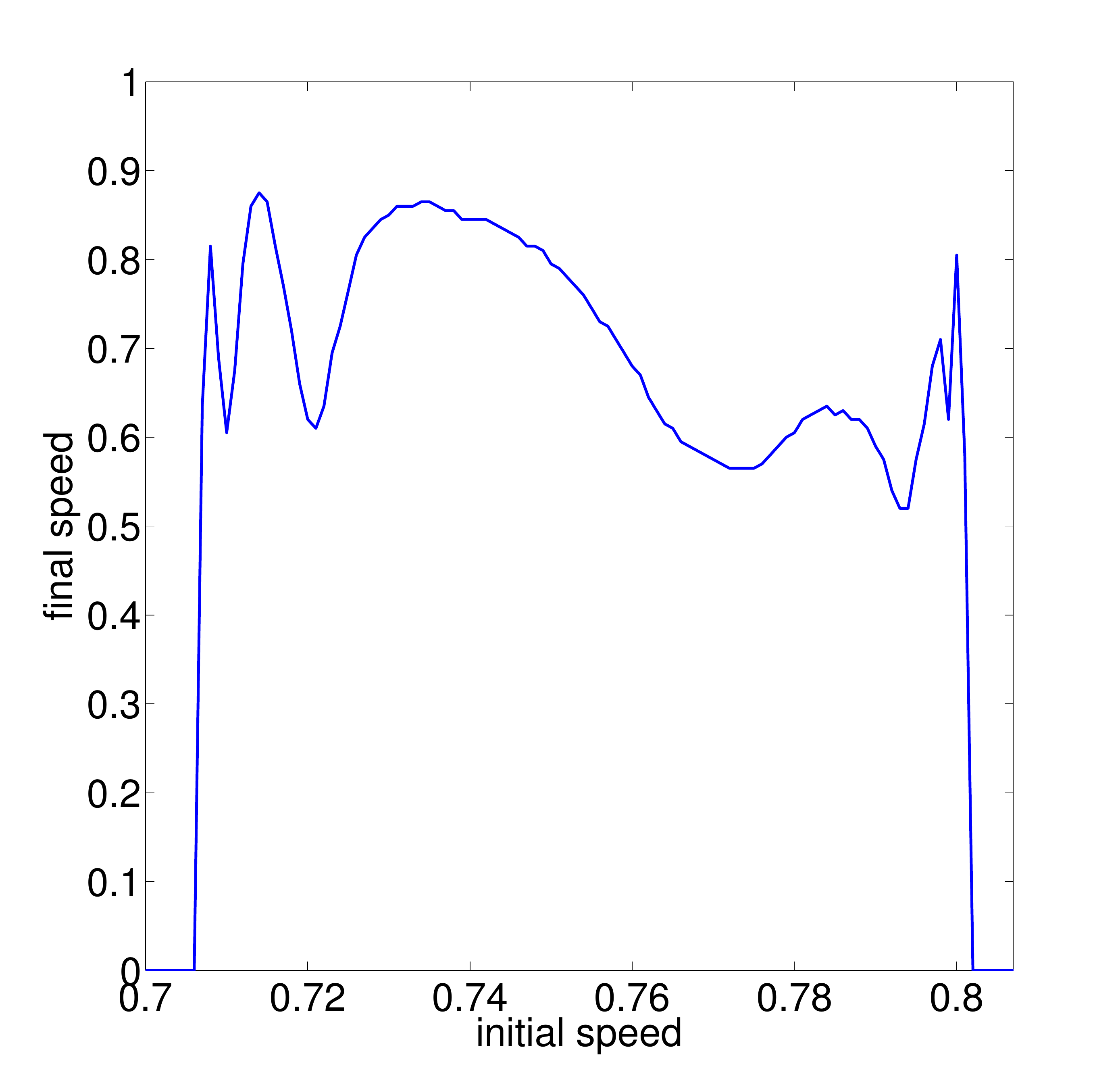}}
\caption{(a) The final velocity after a $AB$ collision as a function of the initial velocity. When the outgoing velocity is plotted to be zero, a false (classically stable) domain wall is formed. Note that false domain walls are created in certain windows in the initial velocity. (b) Zoomed version of the three-bounce window.}
\label{AB_fin}
\end{figure}

We carry out simulations for a wide range of initial velocities $0\le v\le 0.9$. Fig.~\ref{AB_fin} shows final versus initial velocity of the $AB$ kink pair, where the final velocity is calculated as the kink in the positive $x$-axis passes the point $x=50$. 
In our simulations, we observe three very different scattering outcomes: for velocities less than or equal to $0.476$, we find that kinks $A$ and $B$ always repel each other and will accelerate away from each other leaving behind them true vacuum. This is reflected in Fig.~\ref{AB_fin}(a). For $0\le v \le0.476$, there is a velocity regime, where the final velocity is always significantly higher than the initial kink velocities. Within this velocity regime, the final velocity increases with initial velocity.   

\begin{figure}[!htb]
\subfigure[\, ]{\includegraphics[totalheight=7.8cm]{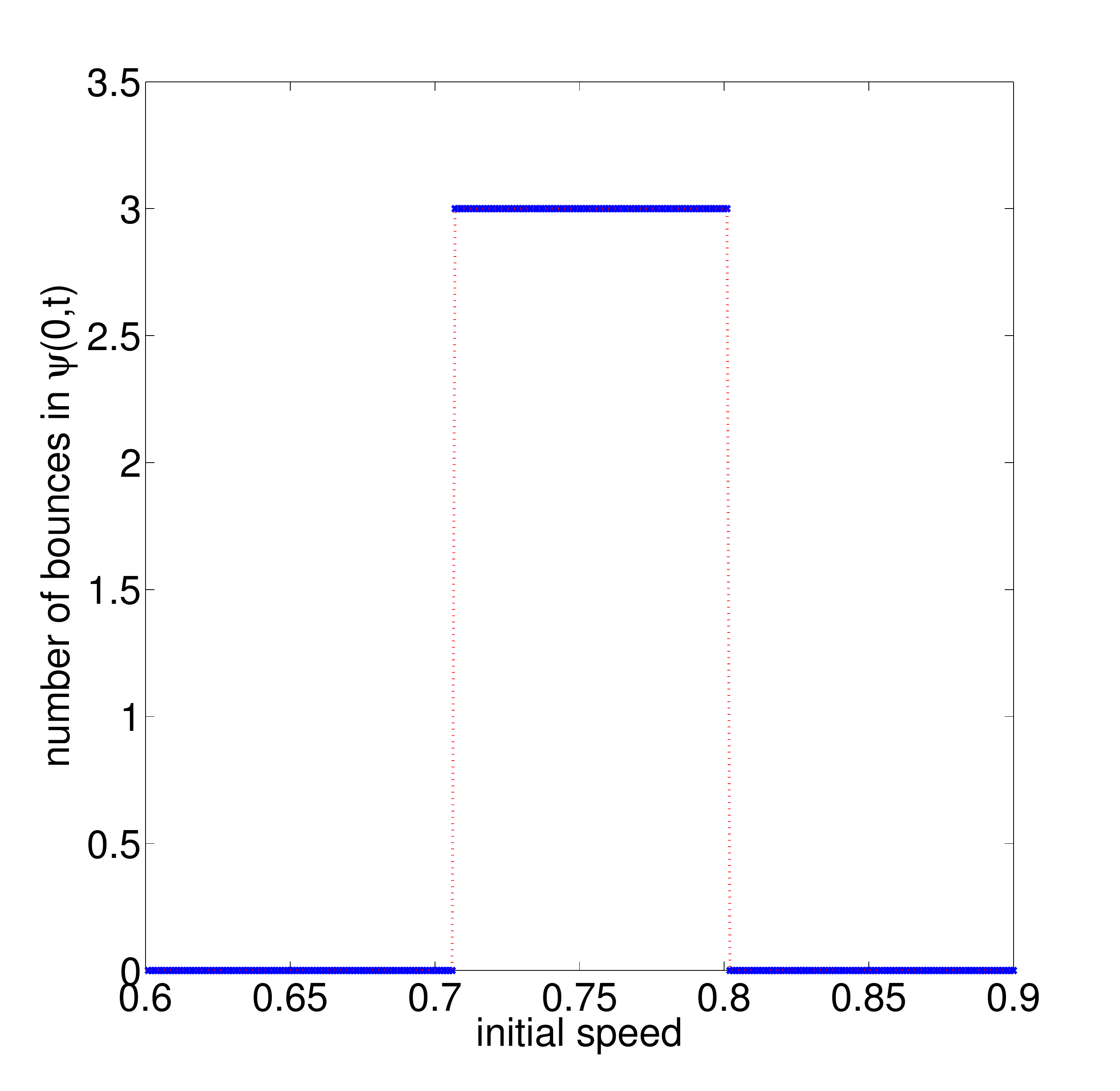}}
\subfigure[\, ]{\includegraphics[totalheight=7.8cm]{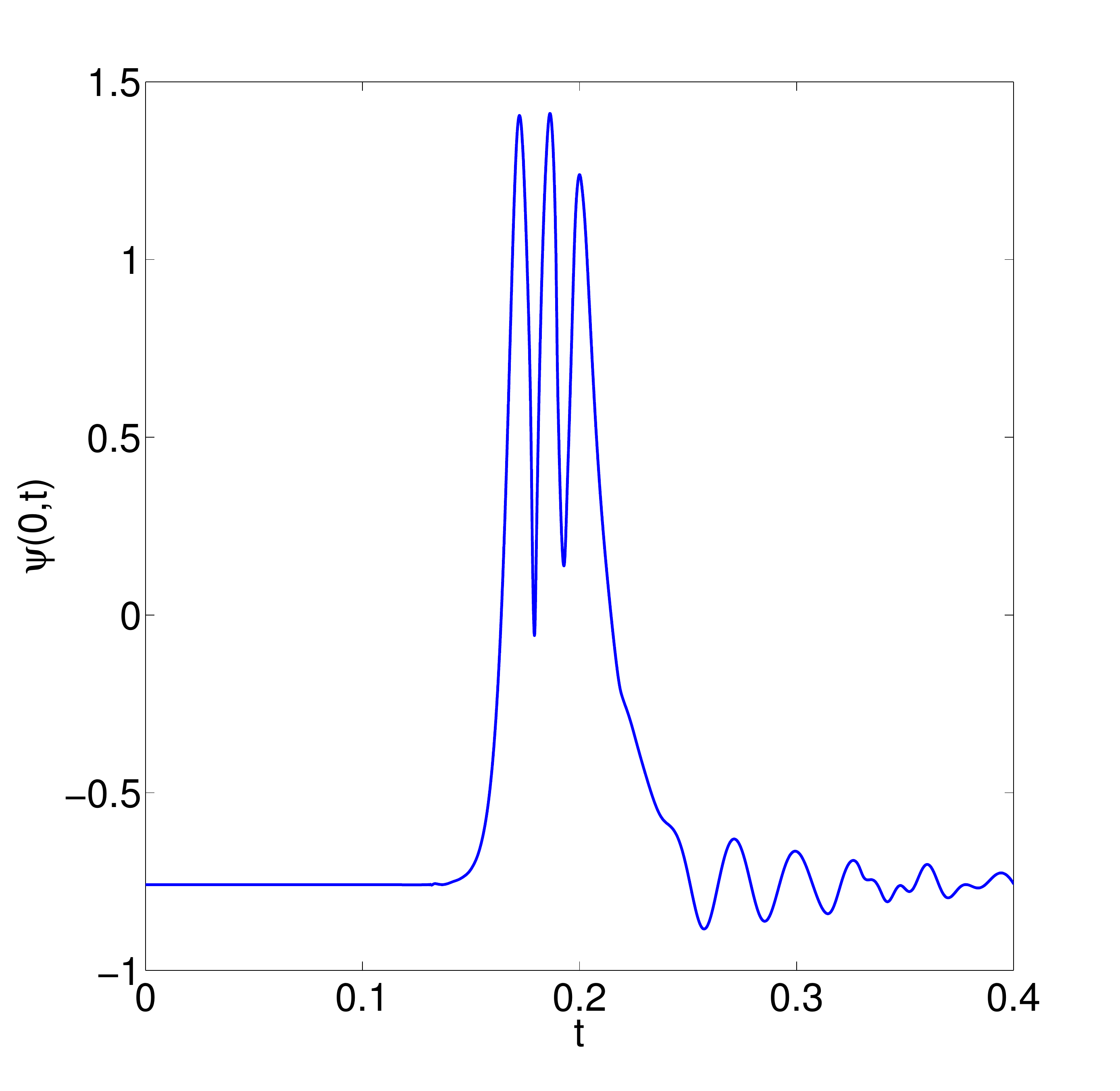}}
\caption{We display for a $AB$ collision: (a) Number of bounces vs initial velocity. Here, the bounce number measures the number of times the shepherd field $\psi$ oscillates at the origin before the kinks separate. (b) Shepherd field $\psi(0,t)$ at the center-of-mass versus time $t$ for the initial velocity $v=0.750$. The three-bounce collison is clearly visible. For reasons of clarity, time is divided by 100.}
\label{AB_bounce}
\end{figure}

For velocities greater than the critical velocity $v_{\text{crit}}=0.476$, the mutual repulsion is overcome and two different types of behaviour are observed: (\emph{i}) formation of a false domain wall and (\emph{ii}) eventual escape back to infinity. Case (\emph{i}) is shown in the snapshots in Fig.~\ref{AB_profile}(a)-(d) taken at four different times during the scattering process for $v=0.7$. The $AB$ pair becomes trapped and oscillates several times, radiating away its energy, before settling into a false domain wall \cite{Haberichter:2015xga}. In this final soliton state, the sheep field $\phi$ is in the false vacuum outside the domain wall, but passes through the  true vacuum inside the domain wall. The $\phi$ field in this configuration cannot separate to infinity because it is trapped by the shepherd field $\psi$ which is in its true vacuum outside the domain wall, but in its false vacuum inside the domain wall. In Fig.~\ref{AB_fin}(a), the outgoing velocity is plotted to be zero whenever a false domain wall is formed. Case (\emph{ii}) is shown in the snapshots plotted in Fig.~\ref{AB_profile} (e)-(h)  for $v=0.8$. Here, the kinks are in an \emph{escape resonance}. Rather than settling into a false domain wall, kinks $A$ and $B$ bounce off each other three times and then, escape back to infinity. The three-bounce collision is clearly visible when plotting the shepherd field $\psi(0,t)$ at the center-of-mass versus time $t$, see Fig.~\ref{AB_bounce}(b) for the initial kink velocity $v=0.750$. Each bounce is represented by a large spike in the shepherd field after which the kinks reflect, recede and then return to bounce off each other again. After the third bounce, kinks $A$ and $B$ escape back to infinity. 

Note that the change from forming a bound state to reflection does not happen at just one critical value of the initial velocity. Typically, there are bands of initial velocity at which domain wall formation occurs, while at other values of the initial velocity the kink pair escapes back to infinity after bouncing off each other three times. This band structure for velocities greater than the critical velocity $v_{\text{crit}}=0.476$ is visible in Fig.~\ref{AB_fin}(a): For initial velocities $v\in[0.707,0.801]$,  kinks $A$ and $B$ are in an escape resonance (see Fig.~\ref{AB_fin}(b) for a zoomed plot of the escape band), while for the velocity ranges $[0.477,0.706]$ and $[0.802,0.9]$ domain walls are formed. In Fig.~\ref{AB_bounce}(a), we display the number of bounces as a function of initial kink velocity. Here, the number of bounces is measured as the number of times the shepherd field $\psi$ oscillates at the origin before the kinks escape back to infinity. With the velocity being accurate up to three significant figures, we count three bounces for all $AB$ collisions with initial velocities $v\in[0.707,0.801]$.

\begin{figure}[!htb]
\subfigure[\, ]{\includegraphics[width=18cm]{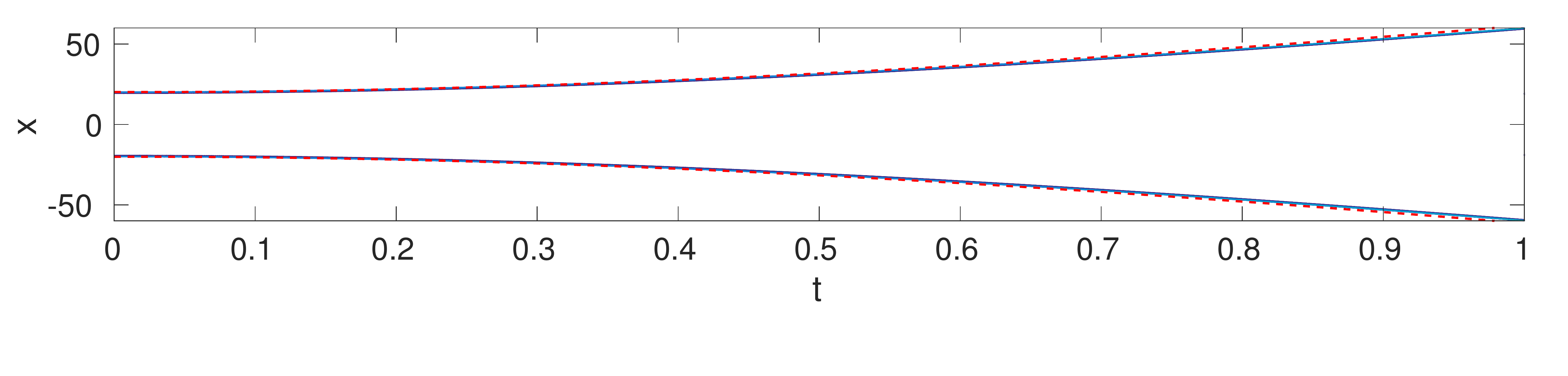}}
\subfigure[\, ]{\includegraphics[width=18cm]{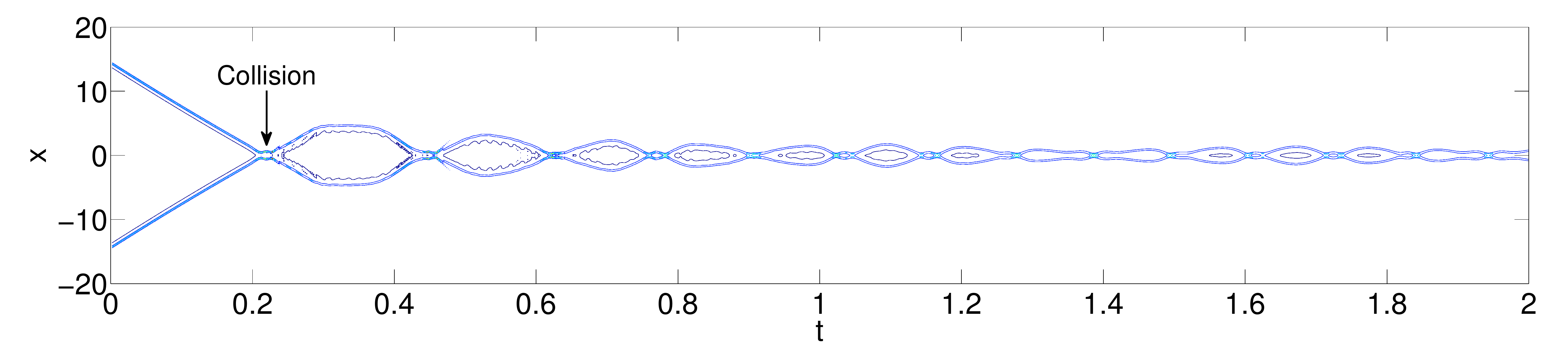}}\\
\subfigure[\, ]{\includegraphics[width=18cm]{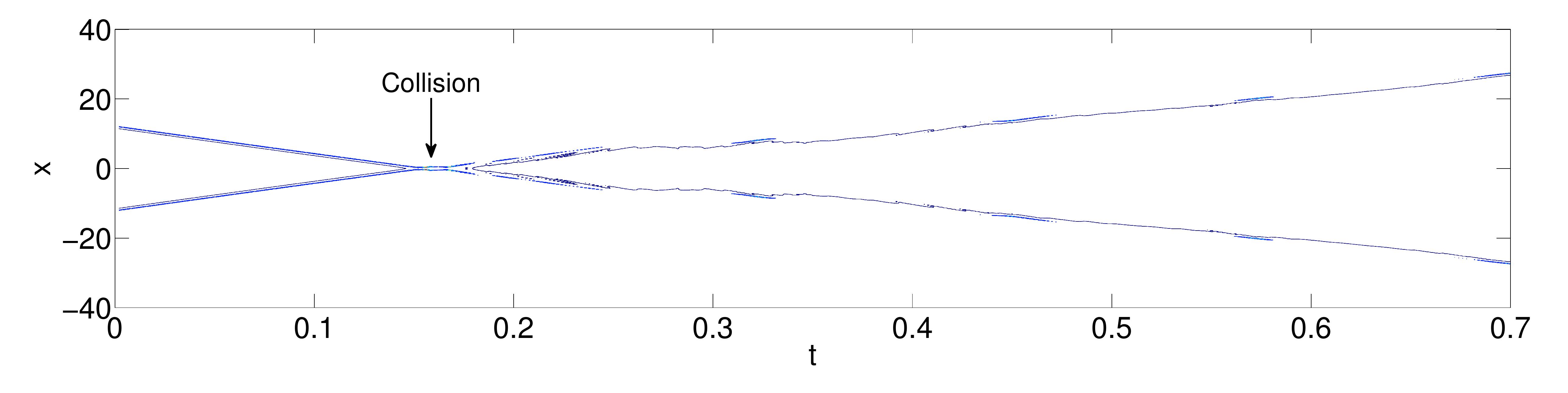}}
\caption{For kink collisions of type $AB$ and for three choices of initial velocity $v$, we display contour plots of the energy density from the sheep field $\phi$. We illustrate three different types of behaviour: (a) From rest, the $AB$ pair repels and the trajectory is well described by a function of the form (\ref{Orbit_fit}) (shown as red dashed line). (b) For $v=0.7$, $A$ and $B$ kink capture each other and form a metastable domain wall configuration \cite{Haberichter:2015xga}. In the following, we describe this type of behaviour as ``sticking". (c) For $v=0.8$, kink $A$ and $B$ are in an escape resonance, that is $A$ and $B$ kink reflect off each other, with three bounces in the shepherd field, and subsequently escape to infinity. For reasons of clarity, time is divided by 100.}
\label{AB_contour}
\end{figure}

As a summary, we track the positions of the solitons in $AB$ collisions by plotting the maxima of the energy density of the sheep field $\phi$. In Fig.~\ref{AB_contour}, we include contour plots of the energy density of the sheep field $\phi$ for three choices of initial velocity. From rest, the kinks $A$ and $B$ repel, see Fig.~\ref{AB_contour}(a). Here, the kink's trajectory can be well approximated by a function of the form 
\begin{equation}
x(t)=\frac{c_1}{2 } t^2+c_2 t+c_3 + c_4\exp\left(-c_5 t\right)\,,
\label{Orbit_fit}
\end{equation}
where $c_1,c_2,c_3,c_4,c_5$ are taken to be fitting parameters. We compare in Fig.~\ref{AB_contour}(a) the kink's trajectory obtained from full field simulations with the function (\ref{Orbit_fit}) with the best fit parameter values $c_1=-0.001$, $c_2=1.346$, $c_3= -122.821$, $c_4=142.979 $ and  $c_5=0.0096$. For small $t$ and short distances, the forces between kinks fall off exponentially fast with the potential being of the form $\exp\left(-mL\right)$, where $m$ is a mass parameter and $L$ denotes the separation between kinks. For larger $t$ and hence larger kink separations, the dominating force between kinks is due to the imbalance in vacuum energies. The $A$ and $B$ kink are pushed apart from each other by the pressure of the true vacuum in between them. The force is constant which gives rise to orbits parabolic in time. For velocities $v\in [0.477,0.706]$ and $v\in [0.802,0.9]$, the $A$ and $B$ kinks capture each other and form a metastable domain wall configuration \cite{Haberichter:2015xga}. In the following, we call this type of behaviour ``sticking". In Fig.~\ref{AB_contour}(b), we display the position plot of such a scattering process for $v=0.7$. For $v=0.7$, we find that the kinks form an excited false domain wall at time $t=20$ (For reasons of clarity, in the figures time is divided by 100.). 
During each oscillation energy is radiated away so that after a few oscillations the kink pair settles into a false domain wall. In Fig.~\ref{AB_contour}(c), we show the position plot for $v=0.8$. Here, kinks $A$ and $B$ are in an escape resonance. Kinks $A$ and $B$ bounce off from each other at $t=14.6, 15.8, 17.1$ and subsequently escape back to infinity. Recall that this type of scattering process is observed for all initial velocities $v\in[0.707,0.801]$.

Note that the appearance of escape bands for kink-antikink collisions in the double well case and in similar models has been explained in the literature \cite{Campbell:1983xu,Dorey:2011yw} by an energy transfer between the translational and vibrational modes of the individual solitons. In the first collision energy is transferred to the vibrational modes, so that below a certain velocity the solitons do not have enough kinetic energy to escape their mutual attraction. Hence, they attract again and collide another time. After two or  three or more kink-antikink collisions enough energy could be transferred back from the vibrational modes to the translational modes, allowing the solitons to escape back to infinity.

\begin{figure}[!htb]
\subfigure[\, ]{\includegraphics[width=12.5cm]{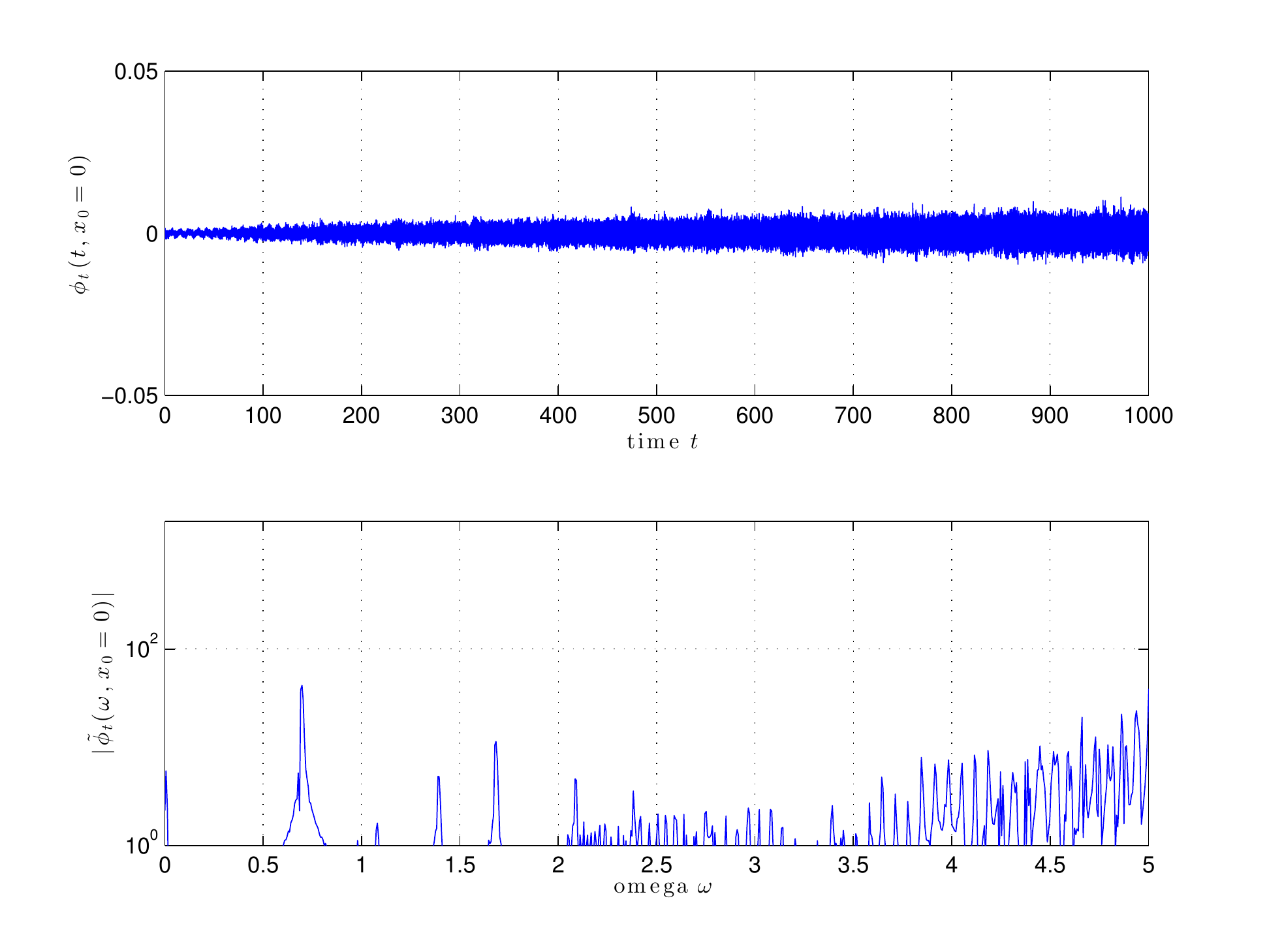}}
\subfigure[\, ]{\includegraphics[width=12.5cm]{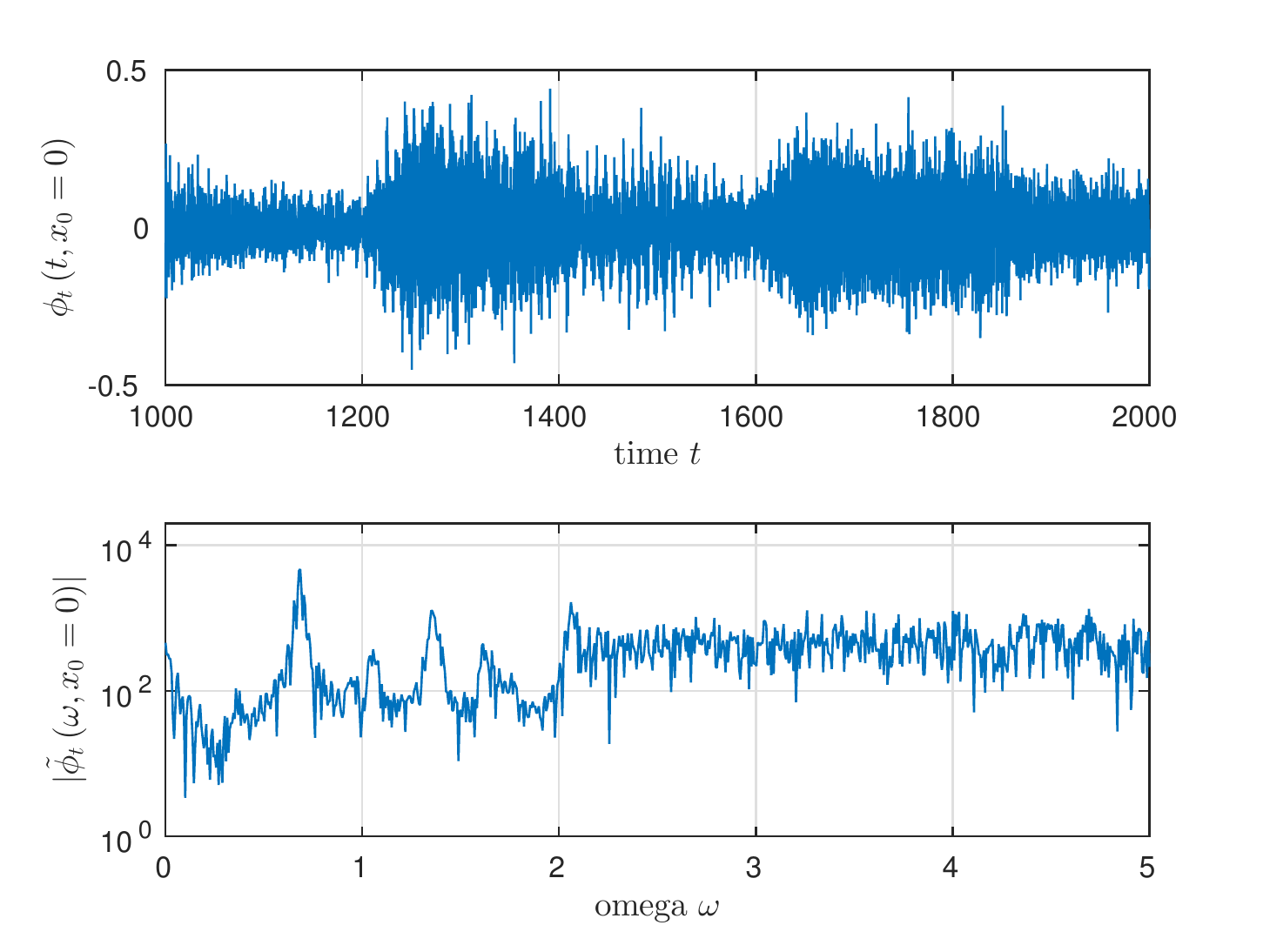}}
\caption{$\phi_t(x=0,t)$ versus $t$ and corresponding power spectra $|\tilde{{\phi}}_t(\omega,x_0=0)|$ versus oscillation frequency $\omega$ when time evolving a false $AB$ domain wall (a) initially squeezed by 0.05\%  and (b) produced in an $AB$ collision for initial kink velocity $0.7$. In (b), we take as initial configuration the ``wobbling'' $AB$ domain wall displayed in Fig.~\ref{AB_profile}~(d). Each initial configuration is time-evolved over $5\times 10^5$ time steps with $\Delta t=0.002$. }
\label{AB_Full_num_AB}
\end{figure}

To get a better understanding of the ``wobbling'' effect seen in $AB$ kink collisions for initial velocities $v\in [0.477,0.706]$ and $v\in [0.802,0.9]$, we want to work out in the following oscillation spectra for $AB$ domain walls. We time evolve two different $AB$ domain walls: (a) 
a fully relaxed $AB$ domain wall (with $x\in[-40,40]$ and $\Delta x=0.01$) initially squeezed by $0.05\%$ and (b) the ``wobbling'' domain wall produced in an $AB$ collision for initial kink velocity $0.7$.  In (b), we take as initial configuration the ``wobbling'' $AB$ domain wall displayed in Fig.~\ref{AB_profile}~(d). Both initial configurations are time evolved for $5\times 10^5$ time steps with $\Delta t=0.002$. In each time step, we measure at $x_0 = 0$ (inside the soliton core) the deviation from the static kink solution by recording $\phi_ṫ(t, x_0 = 0)$. To find the frequency components, we compute the fast fourier transform $|\tilde{{\phi}}_t(\omega,x_0=0)|$ of the recorded data using the MATLAB built-in function f{\tt{ft}}. All model parameters are chosen as in (\ref{Para}). In Fig.~\ref{AB_Full_num_AB}, we display $\phi_t(x=0,t)$ versus $t$ and the resulting power spectra $|\tilde{{\phi}}_t(\omega,x_0=0)|$. We confirm that the fluctuations are governed by a set of discrete frequencies with the dominating frequency found at $\omega\approx0.7$.

\begin{figure}[!htb]
\subfigure[\, ]{\includegraphics[width=8cm]{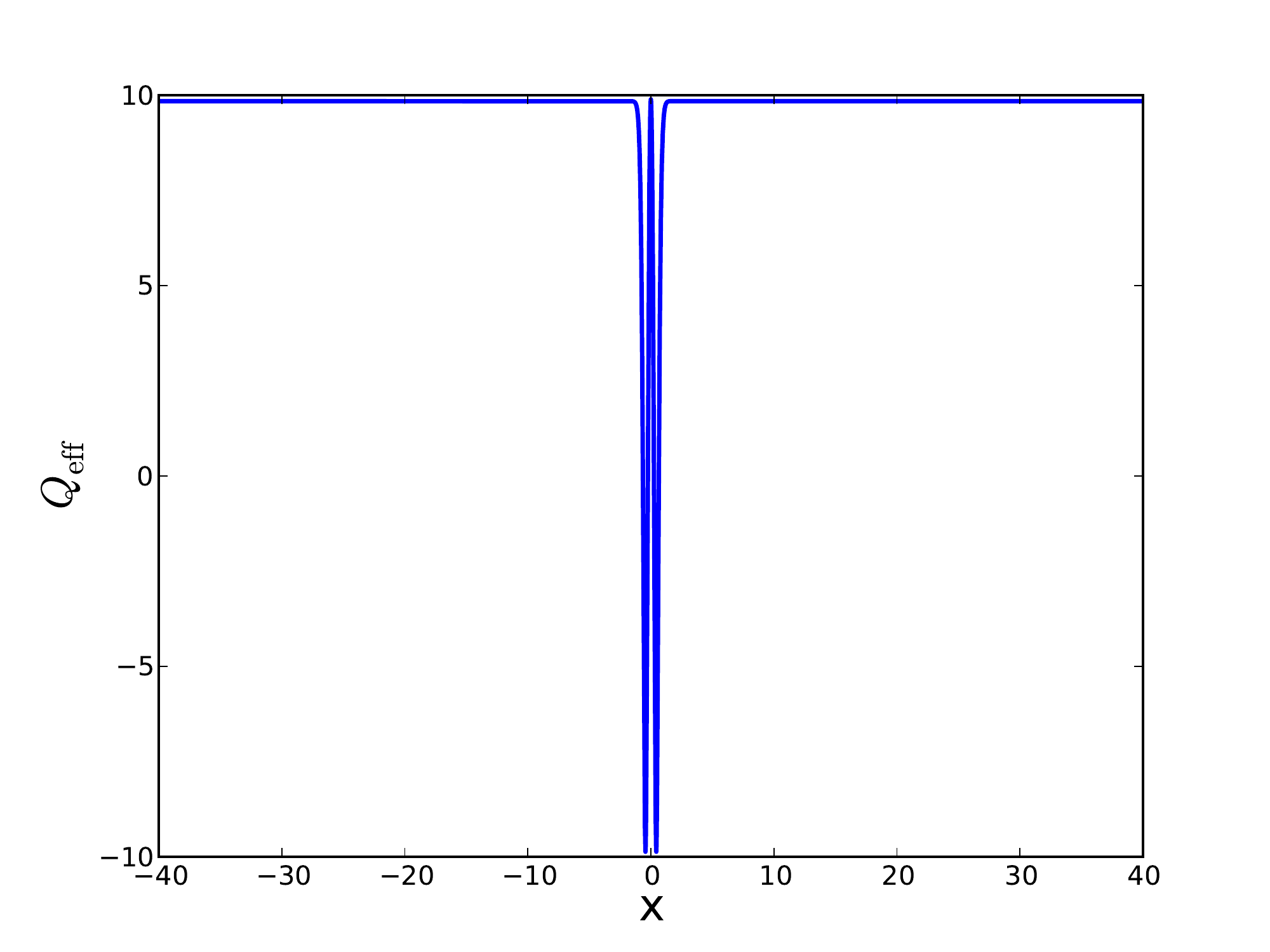}}
\subfigure[\, ]{\includegraphics[width=8cm]{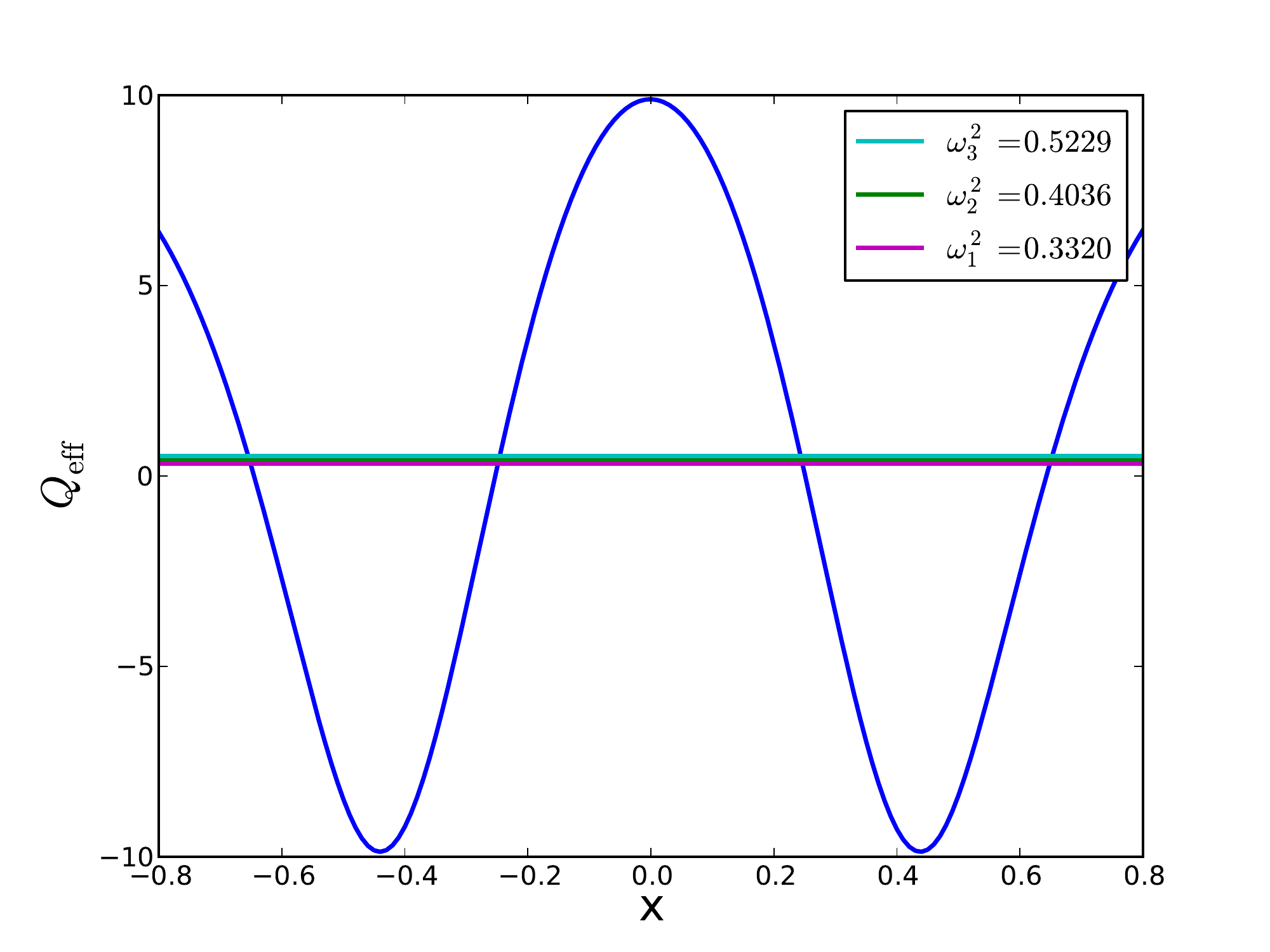}}\\
\caption{Effective potential $Q_\text{eff}$ with first few oscillation frequencies $\omega^2$ for $AB$ domain wall: (a) $Q_\text{eff}$ displayed for $x\in[-40,40]$ ; (b) Zoom.}
\label{AB_Qeff_plot}
\end{figure}

The frequency components observed in the full numerical time evolution can be compared to the shape oscillations obtained in the small perturbation approximation, that is for small fluctuations $\phi= \phi_0(x) + \delta\phi(x, t)$ and $\psi= \psi_0(x) + \delta\psi(x, t)$  around the static $AB$ kink soliton. Here, $\phi_0$ and $\psi_0$ denote sheep and shepherd field of the static $AB$ domain wall and $\delta\phi$ and $\delta\psi$ are the associated fluctuation fields. Expanding the equation of motion for the sheep field $\phi$ up to first order in $\delta\phi$ results in the following linear equation for small oscillations $\delta\phi$
\begin{align}
\delta\ddot{\phi}=\delta\phi^{\prime\prime}-\beta\pi^2\left(2\cos\left(2\pi\phi\right)+\frac{\epsilon_\phi}{2}\cos\left(\pi\phi\right)\right)\delta\phi\,,
\label{eom_phi_lin}
\end{align}
where we neglected the contribution coming from the nonlinear coupling potential $V_{\psi\phi}$. Substituting the ansatz $\delta \phi(x, t) = \cos \omega t \delta \phi(x)$ into the fluctuation equation (\ref{eom_phi_lin}) yields a Schr\"odinger equation
\begin{align}
-\frac{\text{d}^2}{\text{d}x^2}\delta \phi+Q_{\text{eff}}(x)=\omega^2\delta \phi\,,
\label{AB_SL}
\end{align}
with effective potential
\begin{align}
Q_{\text{eff}}(x)=\beta\pi^2\left(2\cos\left(2\pi\phi\right)+\frac{\epsilon_\phi}{2}\cos\left(\pi\phi\right)\right)\,.
\label{Qeff_AB_SL}
\end{align}
The perturbation is required to vanish at spatial infinity for all $t$, hence we impose the boundary conditions $\delta\phi(\pm\infty)=0$. We solve the Schr\"odinger equation (\ref{AB_SL}) on the spatial interval $[-25,25]$ with the {\tt{SLEIGN2}} code \cite{Bailey:2001:ASS:383738.383739}. We display in Fig.~\ref{AB_Qeff_plot} the effective potential (\ref{Qeff_AB_SL}) together with the first few  numerically calculated eigenfrequencies. In particular, we find $\omega_1= 0.58\,,\,\,\omega_2= 0.64$ and $\omega_3=0.71$. 

\subsection{$BA$ scattering}
 
\begin{figure}[!htb]
\subfigure[\, ]{\includegraphics[totalheight=4.cm]{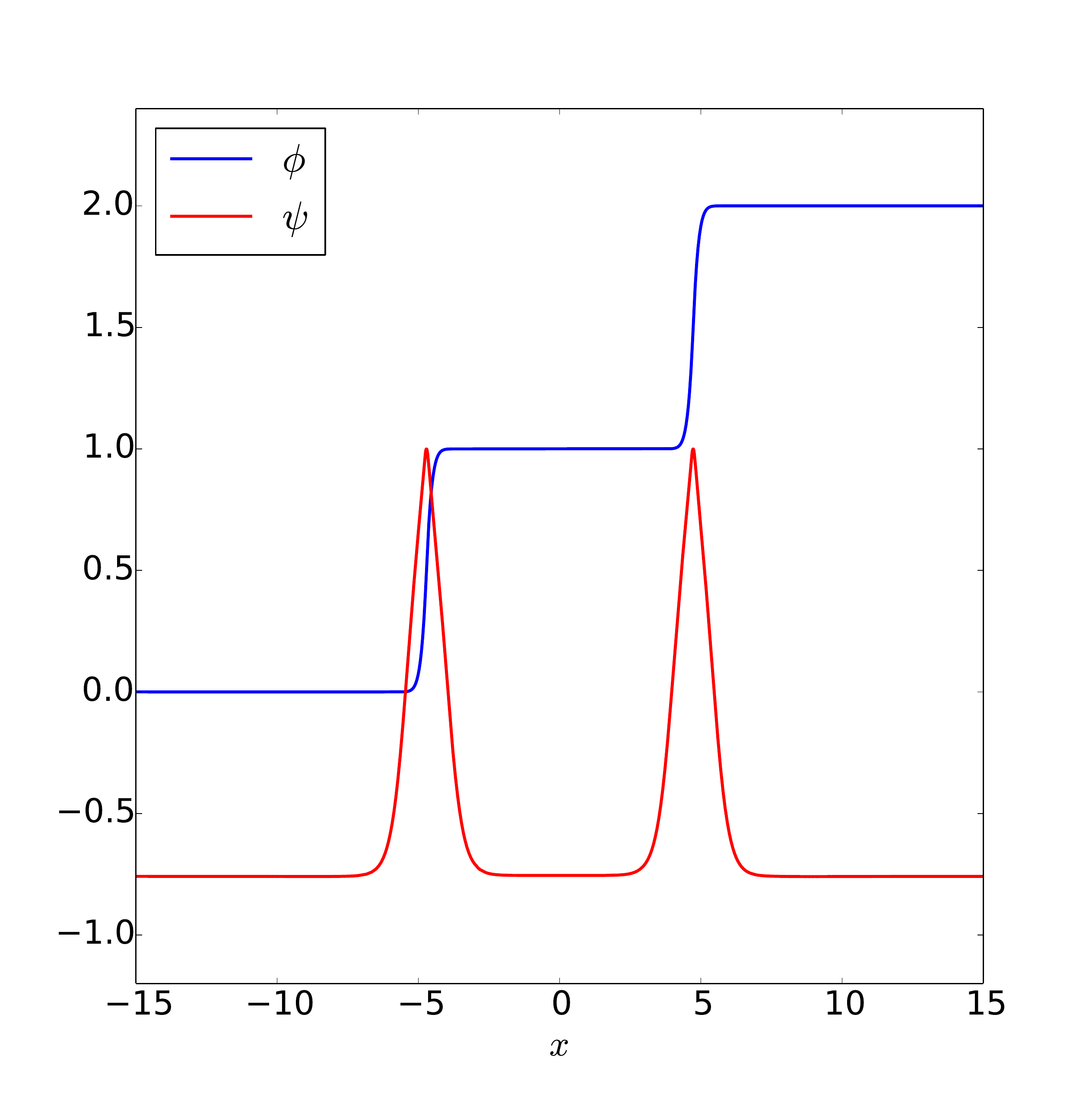}}
\subfigure[\, ]{\includegraphics[totalheight=4.cm]{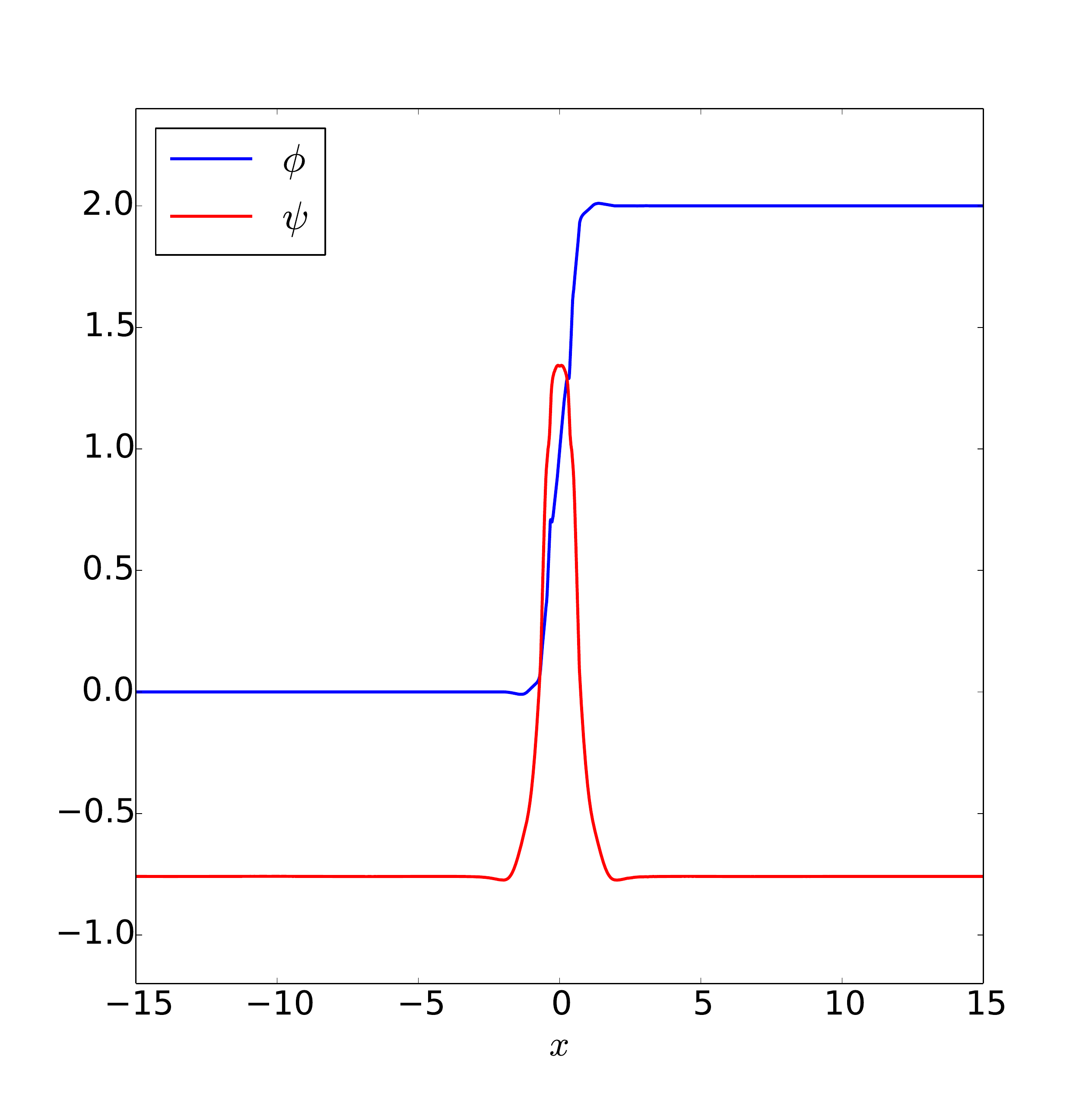}}
\subfigure[\, ]{\includegraphics[totalheight=4.cm]{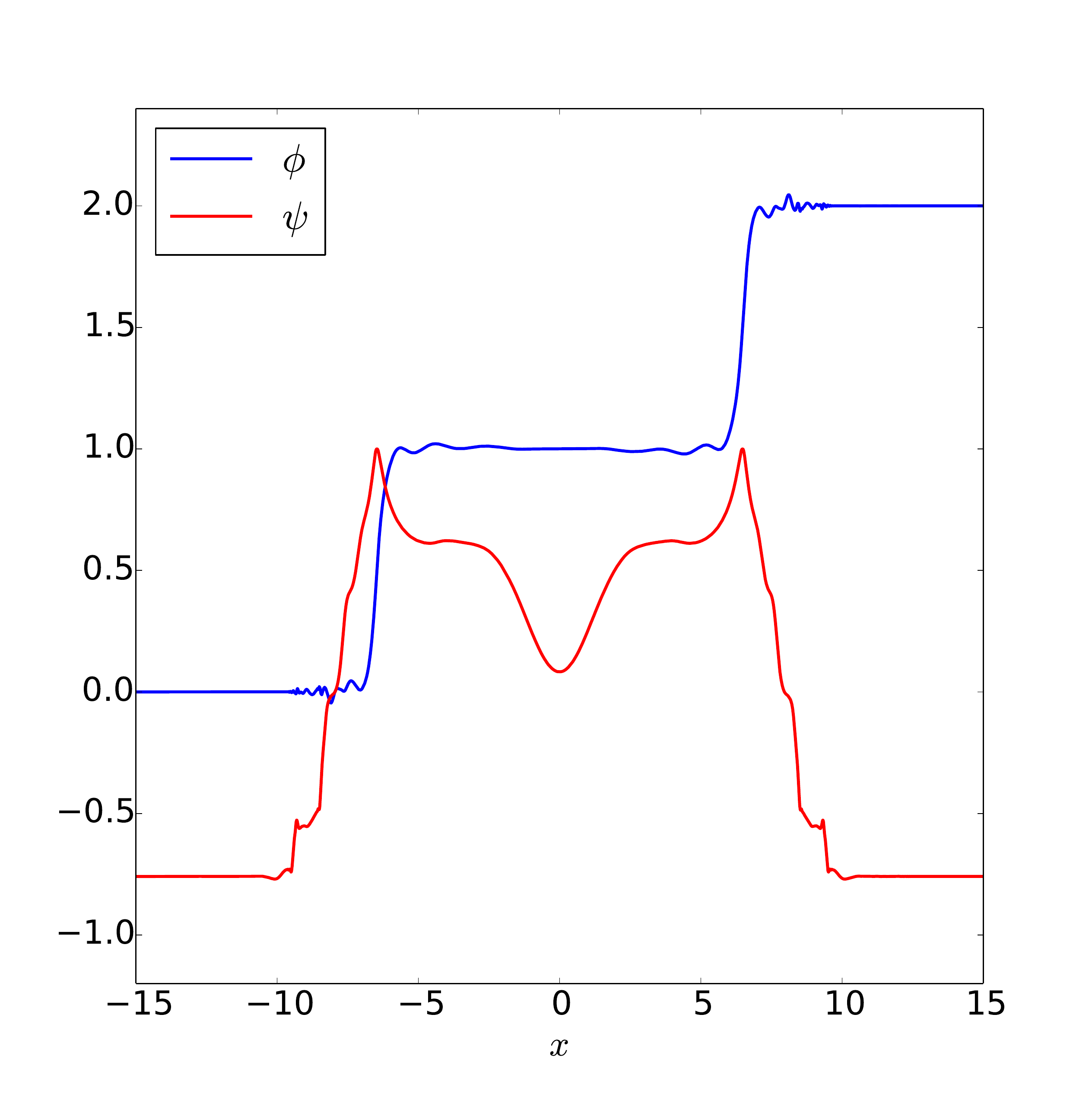}}
\subfigure[\, ]{\includegraphics[totalheight=4.cm]{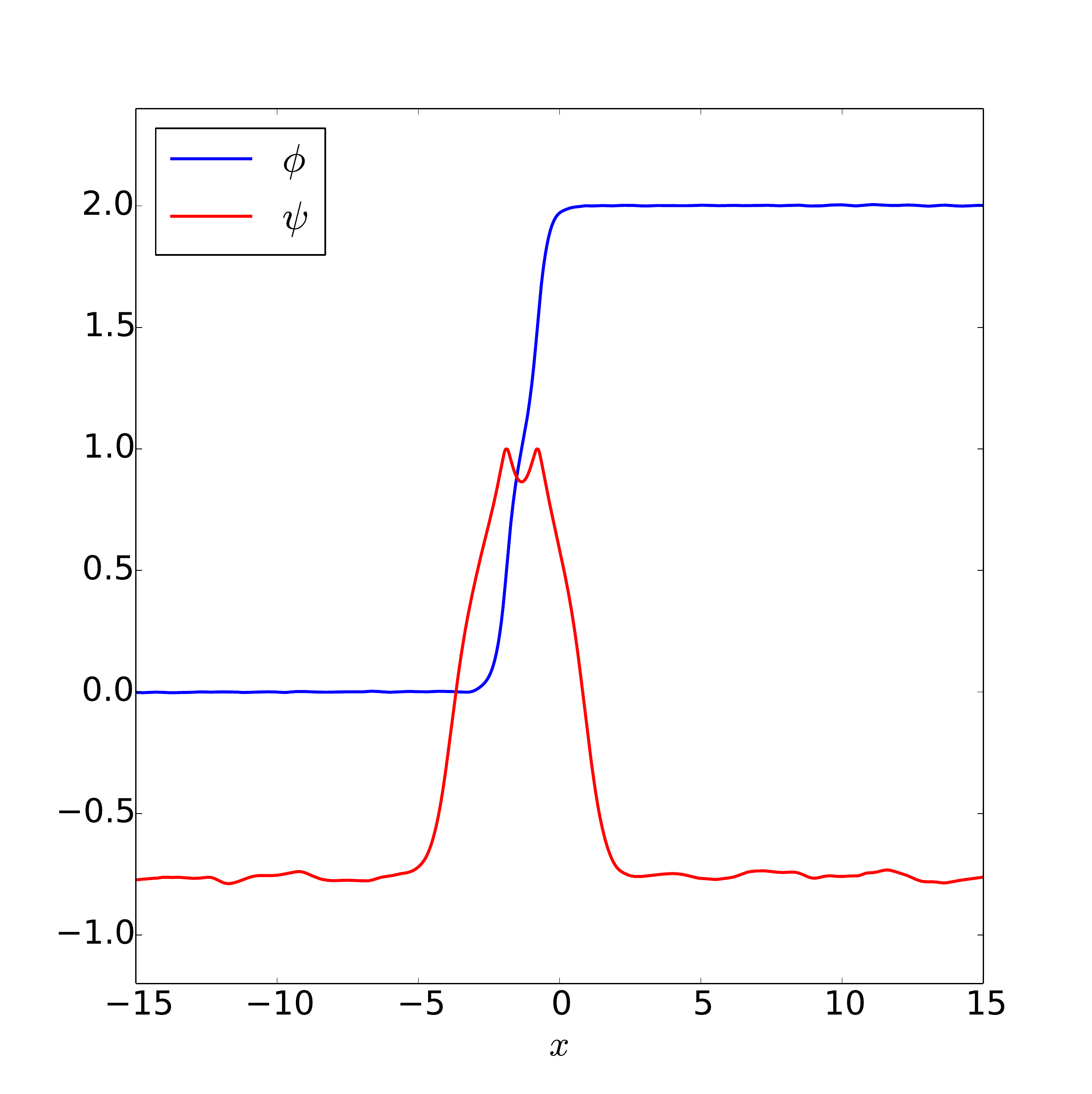}}
\caption{For kink collisions of type $BA$, we display snapshots of the field configuration for the initial velocity $v=0.8$: $A$ and $B$ kinks form a ``true'' domain wall, that is the $\phi$ field interpolates between distinct true vacua.}
\label{BA_profile}
\end{figure}

The next scattering process to investigate is the collision of a kink of type $B$ with one of type $A$. A configuration of type $BA$ can be found in Fig.~\ref{BA_profile}(a). Here, the $B$ sheep field interpolates from the true vacuum at $\phi=0$ to the false vacuum at $\phi=+1$. The $A$ sheep field is then attached to this false vacuum and connects it with the true vacuum at $\phi=+2$. 
Hence, the initial soliton setup is attractive with true vacua outside and with false vacuum in between. This setup results for all initial kink velocities $0\le v\le 0.9$ in the same scattering outcome, that is the creation of a true domain wall interpolating between the two distinct true vacua $\phi=0$ and $\phi=+2$.

In Fig.~\ref{BA_profile}(a)-(d), we plot snapshots of the field configuration for the initial velocity $v=0.8$. The kinks accelerate towards each other converting false vacuum to true vacuum. They form an excited bound state which settles into a true domain wall after a few oscillations.

\subsection{$B\overline{A}$ scattering}

\begin{figure}[!htb]
\subfigure[\, ]{\includegraphics[totalheight=4.cm]{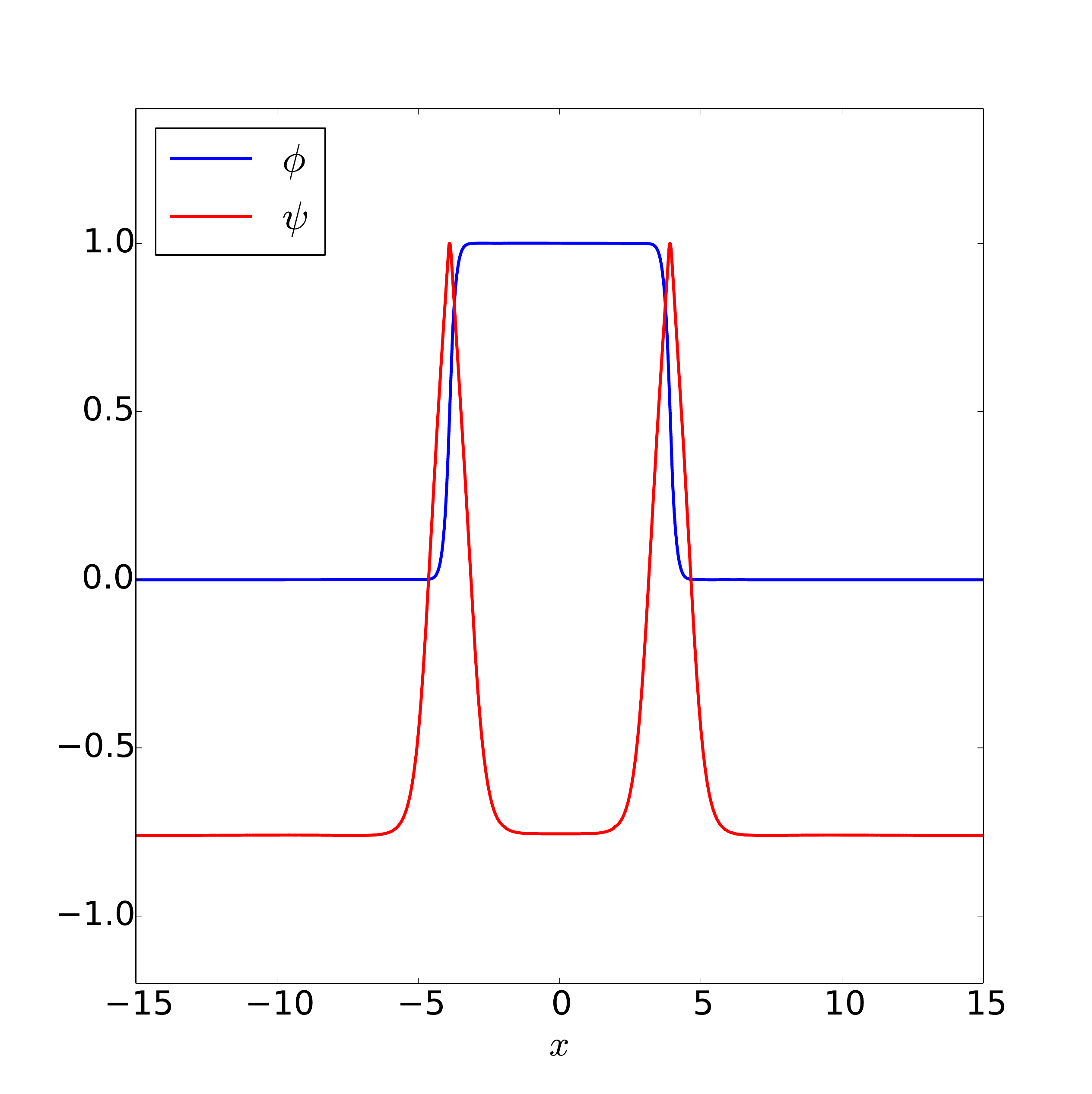}}
\subfigure[\, ]{\includegraphics[totalheight=4.cm]{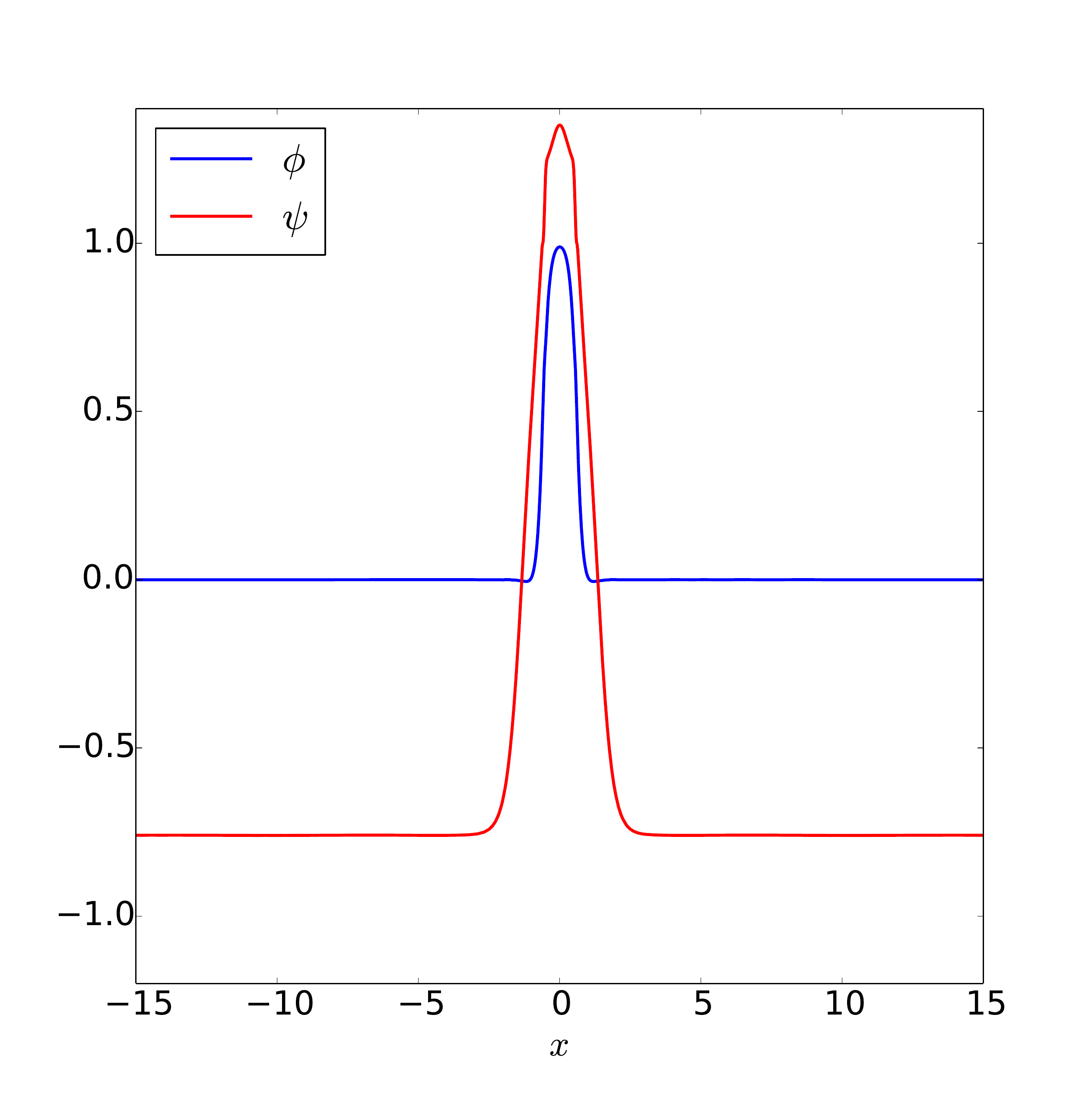}}
\subfigure[\, ]{\includegraphics[totalheight=4.cm]{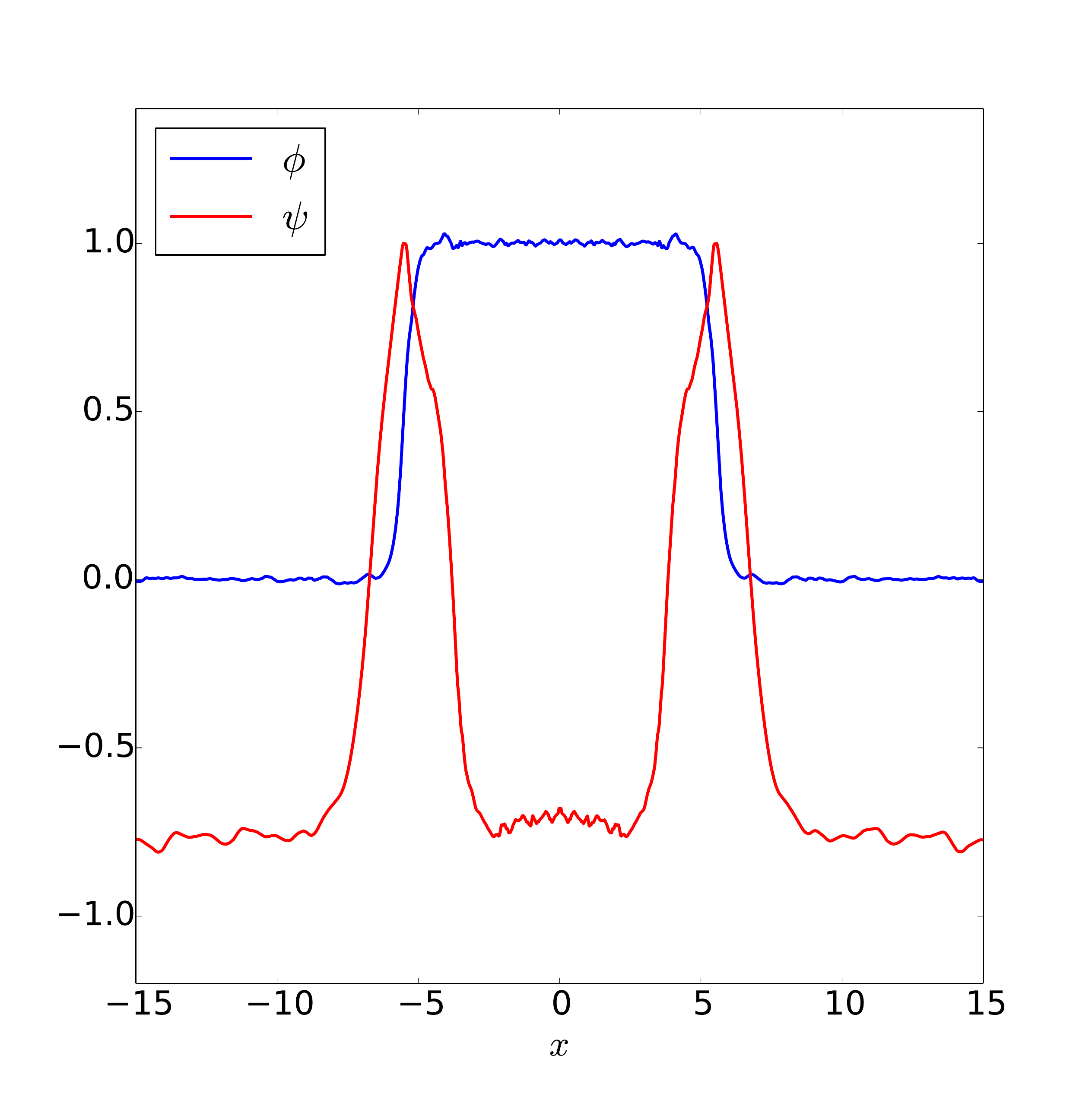}}
\subfigure[\, ]{\includegraphics[totalheight=4.cm]{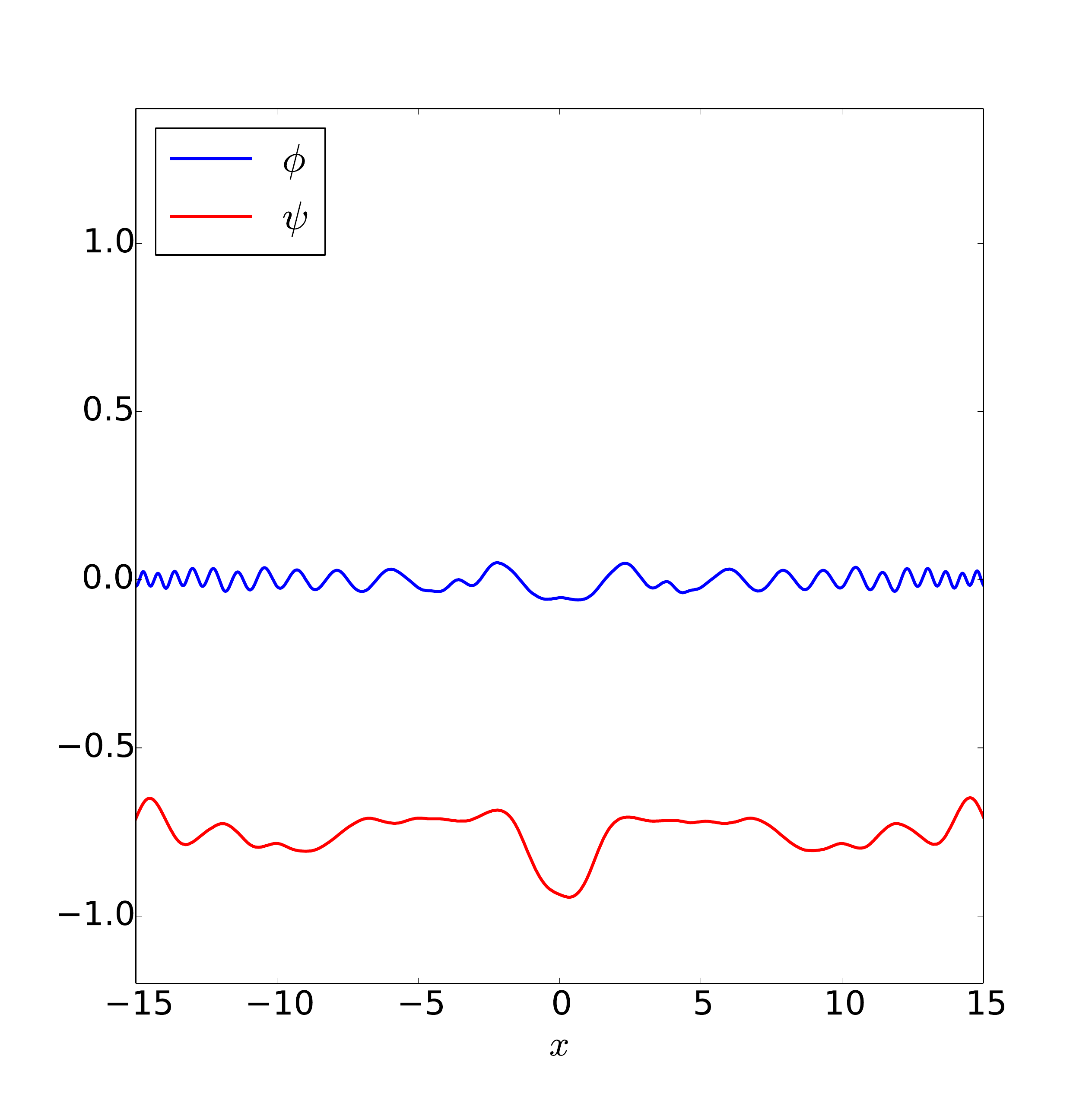}}
\caption{We display snapshots of the field configuration during a $B\overline{A}$ scattering with initial velocity $v=0.8$. During their first collision, the kinks reflect off one another and separate. Then they collide once more, forming an oscillon that will ultimately decay to the true vacuum.  }
\label{BAbar_profiles}
\end{figure}

Recall that in our notation, $B\overline{A}$ is the combination of a kink interpolating from true vacuum to false vacuum with one interpolating from false vacuum to true vacuum in the reverse direction. In the following, we attach a $B$ sheep field interpolating from $\phi=0\rightarrow\phi=+1$ to an $\overline{A}$ sheep field interpolating from $\phi=+1\rightarrow\phi=0$. The initial field configuration is displayed in Fig.~\ref{BAbar_profiles}(a).

Configurations of the type $B\overline{A}$, like those of type $BA$, have true vacua outside and false vacuum in between. Thus the solitons will attract one another from rest. Due to the attraction between the solitons, and the kink-antikink like nature of the configuration,  the final result of any $B\overline{A}$ scattering is the formation of an oscillon that slowly radiates away energy and ultimately annihilates to the true vacuum.  However, we observe that depending on the initial velocity $v$ given to the solitons, this may be achieved in different ways. In some cases, the solitons first reflect from one another and travel apart, before they are drawn back together and collide again. This second collision may result in either the creation of an oscillon, or a repetition of the reflection behaviour. Note that kink-antikink scatterings in which oscillons are formed have been observed before in  a wide range of scalar field theories, see e.g. \cite{Anninos:1991un,Halavanau:2012dv,Braden:2014cra,Gani:2015cda}.  During each oscillation, some of an oscillon's energy is radiated
away, so it will eventually decay. However, the oscillon will persist for a very long time before this occurs.

Snapshots of the field configuration during a $B\overline{A}$ scattering with initial velocity $v=0.8$ are shown in Fig.~\ref{BAbar_profiles}. The sheep field $\phi$ is displayed in blue, and the shepherd field $\psi$ is shown in red. This choice of initial velocity results in two kink collisions. The first collision is seen in Fig.~\ref{BAbar_profiles}(b). During this collision there are three bounces in the shepherd field at the origin before the solitons separate and travel apart. Fig.~\ref{BAbar_profiles}(c) shows the solitons excited by their collision, and well separated by a region of false vacuum. Following this, the solitons attract and collide once more. An oscillon is formed which will ultimately annihilate to the true vacuum. This can be seen in Fig.~\ref{BAbar_profiles}(d).

\begin{figure}[!htb]
\subfigure[\, ]{\includegraphics[width=17cm]{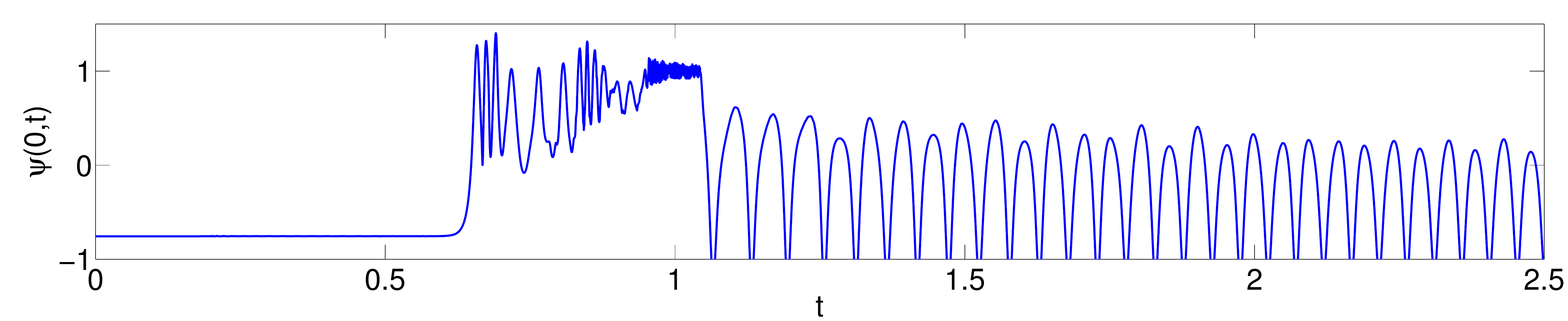}}\\
\subfigure[\, ]{\includegraphics[width=17cm]{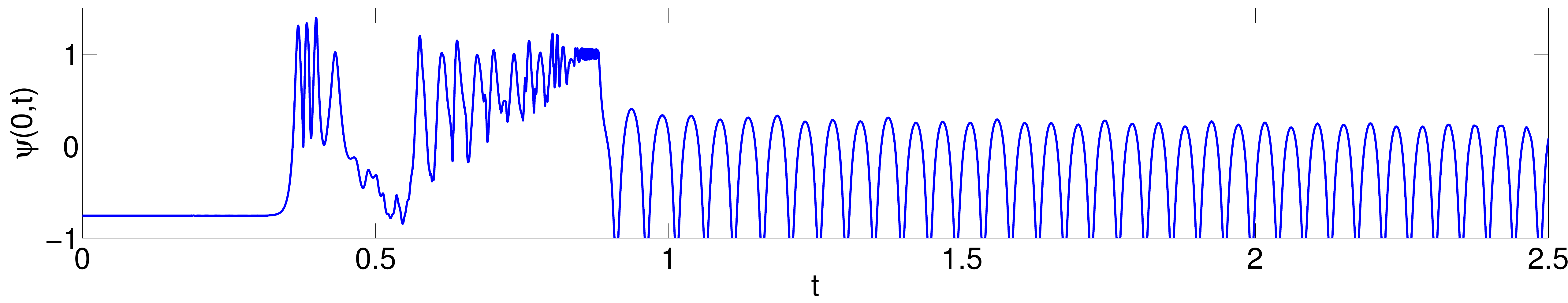}}\\
\subfigure[\, ]{\includegraphics[width=17cm]{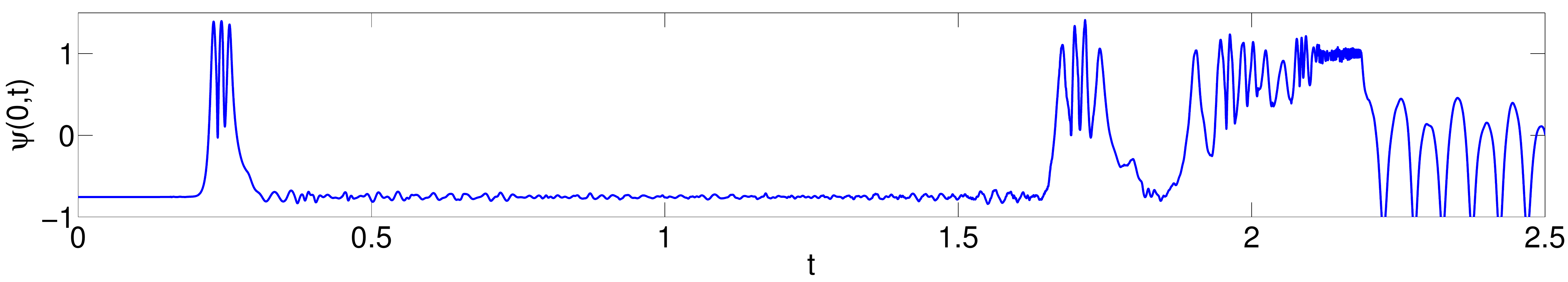}}\\
\subfigure[\, ]{\includegraphics[width=17cm]{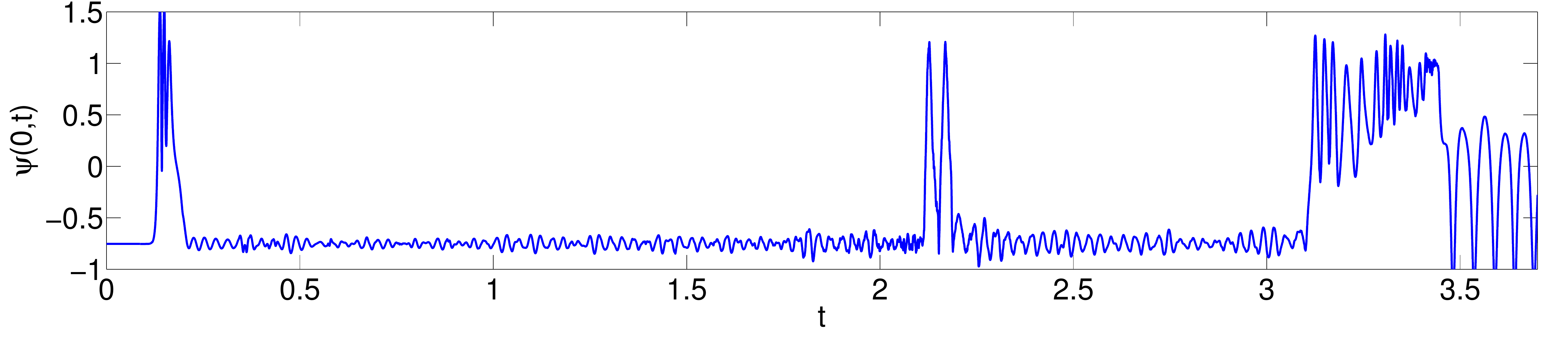}}
\caption{The shepherd field $\psi(0,t)$ as a function of time $t$ during  $B\overline{A}$ scattering for four choices of initial velocity $v$. For reasons of clarity, time is divided by 100.
(a) $v=0$: from rest, the solitons experience an attractive force that draws them together. An oscillon is formed after their first and only collision. This will ultimately decay to the true vacuum.  (b) $v=0.355$: the solitons reflect off each other once, then collide again and capture each other to form an oscillon. (c) $v=0.606$: the solitons reflect off each other two times, before a final collision in which they form an oscillon. (d) $v=0.808$: the solitons reflect off each other twice, with two bounces during the second collision. Finally they annihilate by forming an oscillon.}
\label{BAbar_bounces}
\end{figure}

The kink collisions and oscillon are most clearly seen in the shepherd field $\psi(x,t)$. In Fig.~\ref{BAbar_bounces}, we plot the shepherd field $\psi(0,t)$ at the origin as a function of $t$ for three different choices of initial velocity where the solitons collide once, twice and three times respectively. The simplest case is illustrated in Fig.~\ref{BAbar_bounces}(a). Here the solitons capture each other during their first and only collision at $t=65$ (for reasons of clarity, time is divided by 100). They form an oscillon, which is seen for $t>105$ as a long-lived series of oscillations in the $\psi$ field. Fig.~\ref{BAbar_bounces}(b) corresponds to initial velocity $v=0.355$. During the first collision of the solitons, near $t=40$, there are four bounces in the shepherd field, before it returns to its false vacuum value. The solitons collide again at $t=57$, and this results in the formation of an oscillon at around $t=90$. In Fig.~\ref{BAbar_bounces}(c), the initial velocity is $v=0.606$. The solitons first collide at $t=22$, with the shepherd field bouncing three times before returning to its false vacuum value. The solitons remain separated for some time, before they collide again near $t=165$. During this second collision there are four bounces in the shepherd field. The final soliton collision takes place at $t=187$, and an oscillon is seen for $t>220$ . Another two reflection collision is seen in Fig.~\ref{BAbar_bounces}(d), corresponding to initial velocity $v=0.808$. The solitons collide for the first time at $t=13$, during which we count three bounces. Their second collision occurs at $t=212$. This collision is unusual, as it features only two bounces and the false vacuum is attained in between bounces. Previously we had only observed windows of three bounces or more, and the false vacuum was only attained after the period of bouncing was complete.

\begin{figure}[!htb]
\includegraphics[totalheight=10.cm]{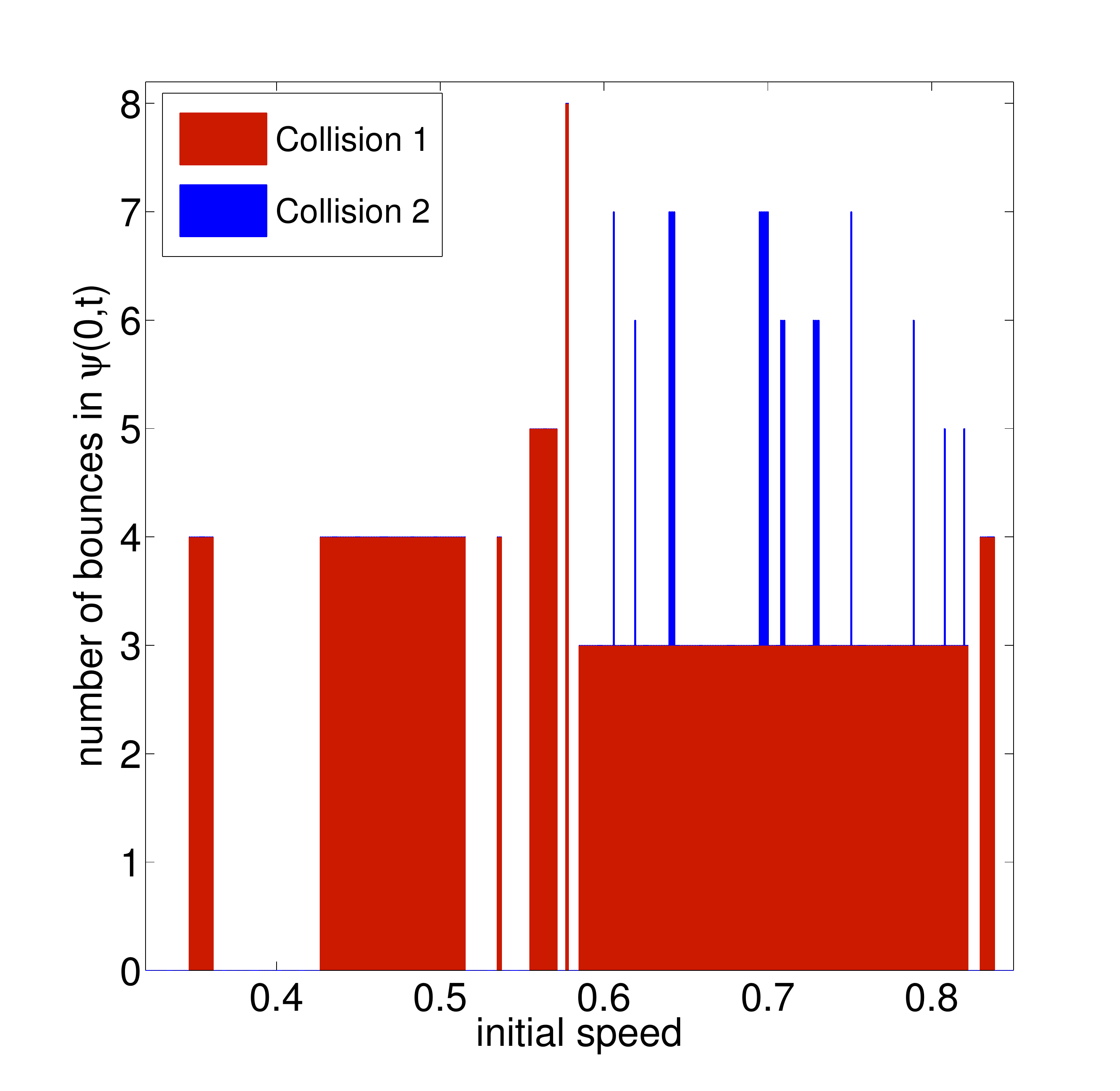}
\caption{Number of bounces as a function of initial velocity $v$ for a $B\overline{A}$ collisions. Here, the bounce number measures the number of times the shepherd field oscillates at the origin before the kinks separate. If no bounces are counted, then the solitons annihilate during their first collision, otherwise we colour in red the number of bounces during the first soliton collision, and in blue the number of bounces during the second collision of the solitons.}
\label{BAbar_summary}
\end{figure}

\begin{figure}[!htb]
\subfigure[\, ]{\includegraphics[width=14.8cm]{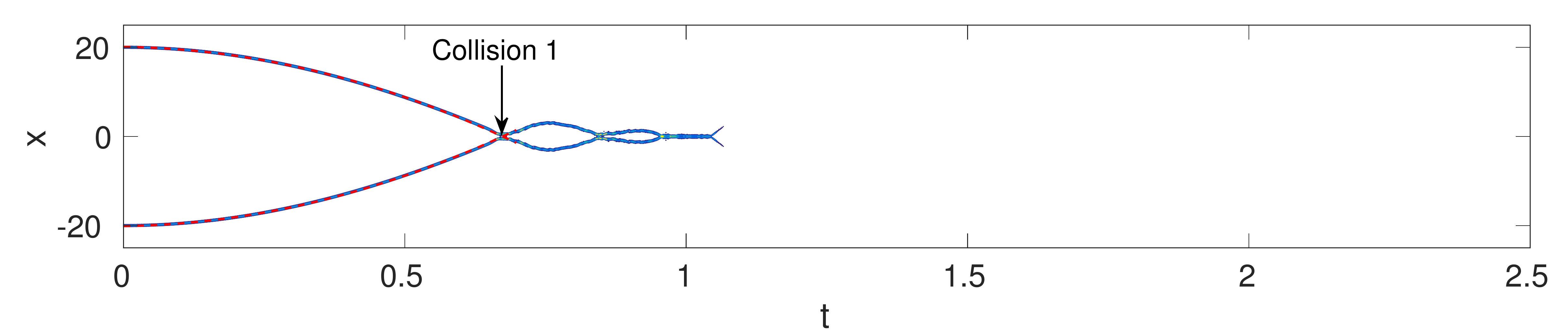}}
\subfigure[\, ]{\includegraphics[width=17cm]{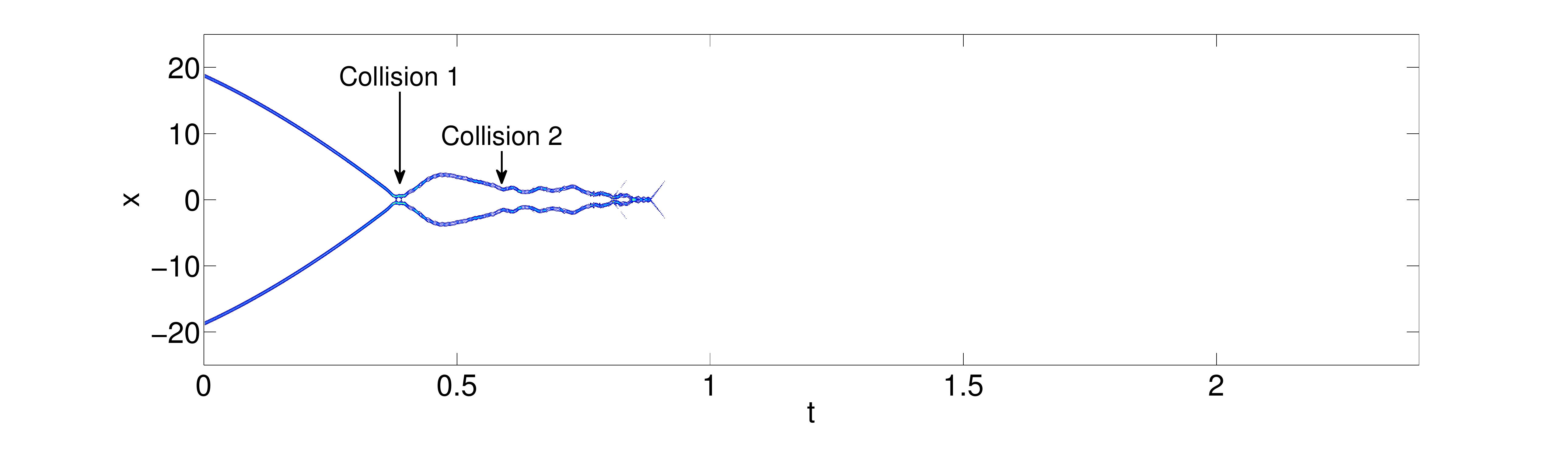}}
\subfigure[\, ]{\includegraphics[width=17cm]{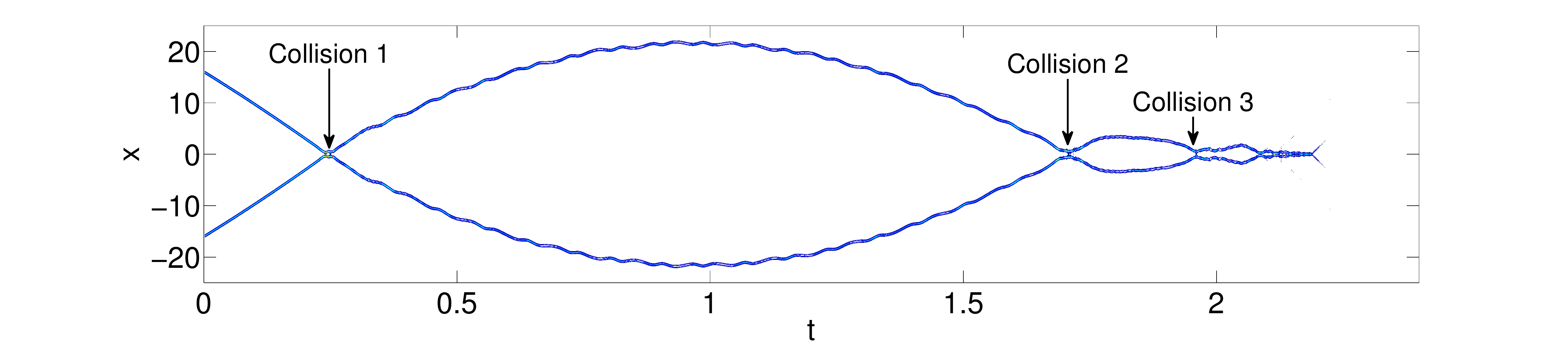}}
\caption{Contour plots of energy density for the sheep field $\phi$ during  $B\overline{A}$ interactions with three different choices of initial velocity $v$. The locations of the soliton collisions are indicated on the plot, and for reasons of clarity, time is divided by 100. (a) $v=0$: From rest the solitons attract and annihilate. Up to the first collision, the kink's trajectory is well described by a function of the form (\ref{Orbit_fit}) (shown as red dashed line). (b) $v=0.355$: the solitons reflect off each other in their initial collision, then collide again and annihilate. (c) $v=0.606$: the solitons reflect off each other twice, before a final collision in which they annihilate. }
\label{BAbar_contours}
\end{figure}

Fig.~\ref{BAbar_summary} summarises the different scattering behaviours that we have observed for initial velocities in the range $0\le v\le 0.9$. It displays the number of bounces in the shepherd field $\psi(0,t)$ at the origin as a function of initial velocity $v$. Where no bounces have been counted, the solitons immediately annihilated, without reflecting away from one another. In all other cases, we find that the solitons reflect away from each other either one or two times before a final collision in which they annihilate by forming an oscillon which decays to the true vacuum. Prior to the formation of the oscillon, we consider the solitons to have reflected off one another if the shepherd field $\psi(0,t)$ takes its false vacuum value after a period of oscillations. We count the number of bounces during the reflections in the same way as we did for those $AB$ collisions in which bouncing occurred. The number of bounces during the first reflection are shown in red, and if there was a second reflection, then the number of bounces during the second reflection are shown in blue. For the first reflection, we find windows of three, four, five  or eight bounces. Of these windows, we only observe a second reflection for particular initial velocities within the three bounce window $v\in[0.585,0.822]$. During the second collision, there are two unusual two bounce reflections for $v=0.808$ and $v=0.820$. The windows for the other values exhibit the more familiar three or four bounce signatures. There is a fractal structure in which windows of the reflection behaviour are separated by regions in which the solitons annihilate. Note that our initial velocity windows are only accurate to three significant figures, and we expect that, were we to investigate with greater resolution, other windows with two or more reflections could be found. 

The positions of the solitons during the scattering process can be understood from the maxima of the energy density of the sheep field $\phi$. In Fig.~\ref{BAbar_contours} we display contour plots of the sheep energy density for three choices of initial velocity: (a) $v=0$, (b) $v=0.355$, and (c) $v=0.606$, with the locations of the soliton collisions indicated in the figures. In Fig.~\ref{BAbar_contours}(a), we see the solitons attract from rest and collide once, resulting in the formation of an oscillon which ultimately decays to the true vacuum. The oscillon is more clearly seen in Fig.~\ref{BAbar_bounces}(a), but the contour plot allows us to track the positions of the solitons away from the origin throughout the process. Up to the first collision point, the parabolic kink orbit in Fig.~\ref{BAbar_contours}(a) is well described by Eq.~(\ref{Orbit_fit}) with the best fit parameter values $c_1=-0.008$, $c_2=-0.017$, $c_3= 20.106$ (and $c_4$ and $c_5$ set to zero).

 Fig.~\ref{BAbar_contours}(b) shows a scattering in which there are two collisions. During the first collision, the solitons reflect off each other. However, they do not travel far before the attractive force draws them back together for a second collision, which creates an oscillon. The first collision takes place near $t=40$, and the second is at $t=57$ (for clarity, time is divided by 100). This can be compared with Fig.~\ref{BAbar_bounces}(b), which uses the same initial velocity. Finally, Fig.~\ref{BAbar_contours}(c) presents a scattering in which the solitons collide and reflect off one another twice before a final collision in which they form an oscillon that decays to the true vacuum. We see that after the first collision, the solitons travel quite far apart before the attraction between them pulls them back together for a second collision. The final collision takes place near $t=180$, ultimately resulting in the annihilation of the solitons. This figure can be compared with Fig.~\ref{BAbar_bounces}(c), which uses the same choice of initial velocity, to see how the locations of the collisions correspond to the bounces in the shepherd field.

\subsection{$A\overline{B}$ scattering}

\begin{figure}[!htb]
\subfigure[\, ]{\includegraphics[totalheight=4.cm]{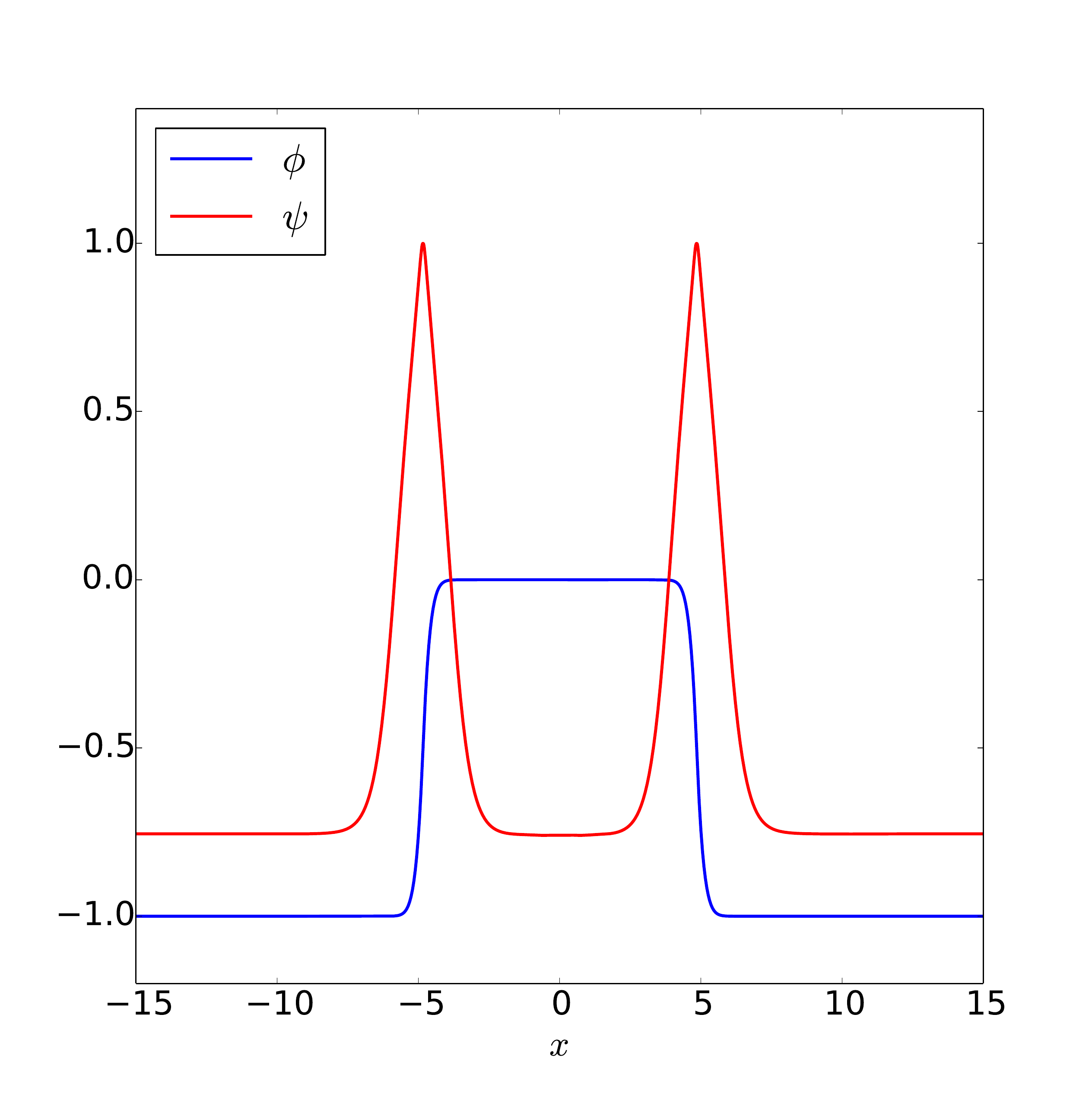}}
\subfigure[\, ]{\includegraphics[totalheight=4.cm]{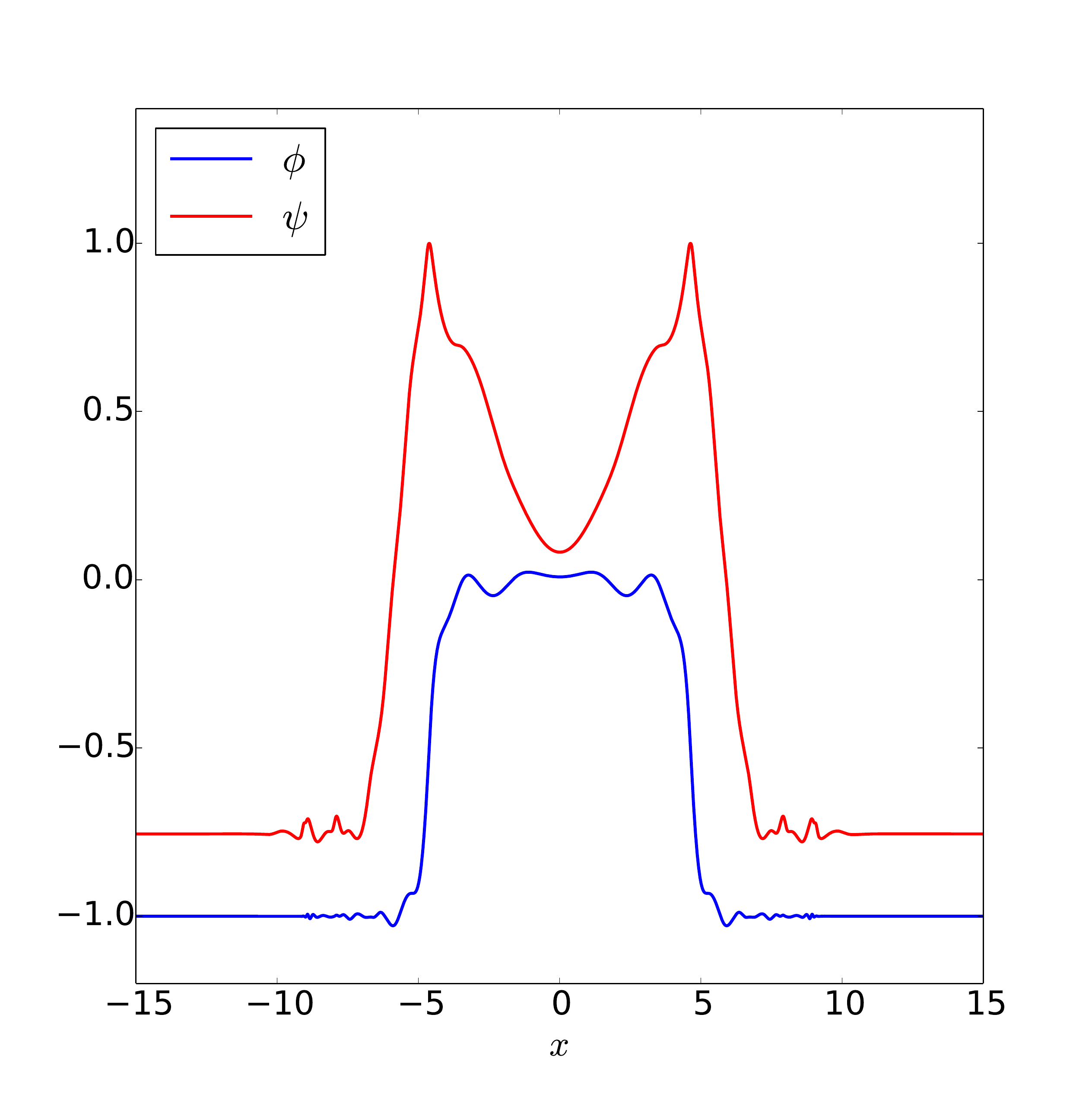}}
\subfigure[\, ]{\includegraphics[totalheight=4.cm]{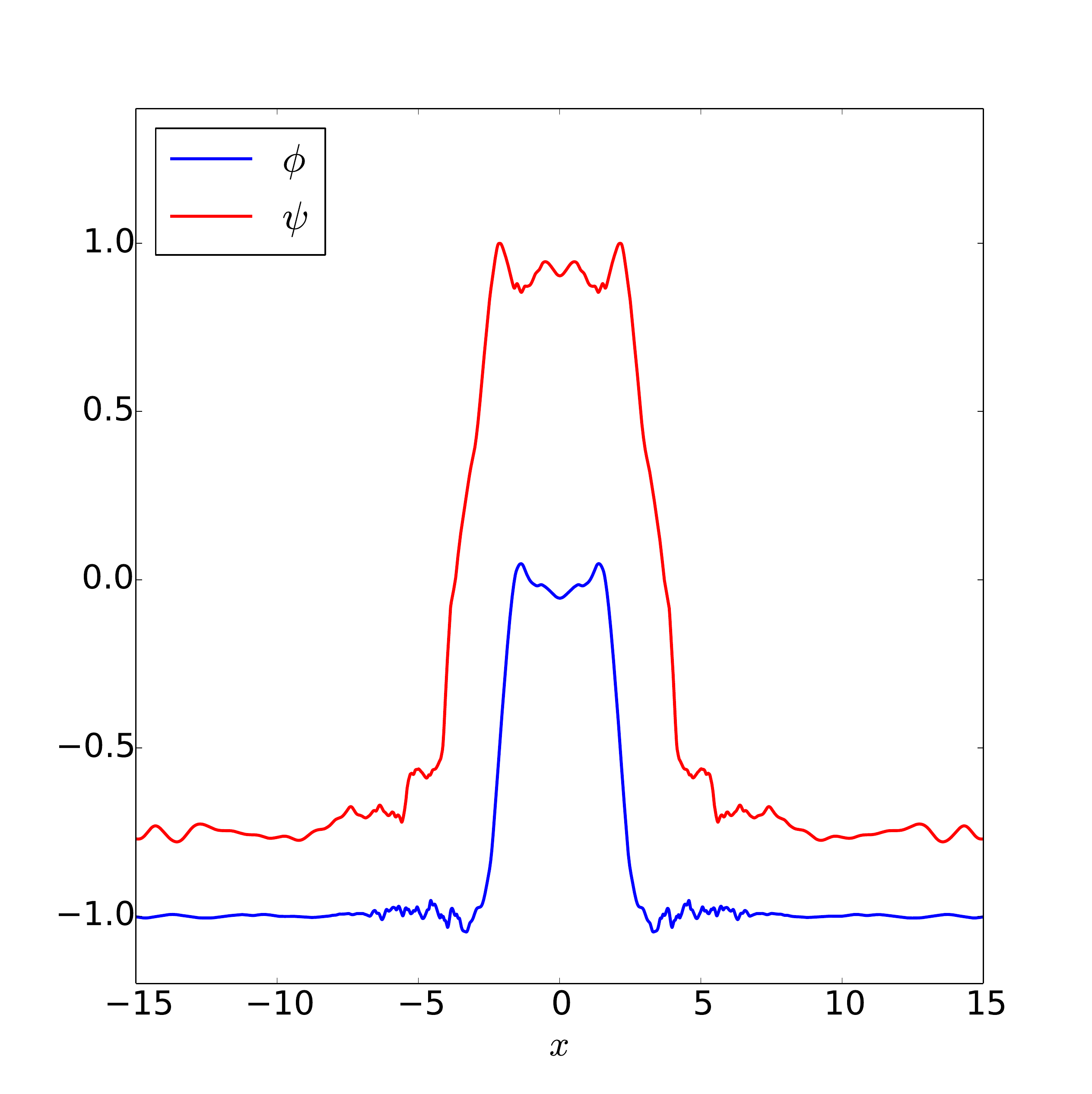}}
\subfigure[\, ]{\includegraphics[totalheight=4.cm]{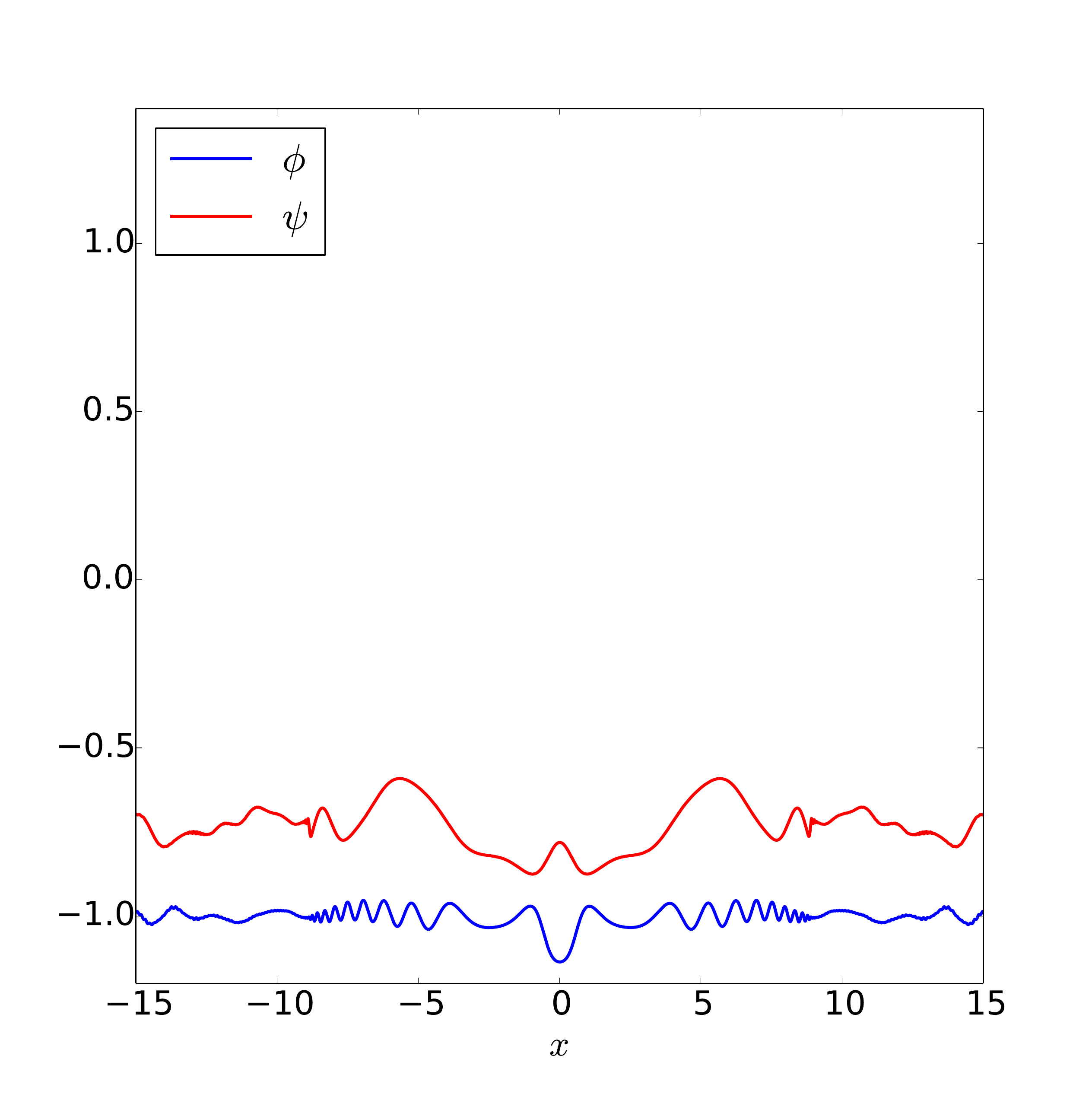}}\\
\subfigure[\, ]{\includegraphics[totalheight=4.cm]{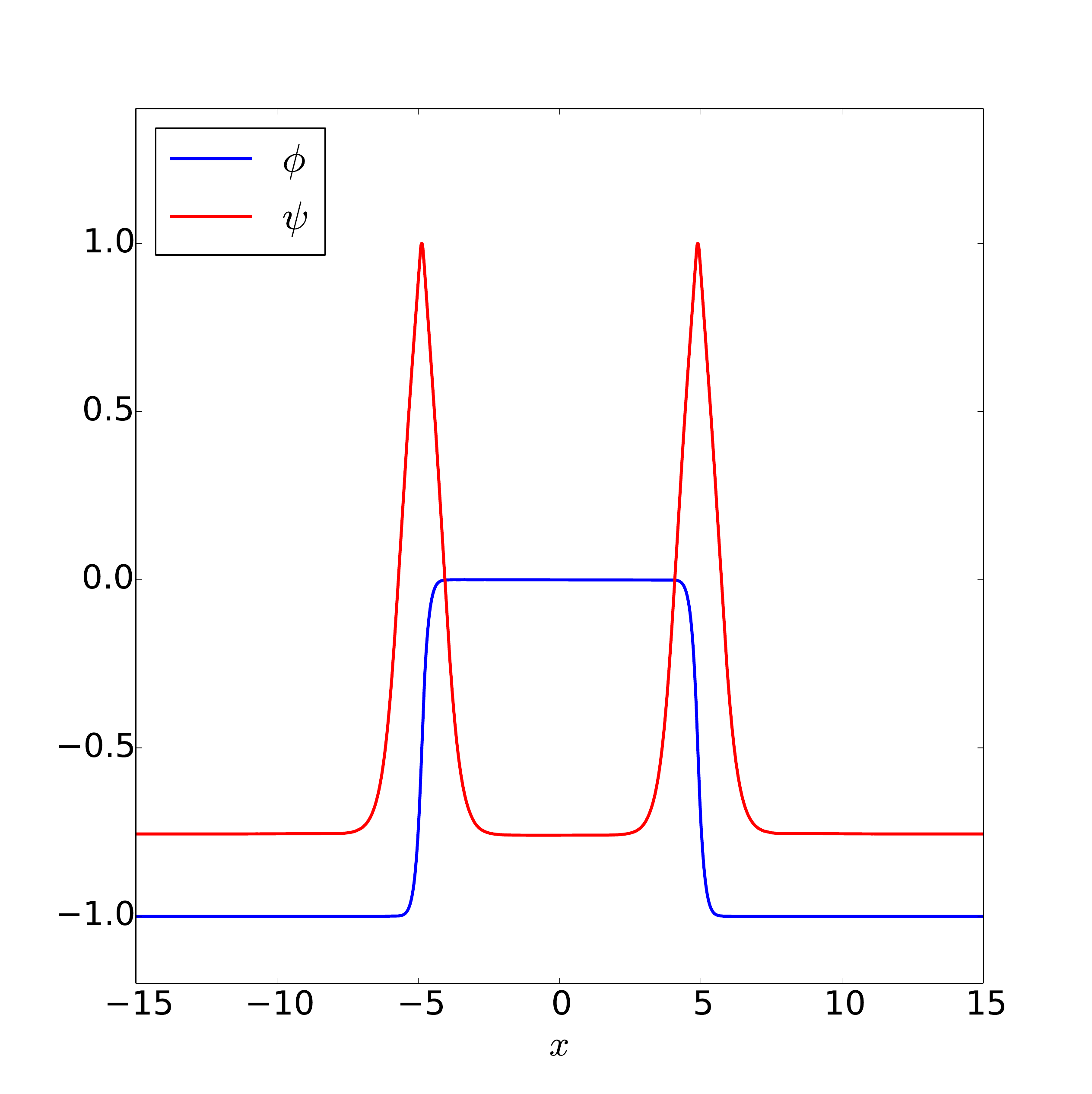}}
\subfigure[\, ]{\includegraphics[totalheight=4.cm]{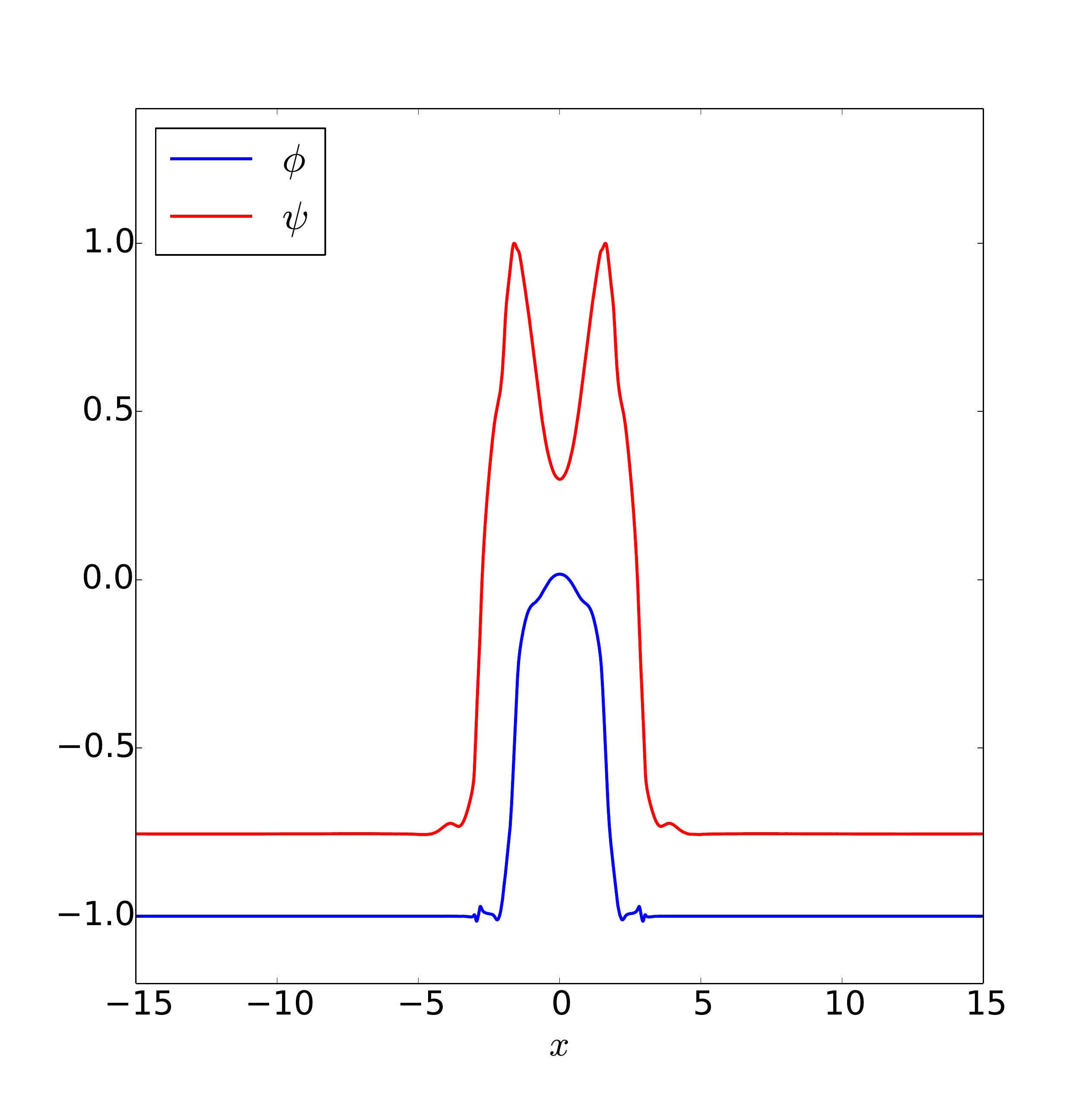}}
\subfigure[\, ]{\includegraphics[totalheight=4.cm]{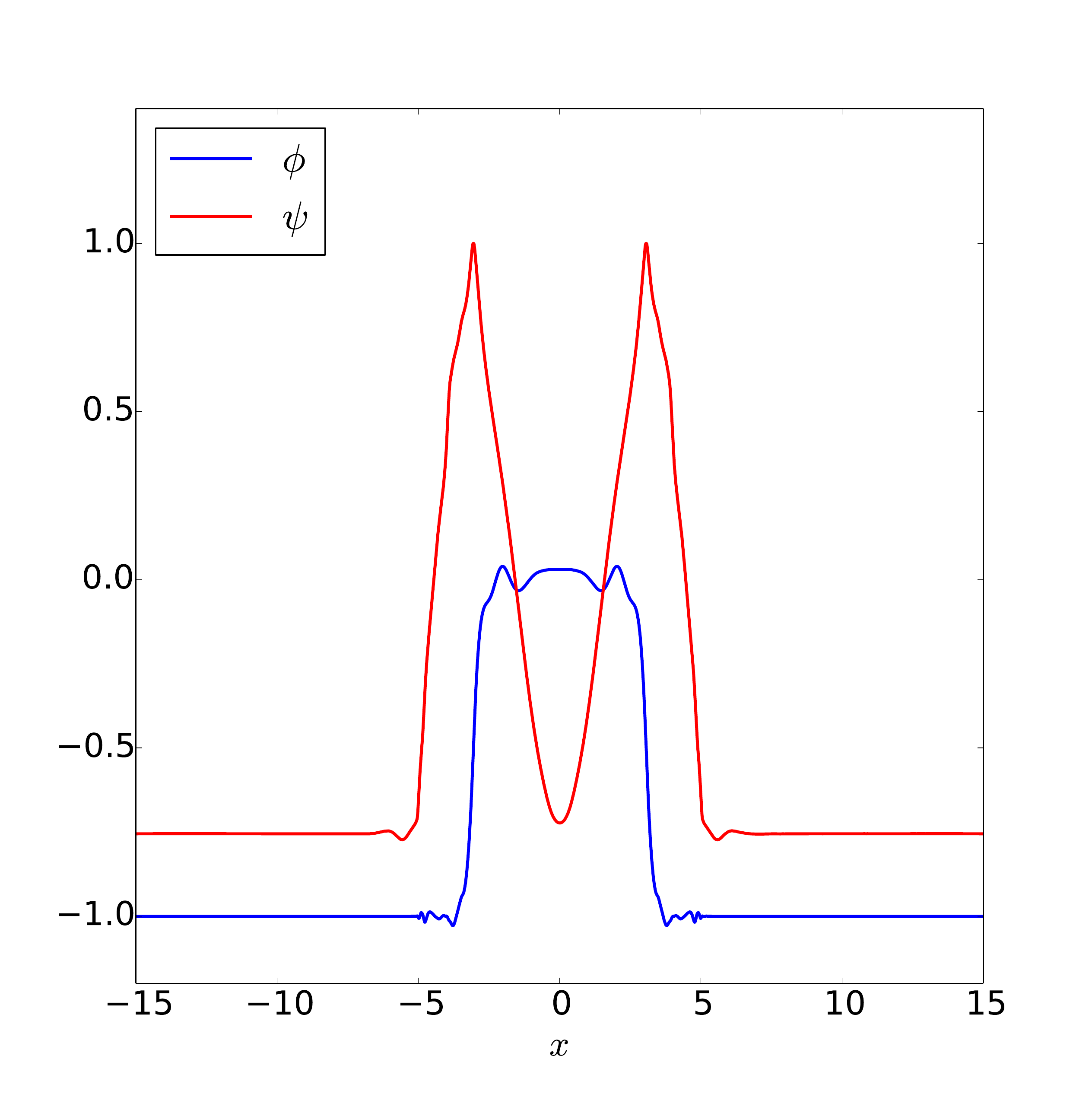}}
\subfigure[\, ]{\includegraphics[totalheight=4.cm]{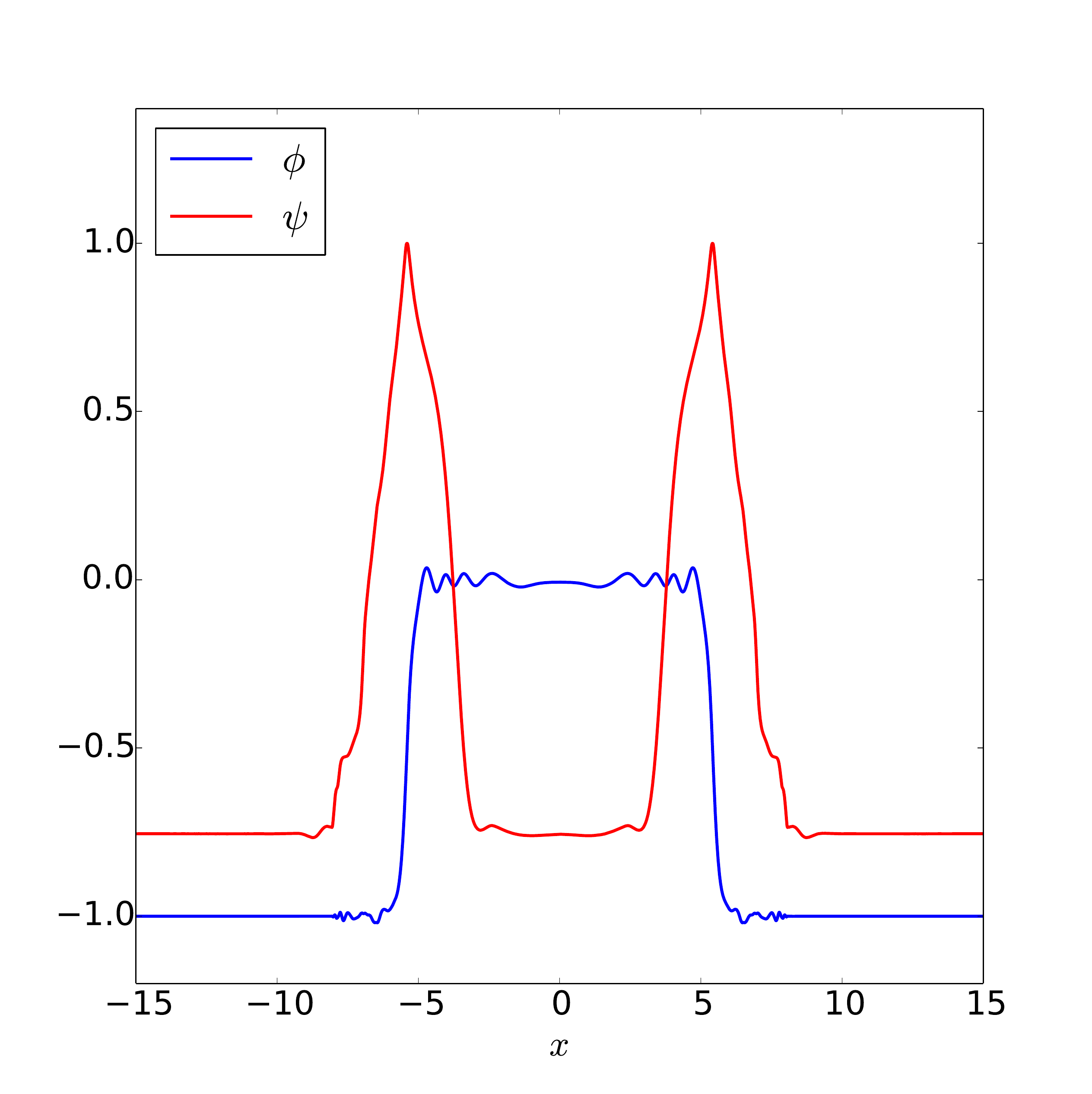}}\\
\caption{Snapshots of the field configuration during  $A\overline{B}$ scatterings for two choices of initial velocity~$v$. (a)-(d)~ $v=0.7$: the kinks collide and capture each other, forming an oscillon which ultimately decays to the false vacuum.
 (e)-(h)~$v=0.8$: the solitons reflect off each other. After three bounces in the shepherd field, the excited solitons separate and travel apart to infinity.}
\label{ABbar_profiles}
\end{figure}

Finally we consider scattering processes of the type $A\overline{B}$, where we attach a kink interpolating from false vacuum to true vacuum to one interpolating from true vacuum to false vacuum in the reverse direction. In our simulations, we choose that the $A$ sheep field $\phi$ interpolates from  $\phi=-1\rightarrow\phi=0$, and the $\overline{B}$ sheep field interpolates from  $\phi=0\rightarrow\phi=-1$. This initial setup of the solitons is shown in Fig.~\ref{ABbar_profiles}(a).

\begin{figure}[!htb]
\subfigure[\, ]{\includegraphics[totalheight=7.8cm]{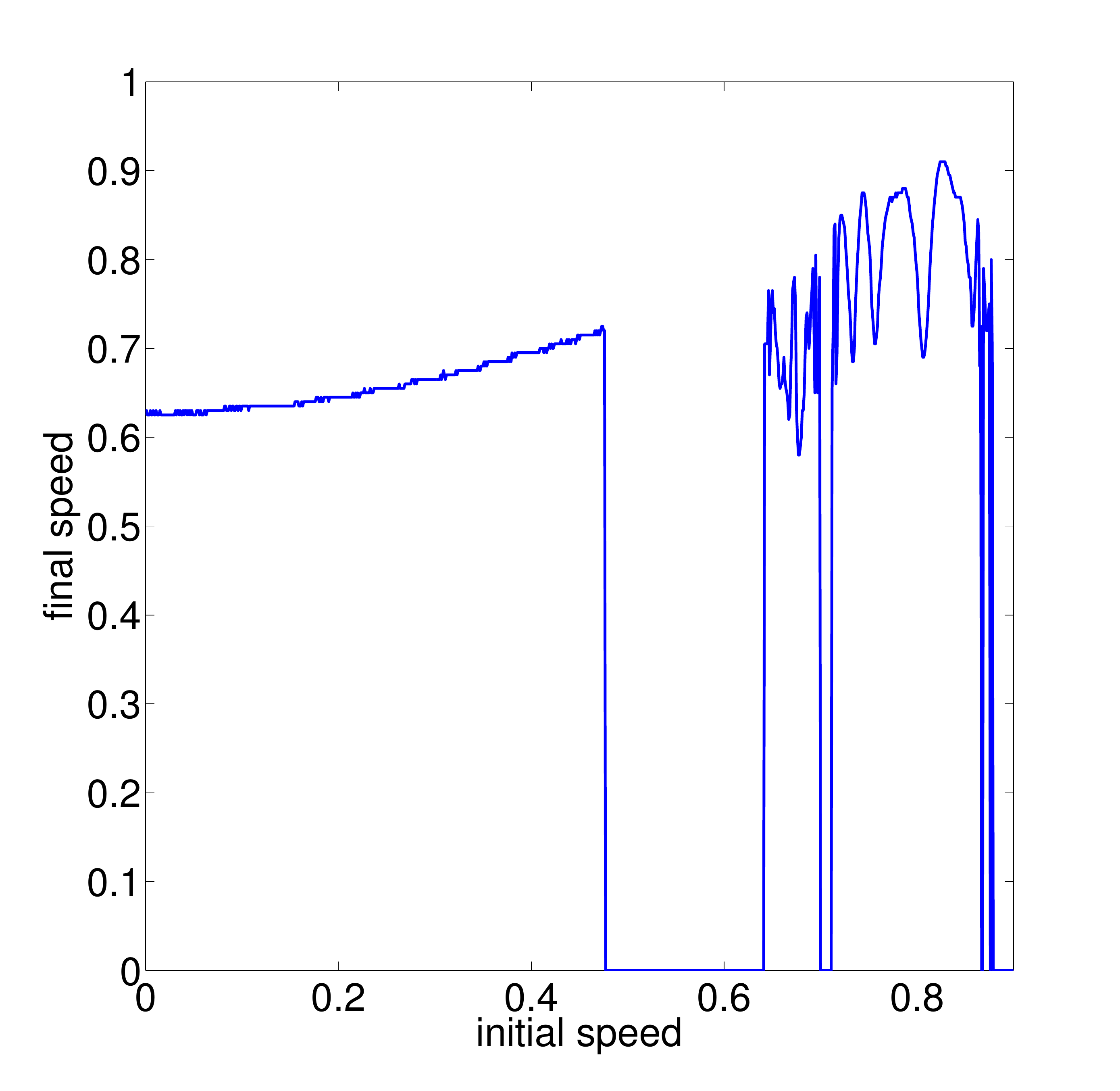}}
\subfigure[\, ]{\includegraphics[totalheight=7.8cm]{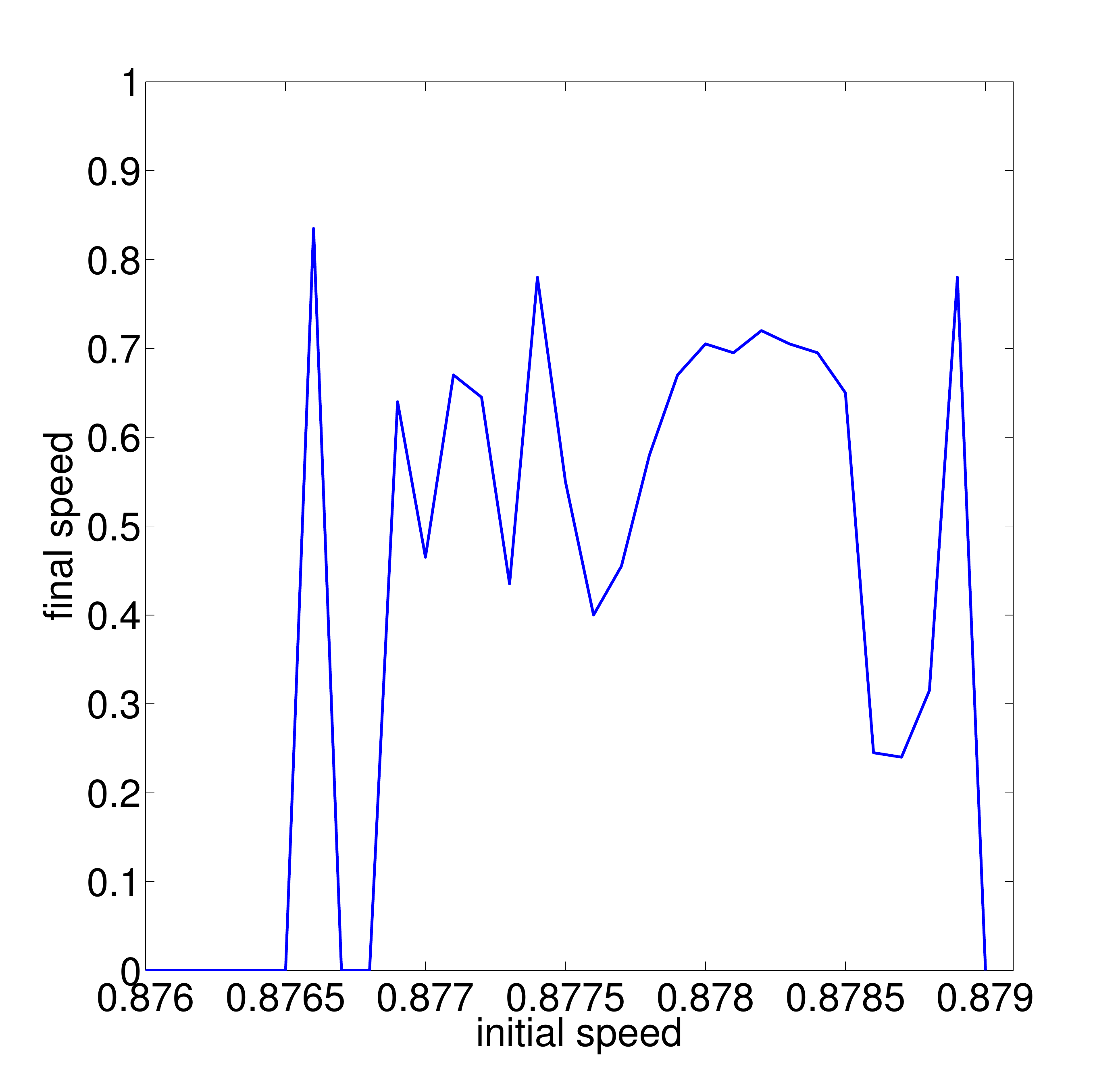}}
\caption{(a) The final velocity $v_{\text{fin}}$ after an $A\overline{B}$ collision as a function of the initial velocity $v$. When the final velocity $v_{\text{fin}}$ is plotted to be zero, we observe the kinks capture each other, forming an oscillon that ultimately decays to the false vacuum. (b) The same figure zoomed into the region $[0.876,0.88]$ to highlight two windows that are not visible in (a).}
\label{ABbar_speed}
\end{figure}

Note that any $A\overline{B}$ configuration has the region of true vacuum in between the kinks. Therefore the solitons will repel from rest. To overcome this repulsion and create soliton collisions, we boost the kinks towards each other with a given initial velocity. We carry out simulations over the range of initial velocities $0\le v\le 0.9$. Fig.~\ref{ABbar_speed} displays the final kink velocity as a function of the initial velocity, where final velocity is measured as the kink in the positive $x$-axis passes $x=50$. Similarly to the $AB$ scatterings discussed earlier, we observe three different scattering outcomes. Firstly, for velocities $0\le v\le 0.476$, the kinks cannot overcome their repulsion and will escape to infinity without ever colliding. Fig.~\ref{ABbar_speed}(a) shows that final velocity is always significantly higher than the corresponding initial velocity in this regime. 

For velocities greater than the critical velocity $v_{\text{crit}}=0.476$, there are two possible outcomes: (i) the solitons capture each other and form an oscillon which ultimately decays to the false vacuum, or (ii) the solitons reflect and recede from each other, escaping to infinity. Case (i), henceforth referred to as \emph{annihilation}, is illustrated by the snapshots in Fig.~\ref{ABbar_profiles}(a)-(d), which correspond to the initial velocity $v=0.7$. The $A\overline{B}$ pair collide and trap each other, as seen in (b) and (c).  Ultimately the configuration annihilates to the false vacuum, shown in (d). In Fig.~\ref{ABbar_speed}, the final velocity is plotted to be zero whenever the kinks annihilate. Case (ii), which we will refer to as \emph{escape}, is shown by the snapshots in Fig.~\ref{ABbar_profiles}(e)-(h), where the initial velocity is $v=0.8$. The kinks collide and separate, seen in (f) and (g). After three bounces in the shepherd field, the excited solitons travel apart to infinity, which is shown in (h). 

In the region $v> 0.476$, Fig~\ref{ABbar_speed}(a) shows that a fractal structure emerges in which there are several distinct windows of the escape behaviour separated by regions of annihilation. This is similar to the kind of fractal structure observed in other soliton models \cite{Campbell:1983xu,Peyrard:1984qn,Anninos:1991un,Goodman:2005,
PhysRevLett.98.104103}. In Fig.~\ref{ABbar_speed}(a), we increment the initial velocity by $10^{-3}$, and we expect that with greater resolution, more windows would be observed. Fig.~\ref{ABbar_speed}(b) zooms in on the interval $v\in[0.876,0.88]$, with initial velocity incremented by $10^{-4}$. This clarifies that there are at least two windows in this interval, though in Fig.~\ref{ABbar_speed}(a) there appeared to be only one.

\begin{figure}[!htb]
\includegraphics[totalheight=10.cm]{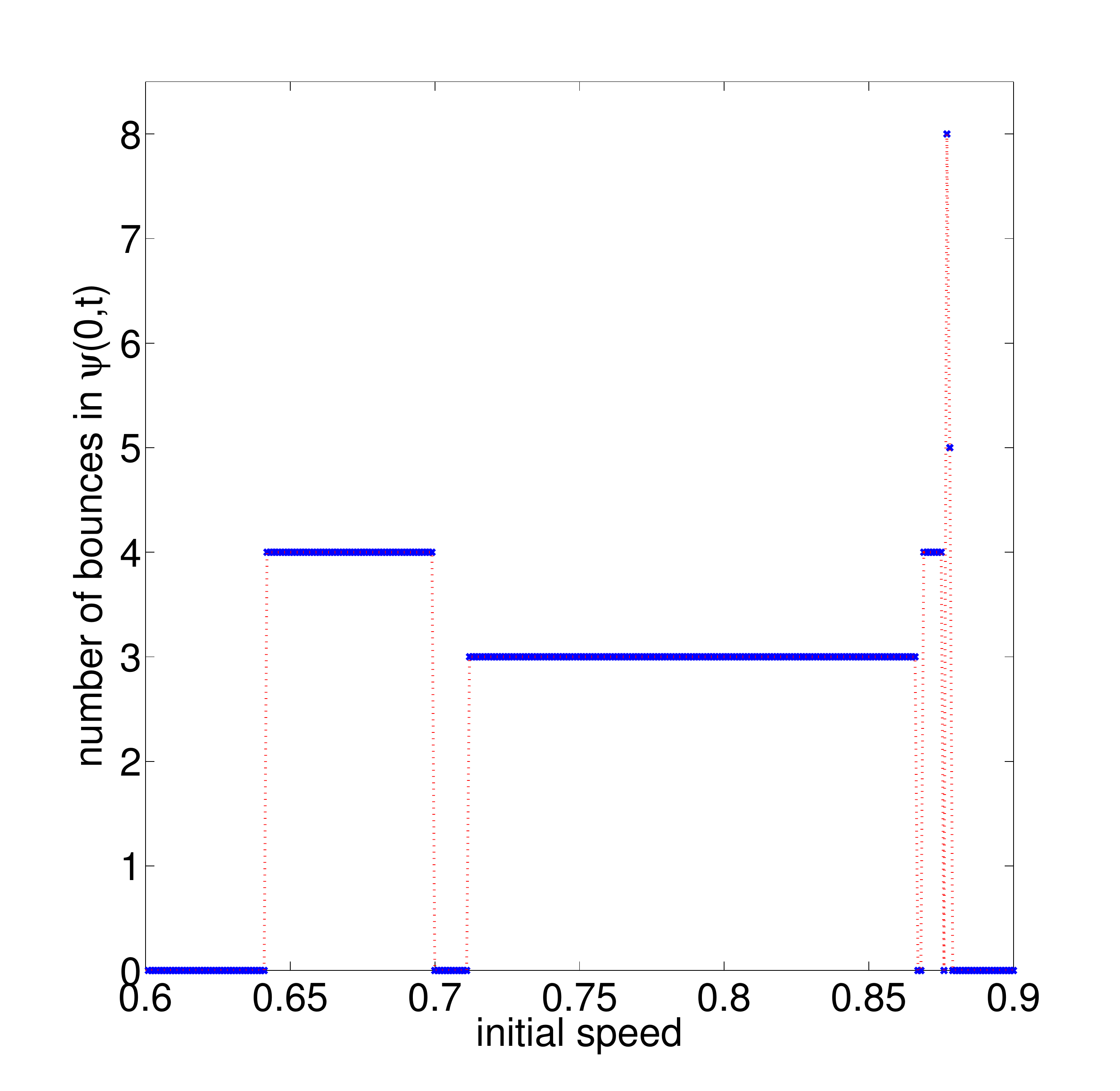}
\caption{Number of bounces as a function of initial velocity $v$ for collisions of type $A\overline{B}$. Here, the bounce number measures the number of times the shepherd field $\psi$ oscillates at the origin before the solitons separate. If no bounces are counted, then the solitons do not separate, but annihilate to the false vacuum.}
\label{ABbar_bounces}
\end{figure}

The windows of escape behaviour can be classified by the number of bounces in the shepherd field $\psi$ at the origin before the solitons separate and escape to infinity. Fig.~\ref{ABbar_bounces} shows the number of bounces in $\psi(0,t)$ as a function of the initial velocity over the interval where these windows appear. We have observed only windows of three bounces or more, with the three bounce window being by far the longest. For $v>0.866$, the windows are of a much shorter length. In this region, we expect to see more windows emerge if data is taken with a greater resolution. The shortest window is that with 8 bounces, which is the largest number of bounces that we have seen. It is structured as a group of six bounces, with two further bounces occurring after some time. This is unusual compared to the other scattering events, in which all of the bounces in the shepherd field happened close together.

\begin{figure}[!htb]
\subfigure[\, ]{\includegraphics[width=17cm]{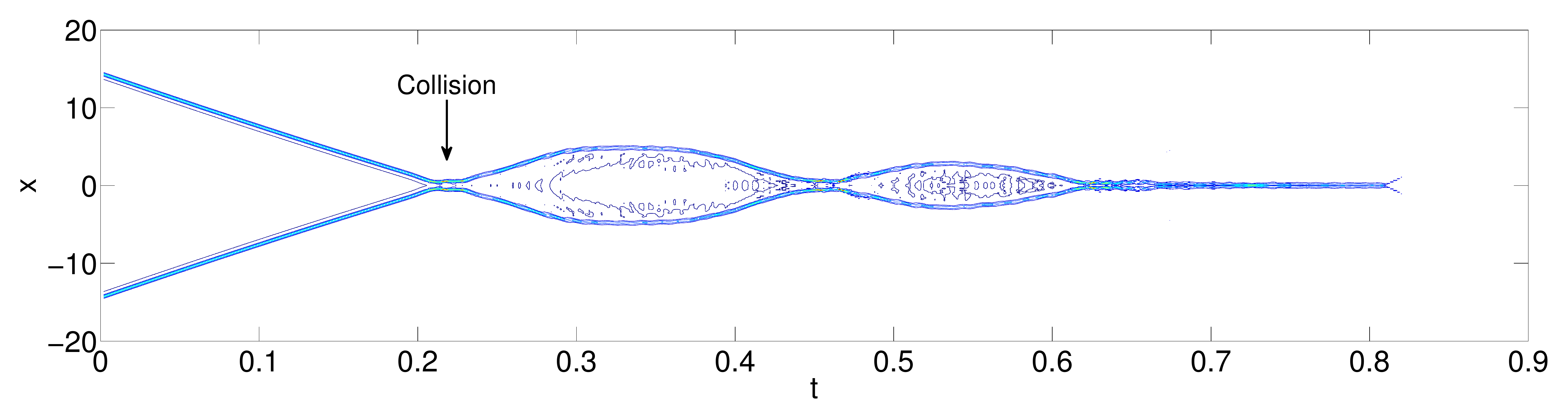}}
\subfigure[\, ]{\includegraphics[width=17cm]{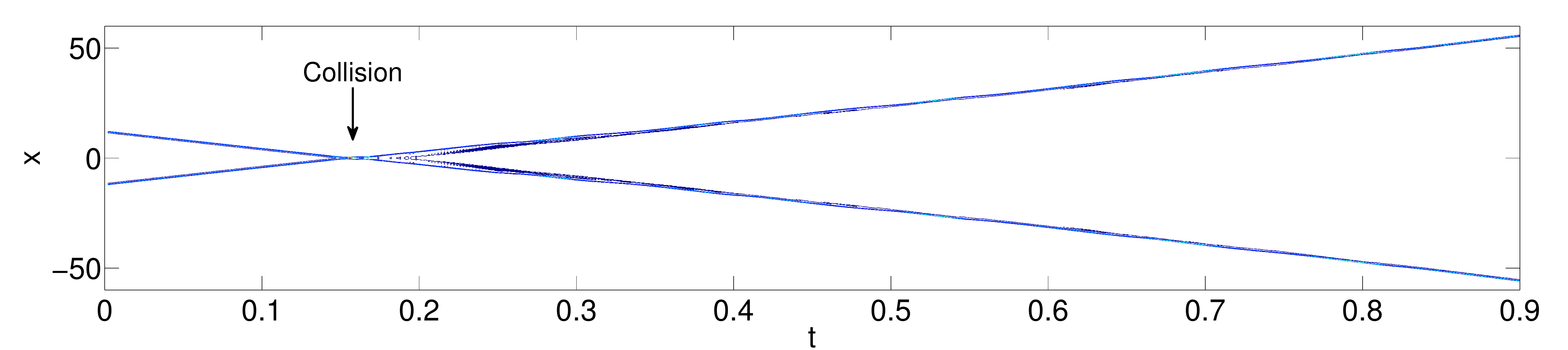}}
\caption{Contour plots of energy density for the sheep field $\phi$ during  $A\overline{B}$ interactions with two different choices of initial velocity $v$. We indicate the locations of the collisions in the plots. For reasons of clarity, time is divided by 100. (a) $v=0.7$: the solitons collide and form an oscillon which decays to the false vacuum. (b) $v=0.8$: the solitons collide and reflect off each other. Once they have separated, the solitons travel apart and escape to infinity.}
\label{ABbar_contours}
\end{figure}

To compare the paths taken by the solitons during $A\overline{B}$ scattering events resulting in annihilation or escape, contour plots of the energy density for the sheep field $\phi$ with two choices of initial velocity are contrasted in Fig.~\ref{ABbar_contours}. In Fig.~\ref{ABbar_contours}(a), where $v=0.7$, the collision of the solitons results in the formation of an oscillon at $t=80$ (for clarity, time is divided by 100). This gradually loses energy, and ultimately annihilates to the false vacuum. 
The escape behaviour is seen in Fig.~\ref{ABbar_contours}(b), where $v=0.8$.  Here the solitons reflect off each other, and once separated they travel apart and escape to infinity.

\section{Conclusions}\label{Sec_Con}
We have investigated kink collisions in a $1+1$ dimensional scalar field theory with multiple vacua. We summarize in Table~\ref{Tab_sum} the different scattering behaviours observed for $AB$, $BA$, $B\overline{A}$ and  $A\overline{B}$ kink collisions: repulsion, annhilation, formation of a true or false domain wall (``sticking") and reflecting off each other. Here, the tick symbol  indicates that the respective behaviour is observed, whereas the cross symbol denotes that the respective behaviour is not exhibited.

\begin{table}[htb]
\caption{Summary of the different scattering behaviours observed for $AB$, $BA$, $B\overline{A}$ and  $A\overline{B}$ kink collisions: repulsion, annhilation, formation of a true or false domain wall (``sticking") and reflecting off each other. Here, the symbol ``\color{green}\tickYes\color{black}"  indicates that the respective behaviour is observed, whereas ``\color{red}\tickNo\color{black}\,'' denotes that the respective behaviour is not exhibited. }
\begin{tabular}{c|cccc}
\hline\hline
& Repel & Annihilate & Stick & Reflection\\\hline
$AB$ &\,\color{green}\tickYes\color{black}&\, \color{red}\tickNo\color{black}&\, \color{green}\tickYes\color{black}&\, \color{green}\tickYes\color{black}\\
$BA$ &\, \color{red}\tickNo\color{black}&\, \color{red}\tickNo\color{black}&\,\color{green}\tickYes\color{black}&\, \color{red}\tickNo\color{black}\\
$B\overline{A}$ &\, \color{red}\tickNo\color{black}&\, \color{green}\tickYes\color{black}&\, \color{red}\tickNo\color{black}&\, \color{green}\tickYes\color{black}\\
$A\overline{B}$ &\, \color{green}\tickYes\color{black}&\, \color{green}\tickYes\color{black}&\, \color{red}\tickNo\color{black}&\, \color{green}\tickYes\color{black}\\\hline\hline
\end{tabular}
\label{Tab_sum}
\end{table}

Whenever there is true vacuum in between the kink pair (see $AB$, $A\overline{B}$), the kinks are found to be repulsive up to a critical value of the initial velocity at which they can overcome their mutual repulsion. Kink configurations with false vacuum in between (see $BA$, $B\overline{A}$) feel an attractive force and always capture each other after colliding.  

In $AB$ collisons, we observe ``sticking" and reflection velocity regimes. In the sticking velocity regimes, false domain walls \cite{Haberichter:2015xga} are formed, whereas for velocities within the reflection regime the kinks reflect off each other and then escape back to infinity. Within the reflection regime, there are always three bounces counted in the shepherd field before the kinks escape. The appearance of these escape bands in which the two kinks ultimately escape back to infinity has been related in similar kink models to an energy exchange between translational and vibrational modes \cite{Campbell:1983xu,Dorey:2011yw}. 

For $BA$ collisions, we find for all initial velocities the same scattering outcome, the formation of a true domain wall. 

Scattering processes of the type $A\overline{B}$ exhibit three different types of behaviour: repulsion at low initial velocities, annhilation and reflecting off each other with subsequent escape back to infinity. Recall that the annihilation does not occur immediately. The $A\overline{B}$ pair forms an oscillon which eventually decays to the true vacuum. Furthermore, we find regions of initial velocities or windows for which reflection and annhilation alternate. Note that the final kink velocity versus initial velocity graph has a fractal structure \cite{Anninos:1991un}. When we zoom into the initial kink velocity near to an interface between e.g. a three-bounce window and an annihilation window one sees the appearance of more four-bounce and annihilation windows. 

The final scattering outcome of any $B\overline{A}$ collision is the formation of an oscillon that radiates away energy and ultimately annihilates to the true vacuum. Note that there are two very different ways in which an oscillon can be formed. The $B\overline{A}$ kinks first reflect from one another and travel apart, before they are drawn back together and collide again. Depending on the initial velocity, the second collision results in either annhilation of the kink pair or in a number of bounces in the shepherd field. When refining the resolution, we expect to find other windows with two or more reflections.

\begin{figure}[!htb]
\subfigure[\, ]{\includegraphics[totalheight=4.0cm]{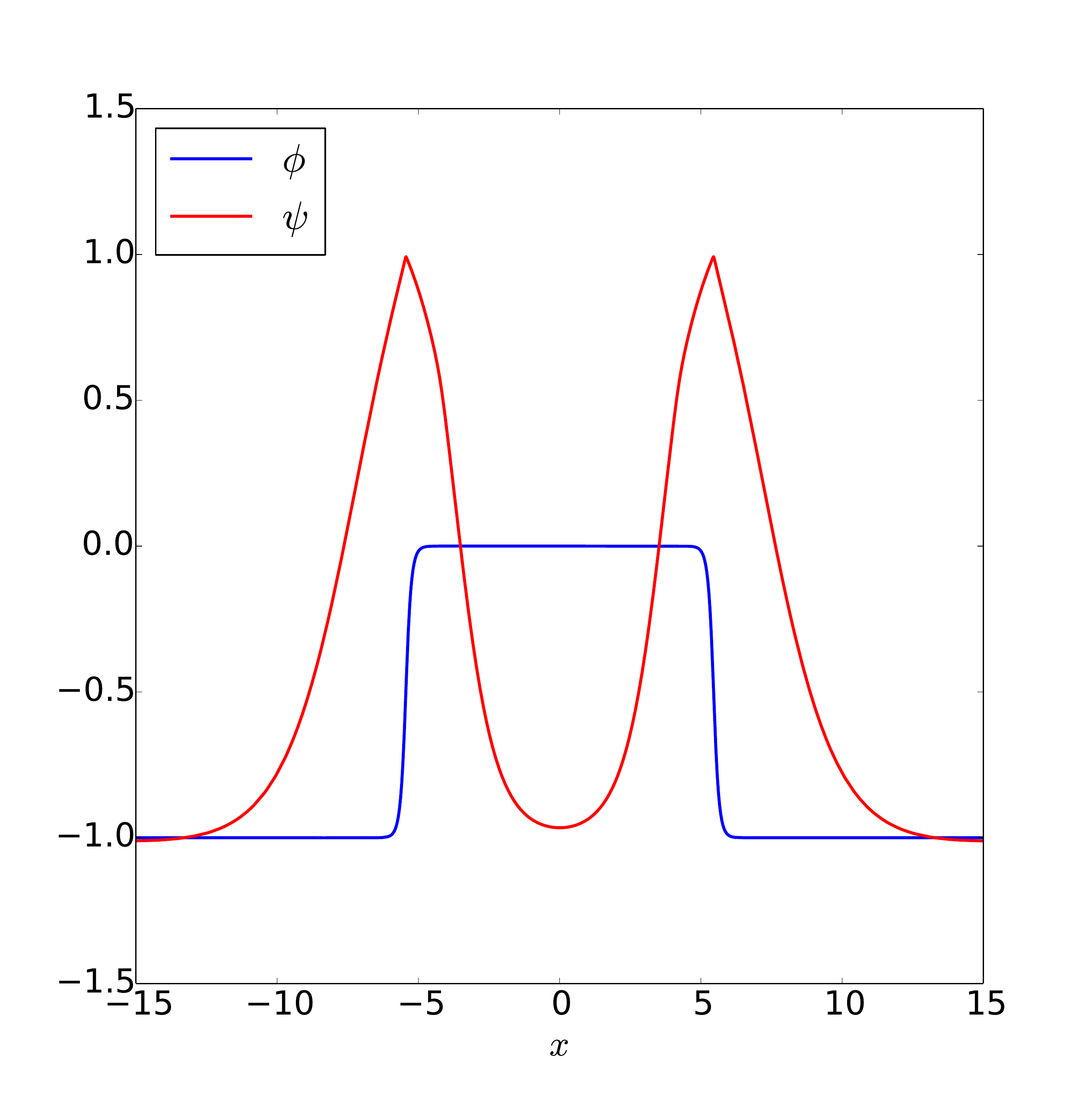}}
\subfigure[\, ]{\includegraphics[totalheight=4.0cm]{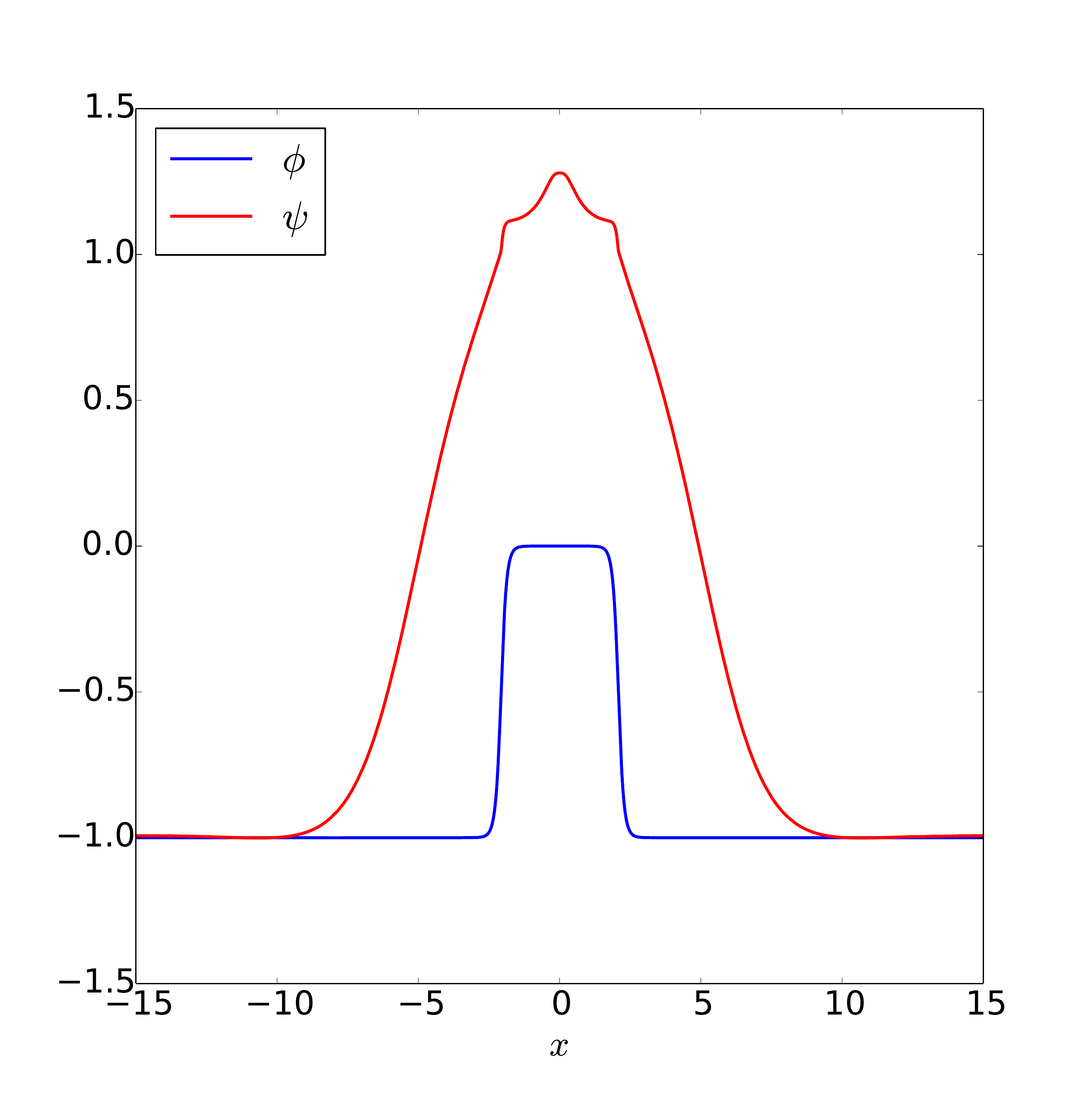}}
\subfigure[\, ]{\includegraphics[totalheight=4.0cm]{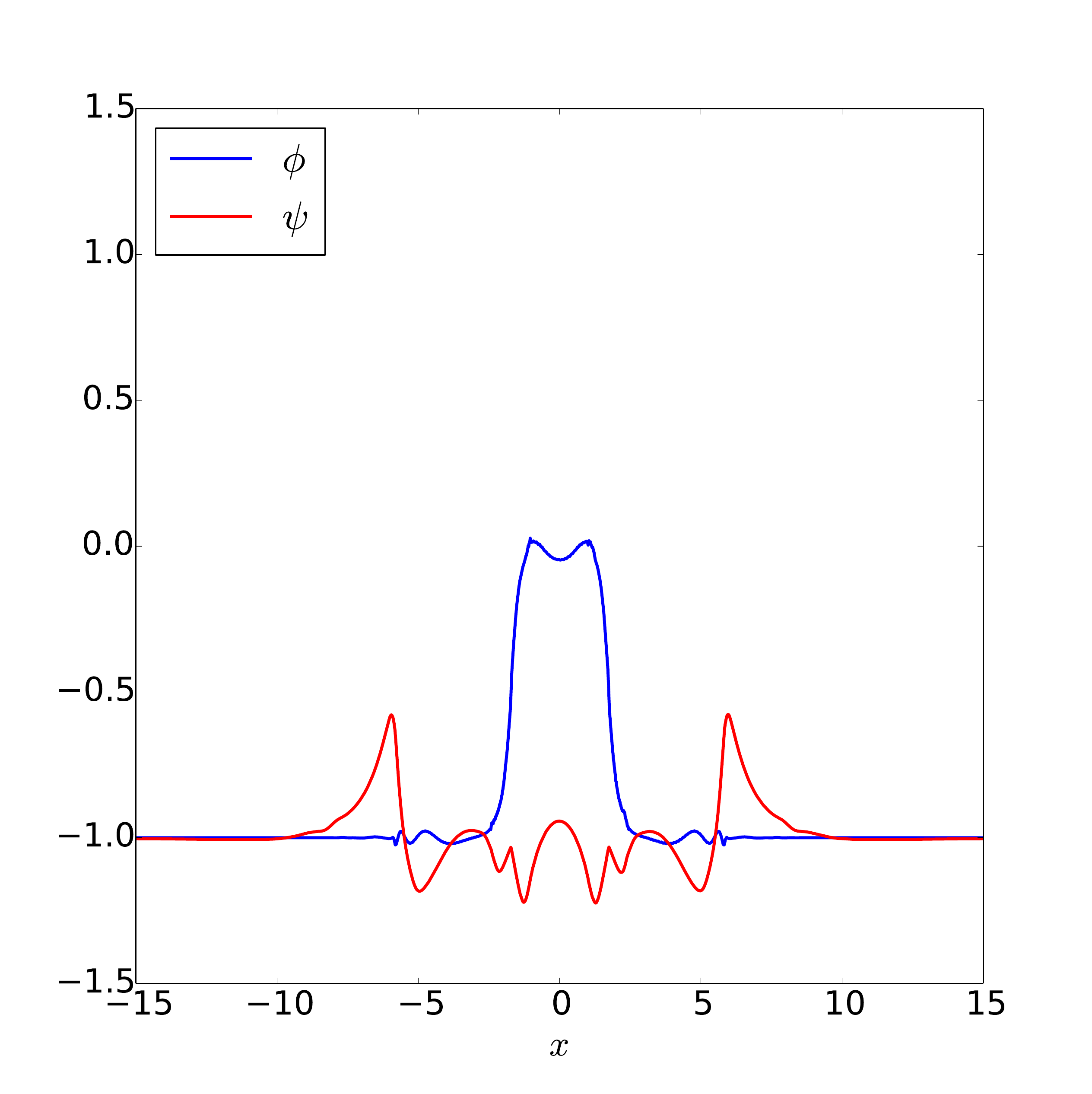}}
\subfigure[\, ]{\includegraphics[totalheight=4.0cm]{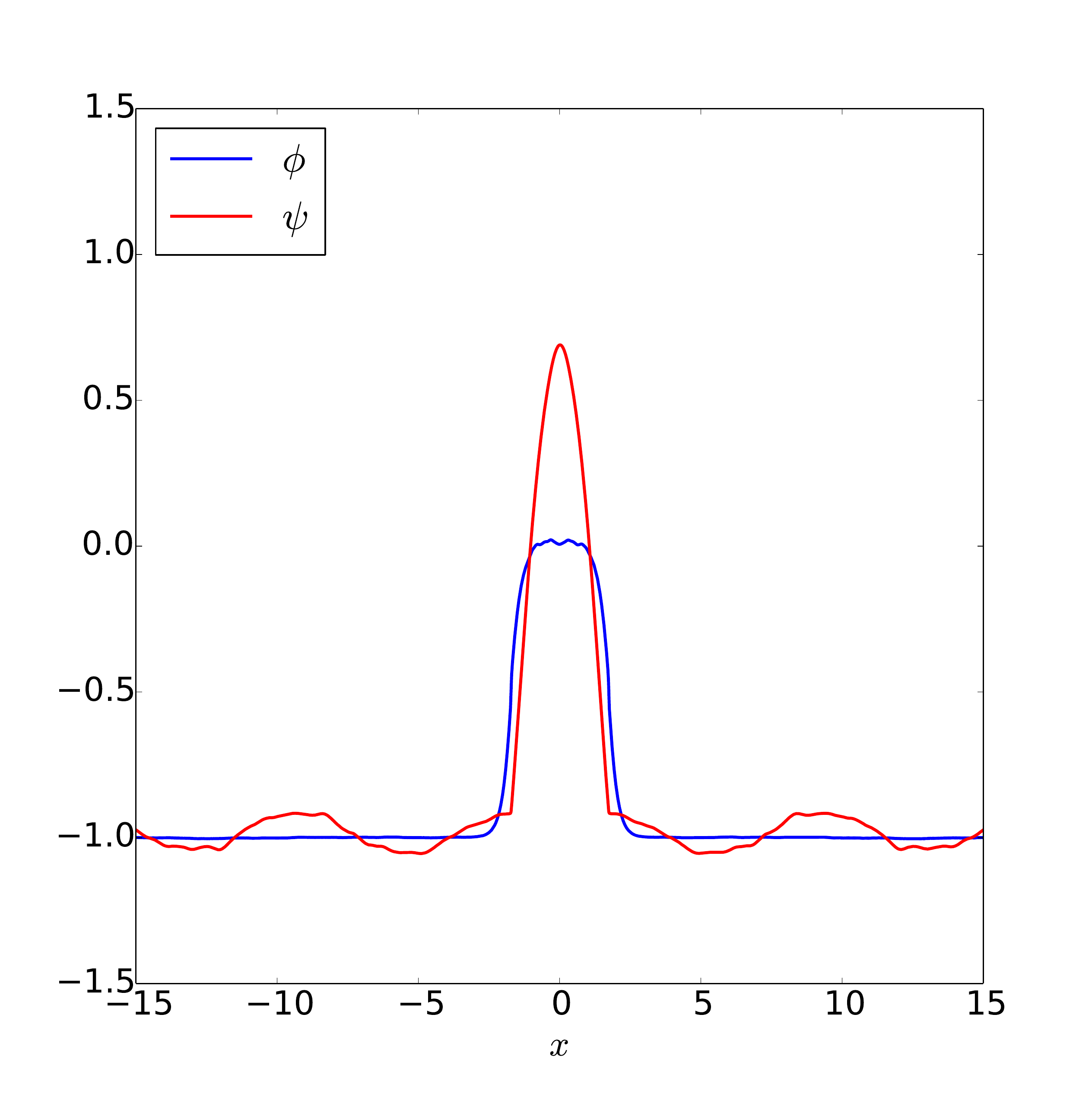}}
\caption{Example of a $A\bar{B}$ kink collision with coupling constants (\ref{const_SGlike}) chosen closer to the ones of the sine-Gordon model. We display snapshots of the field configuration for the initial velocity $v=0.8$:  An oscillon in the shepherd field is formed preventing the sheep fields to annihilate or to pass through each other.}
\label{Kinks_SGlike}
\end{figure}

Note that for the specific parameter choice (\ref{Para}), in all the scattering events we have examined so far the scattered kinks never pass through each other. Even when choosing the coupling constants so that the model approaches the sine-Gordon model, the kinks do not pass through each other. For example, scattering $A$ and $\bar{B}$ kinks in model (\ref{Lag_false}) with parameters 
\begin{align}
\alpha=1\,,\quad \beta=0.05\,,\quad \gamma=0.01\,,\quad a=1\,,\quad \epsilon_\psi=0.1\,,\quad \epsilon_\phi =0\, ,\quad \lambda=0.01\,.
\label{const_SGlike}
\end{align}
results in an oscillon in the shepherd field which prevents the sheep fields from passing through each other, see Fig.~\ref{Kinks_SGlike}. Note that recently it has been found that monopoles and antimonopoles prefer to annihilate rather than scatter \cite{Vachaspati:2015ahr}. Possible lines of further investigation could be the study of multi-kink solutions and their scattering behaviour.

\section*{Acknowledgements}

J.~A. and M.~H.  would like to thank Steffen Krusch for useful discussions throughout the project. M. H. thanks Andrzej Wereszczynski and the Jagiellonian University, Krakow for hospitality. 

M.~P. acknowledges financial support by NSERC of Canada. J. A. acknowledges the UK Engineering and Physical Science Research Council (Doctoral Training Grant Ref. EP/K50306X/1) and the University of Kent School of Mathematics, Statistics and Actuarial Science for a PhD studentship. The work of M. E. is supported in part by a JSPS Grant-in-Aid for Scientific Research (KAKENHI Grant No.~26800119).
The work of M.~E. and M.~N. is supported in part by a Grant-in-Aid for Scientific Research (KAKENHI Grant No.16H03984). The work of M.~N is also supported in part by a Grant-in-Aid for Scientific Research on Innovative Areas ``Topological Materials Science''  (KAKENHI Grant No. 15H05855) and ``Nuclear Matter in neutron Stars investigated by experiments and astronomical observations" (KAKENHI Grant No. 15H00841) from the Ministry of Education, Culture, Sports, Science, and Technology (MEXT) of Japan. The work of M.~E., M.~H. and M.~N. is supported by MEXT-Supported Program for the Strategic Research Foundation at Private Universities ``Topological Science" (Grant No. S1511006).

\appendix

\section{Point Particle Approximation}
In this appendix, we discuss a toy model which mimics some of the features of the scattering of our kinks. Let $m$ be the reduced mass of $A$ and $B$ kinks, and let $r(t)$ be their relative position at time $t$. We try to model the sheep scattering using a point particle approximation with equation of motion
\begin{align}\label{eqm_point}
m\ddot{r} = 2mg\left(\tanh^2(r)-f\right) + \frac{2a}{r}e^{-br} - d\dot{r}e^{-cr}  ,
\end{align}
where $m,~g,~a,~b,~c,~d,~f$ are constants. Here, the first term in (\ref{eqm_point}) allows for the existence of metastable bound state, while the second term acts as a contact interaction term preventing $A$ and $B$ kinks from passing through each other. To allow for radiation, we add a friction term of the form  $d\dot{r}e^{-cr}$. We solve (\ref{eqm_point}) subject to the conditions
\begin{align}\label{ICs_point}
r(0)=2,\quad\dot{r}(0)=v_0,
\end{align}
so that the points start at positions $\pm 1$, and where $v_0$ is used to specify the initial velocity. The signs of $a$ and $g$ are used to determine which type of configuration we study.

\begin{figure}[!htb]
\centering
\subfigure[\,]{\includegraphics[height=4cm]{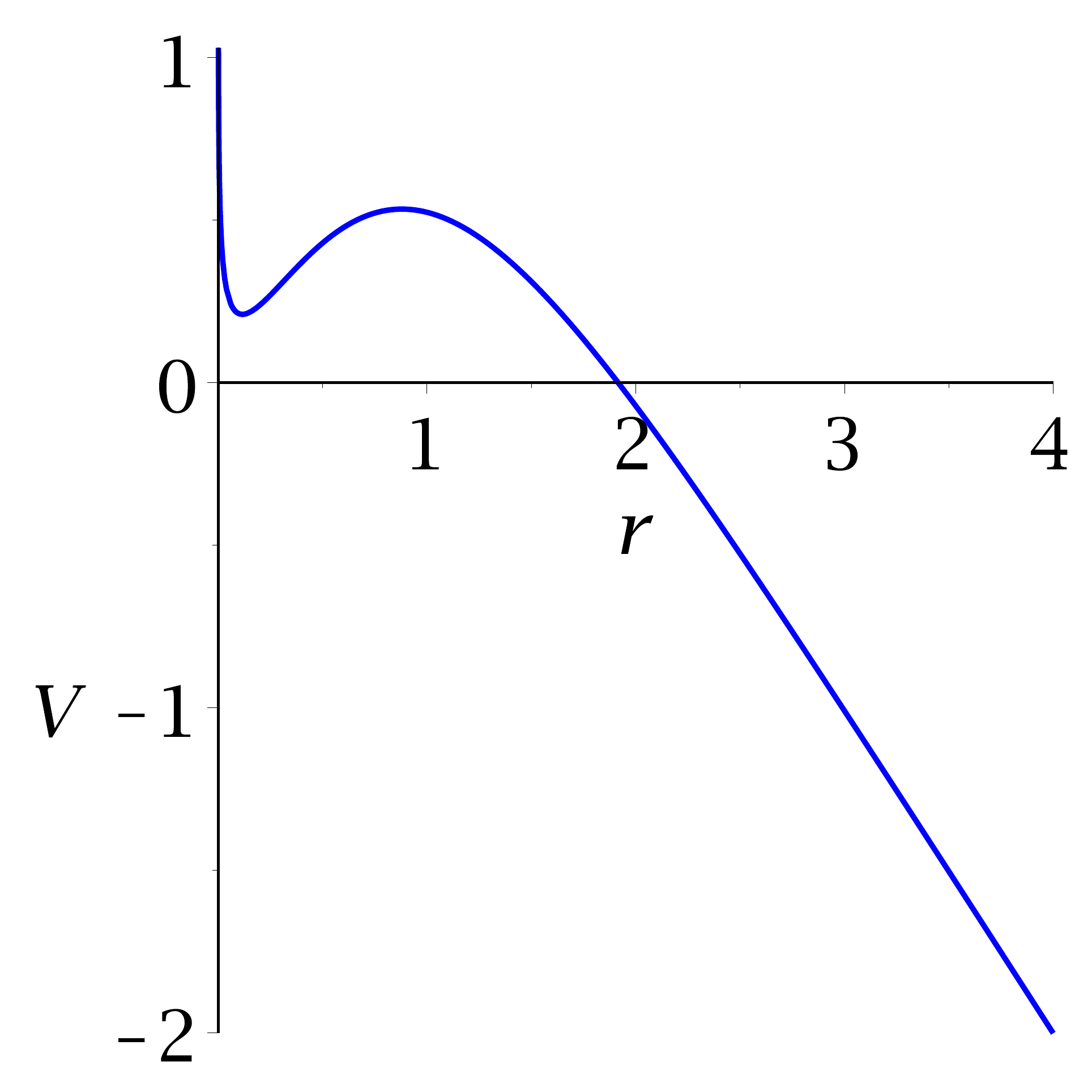}}\qquad
\subfigure[\,]{\includegraphics[height=4cm]{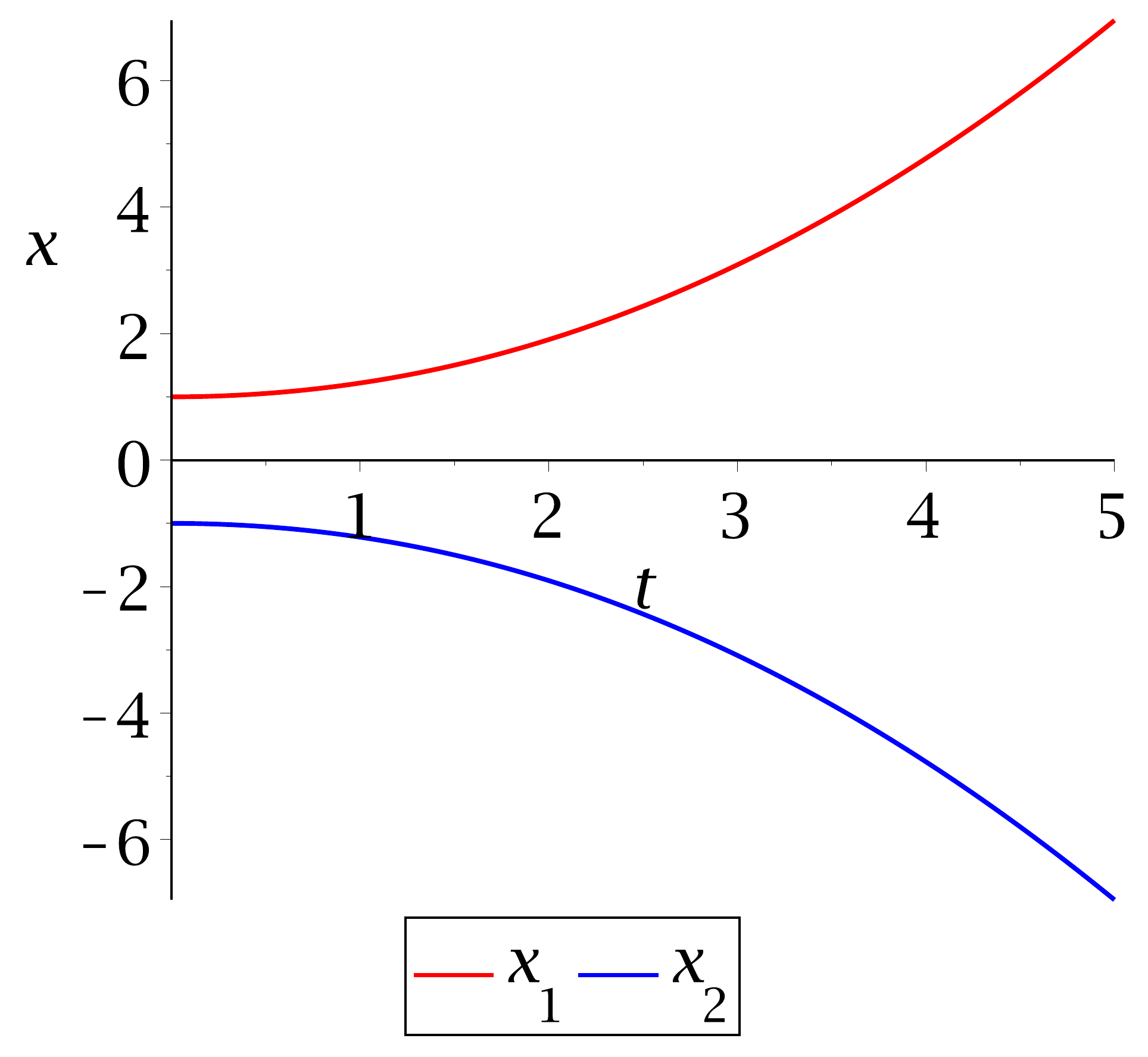}} \\
\subfigure[\,]{\includegraphics[height=4cm]{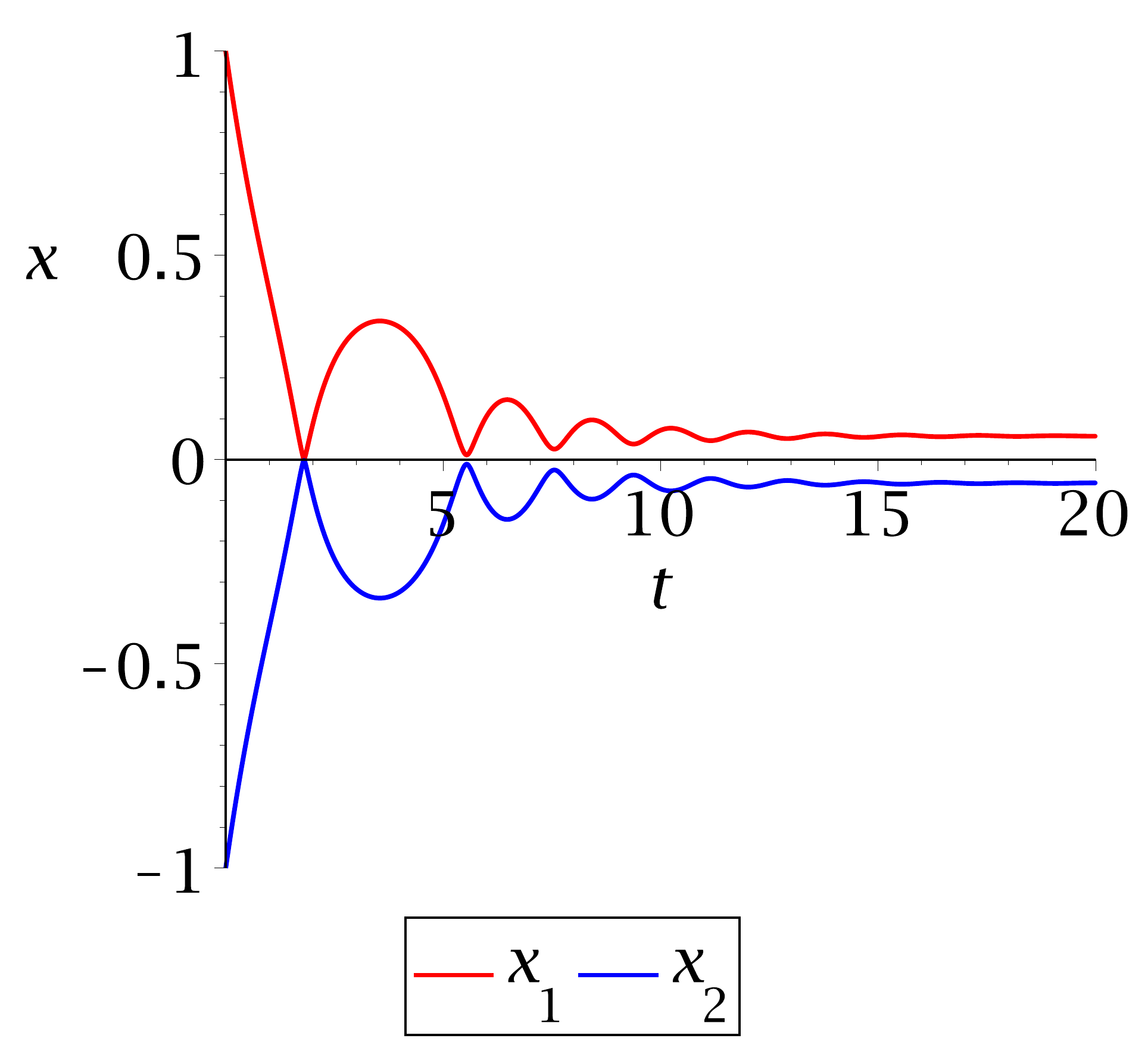}}\qquad
\subfigure[\,]{\includegraphics[height=4cm]{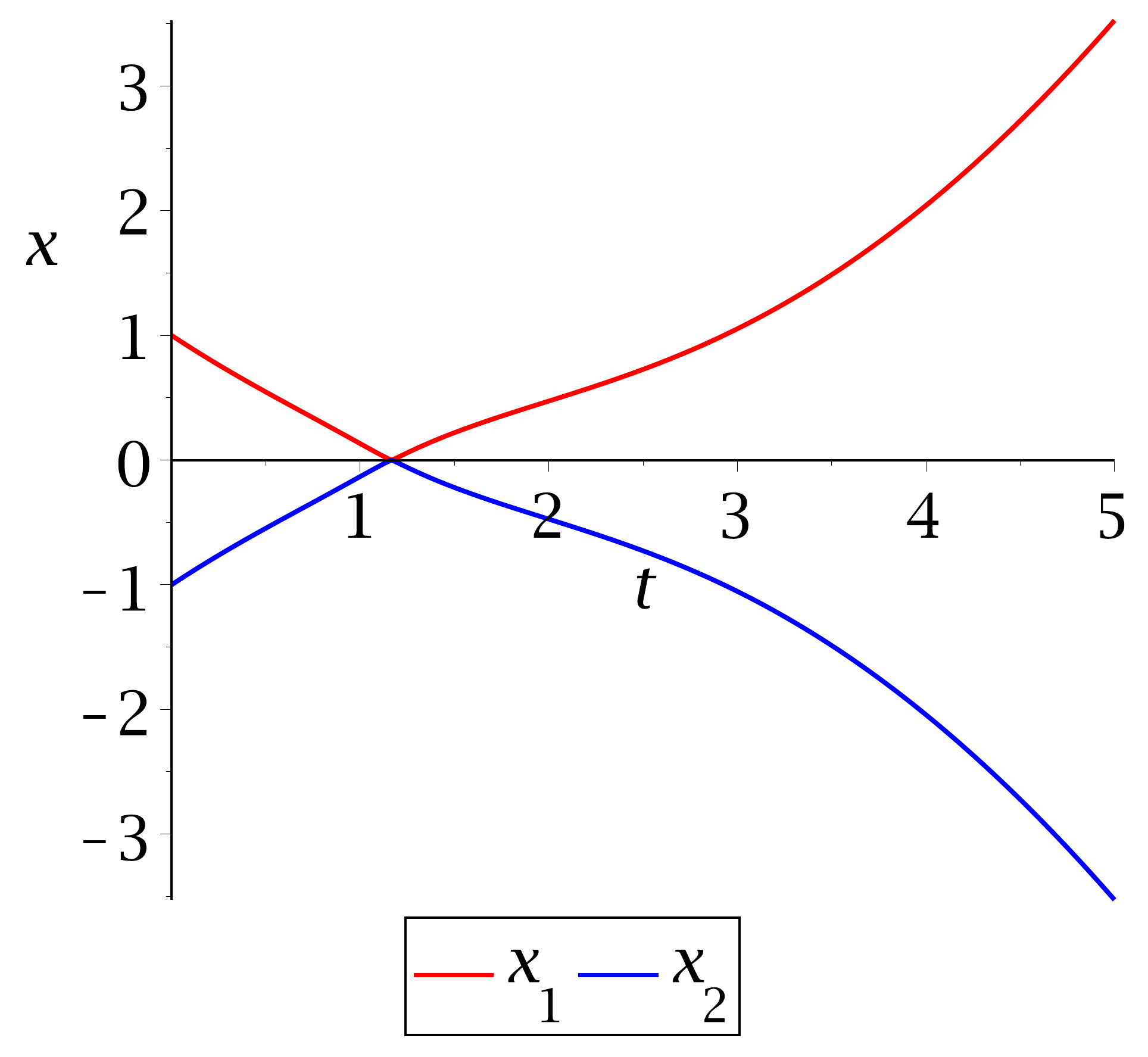}}
\caption{(a) The potential $V(r)$ used to mimic $AB$ scattering. (b) Solutions $x_1(t)$ in red and $x_2(t)$ in blue, for initial velocity $v_0=0$. (c) Solutions $x_1(t)$ and  $x_2(t)$ for $v_0=-1.5$. (d) Solutions  $x_1(t)$ and  $x_2(t)$ for $v_0=-2$.}
\label{AB}
\end{figure}

By choosing $a>0$ and $g>0$, we can replicate the three scattering outcomes of the $AB$ configuration. For example, set the parameters to be
\begin{align}
 m=1,\quad g=1,\quad a=0.1, \quad b=5,\quad c=3,\quad d=1,\quad f=\tfrac{1}{2} .
 \end{align}
The corresponding potential $V(r)$ is displayed in Fig.~\ref{AB}(a). It has one minimum, which is located near the origin. For different values of the initial velocity $v_0$, we observe different scattering outcomes. In Fig~\ref{AB}(b), we show the solutions $x_1(t)$ and $x_2(t)$ for $v_0=0$, where these are related to the solution of \eqref{eqm_point} by $r(t)=x_1(t)-x_2(t)$. This mimics the behaviour of the $AB$ kinks for low initial velocities, where they cannot overcome their repulsion, and separate to infinity. Fig~\ref{AB}(c) displays the solutions for initial velocity $v_0=-1.5$. Here we see the formation of a metastable bound state, akin to the formation of a false domain wall in the $AB$ ``sticking" behaviour. Finally, in Fig.~\ref{AB}(d), we take initial velocity $v_0=-2$, and observe the points reflect off each other and separate to infinity.

\begin{figure}[!htb]
\centering
\subfigure[\,]{\includegraphics[height=4cm]{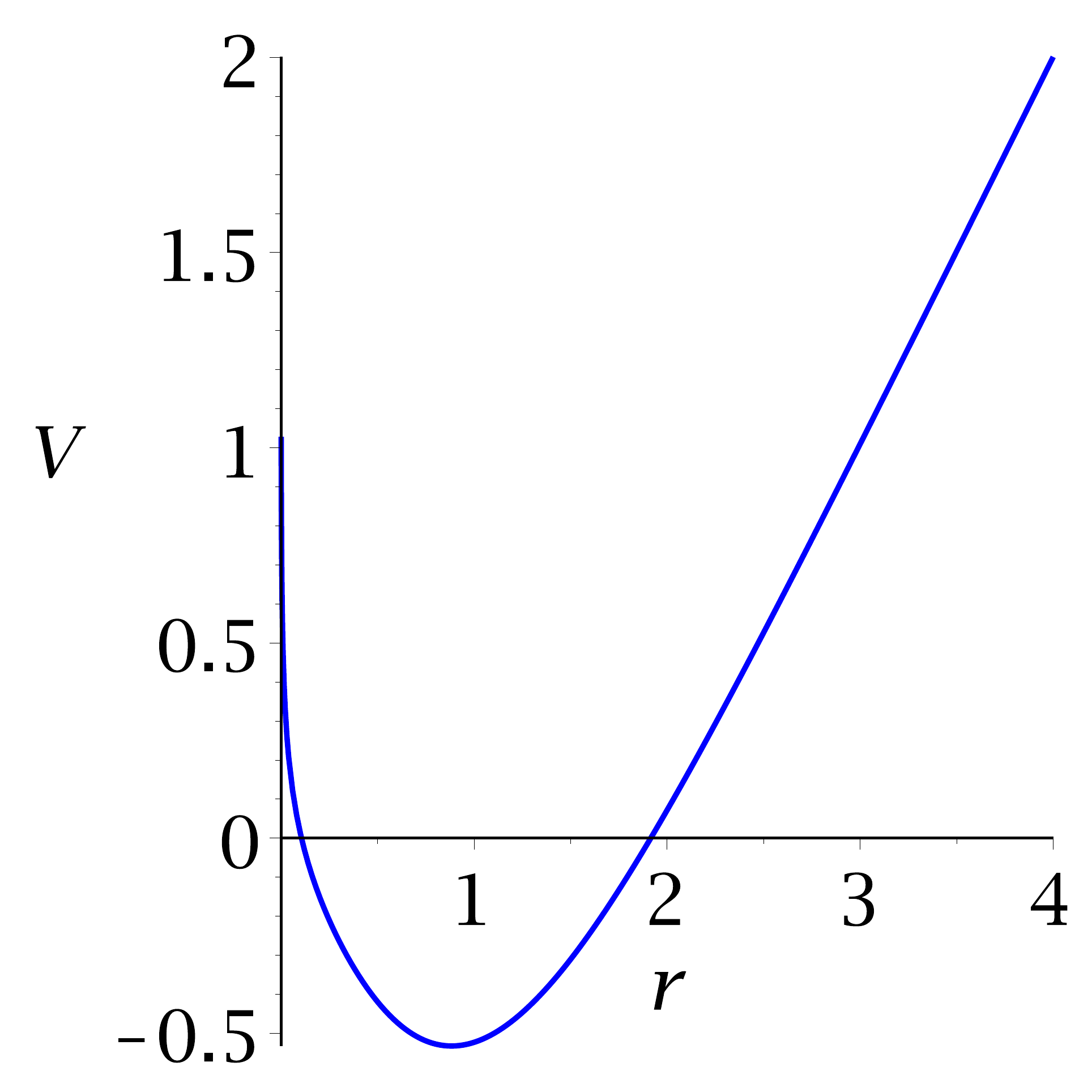}}\qquad
\subfigure[\,]{\includegraphics[height=4cm]{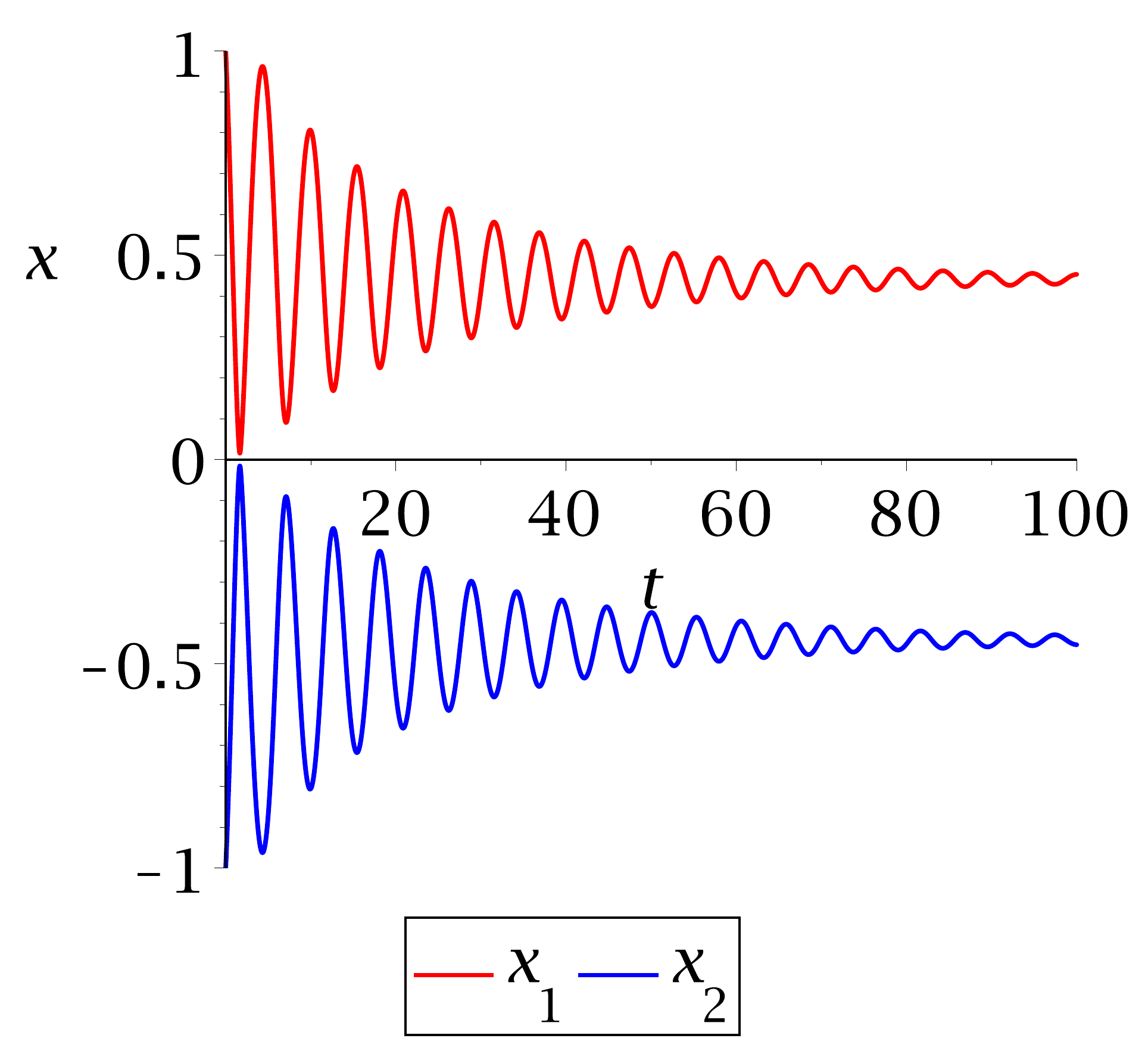}}
\caption{(a) The potential $V(r)$ used to mimic $BA$ scattering. (b) Solutions $x_1(t)$ in red and $x_2(t)$ in blue for $v_0=-1$.}
\label{BA}
\end{figure}

To model the $BA$ configuration, we take $a>0$ and $g<0$. For example,
\begin{align}
 m=1,\quad g=-1,\quad a=0.1, \quad b=5,\quad c=3,\quad d=1,\quad f=\tfrac{1}{2} .
\end{align}
We display the corresponding potential $V(r)$ in Fig.~\ref{BA}(a). It has one minimum, located near $r=1$. When solving the equation of motion \eqref{eqm_point}, the result is always the formation of a bound state. This is also true of the $BA$ configurations, which will ultimately form a true domain wall regardless of the initial velocity. Fig.~\ref{BA}(b) shows the solutions $x_1(t)$ and $x_2(t)$ obtained from the initial velocity $v_0=-1$, which form a bound state over time.

\begin{figure}
\centering
\subfigure[\,]{\includegraphics[height=4cm]{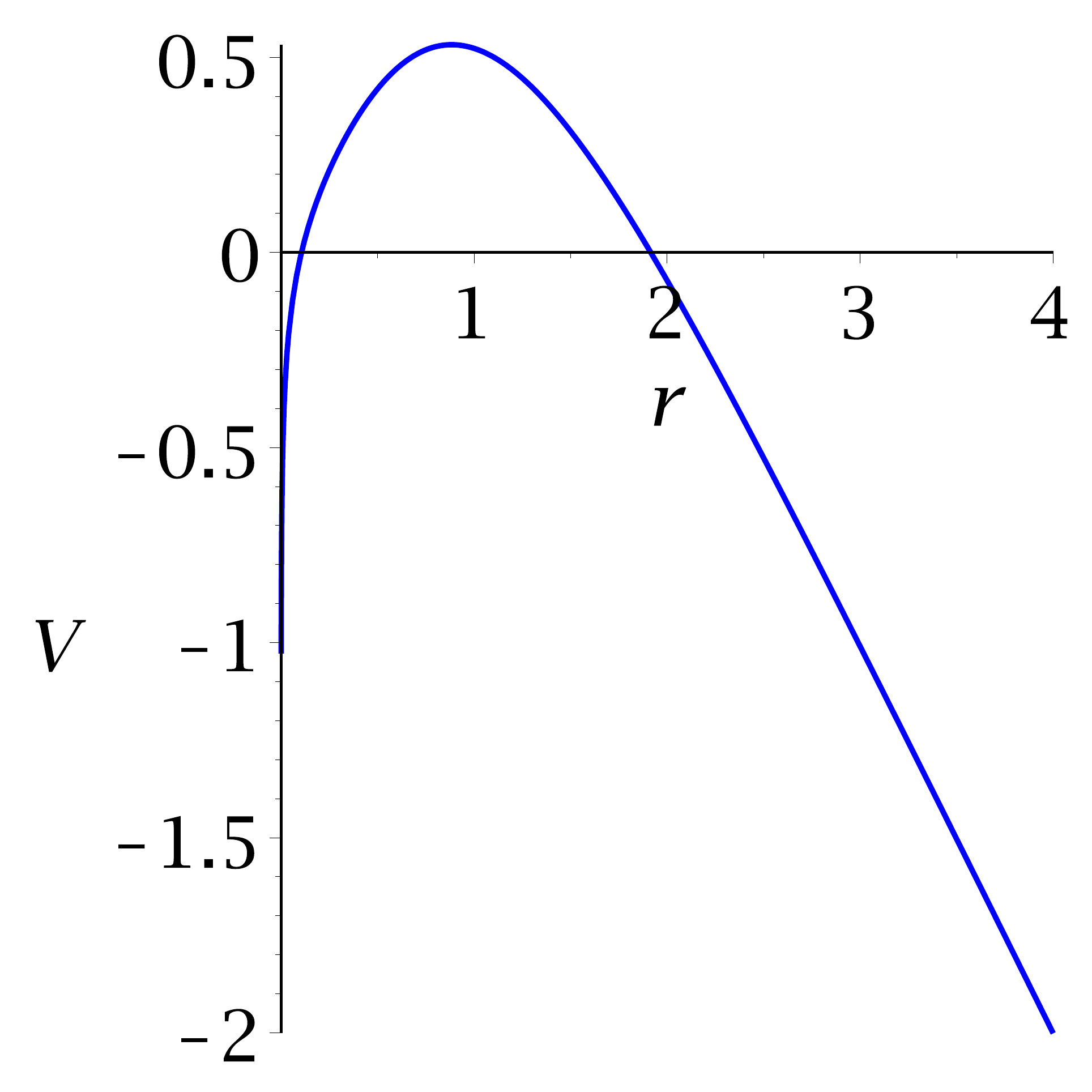}}\quad
\subfigure[\,]{\includegraphics[height=4cm]{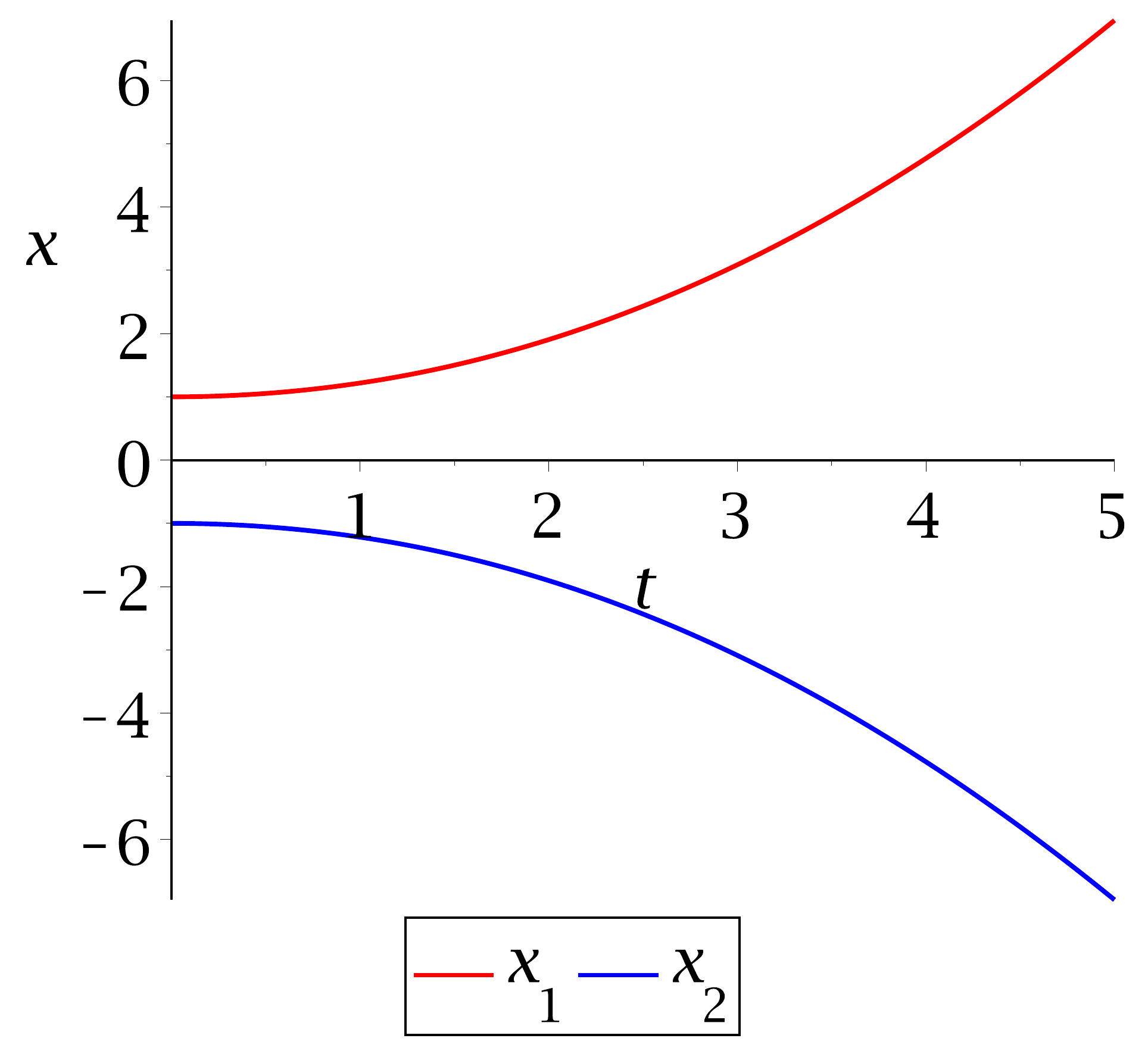}}\quad
\subfigure[\,]{\includegraphics[height=4cm]{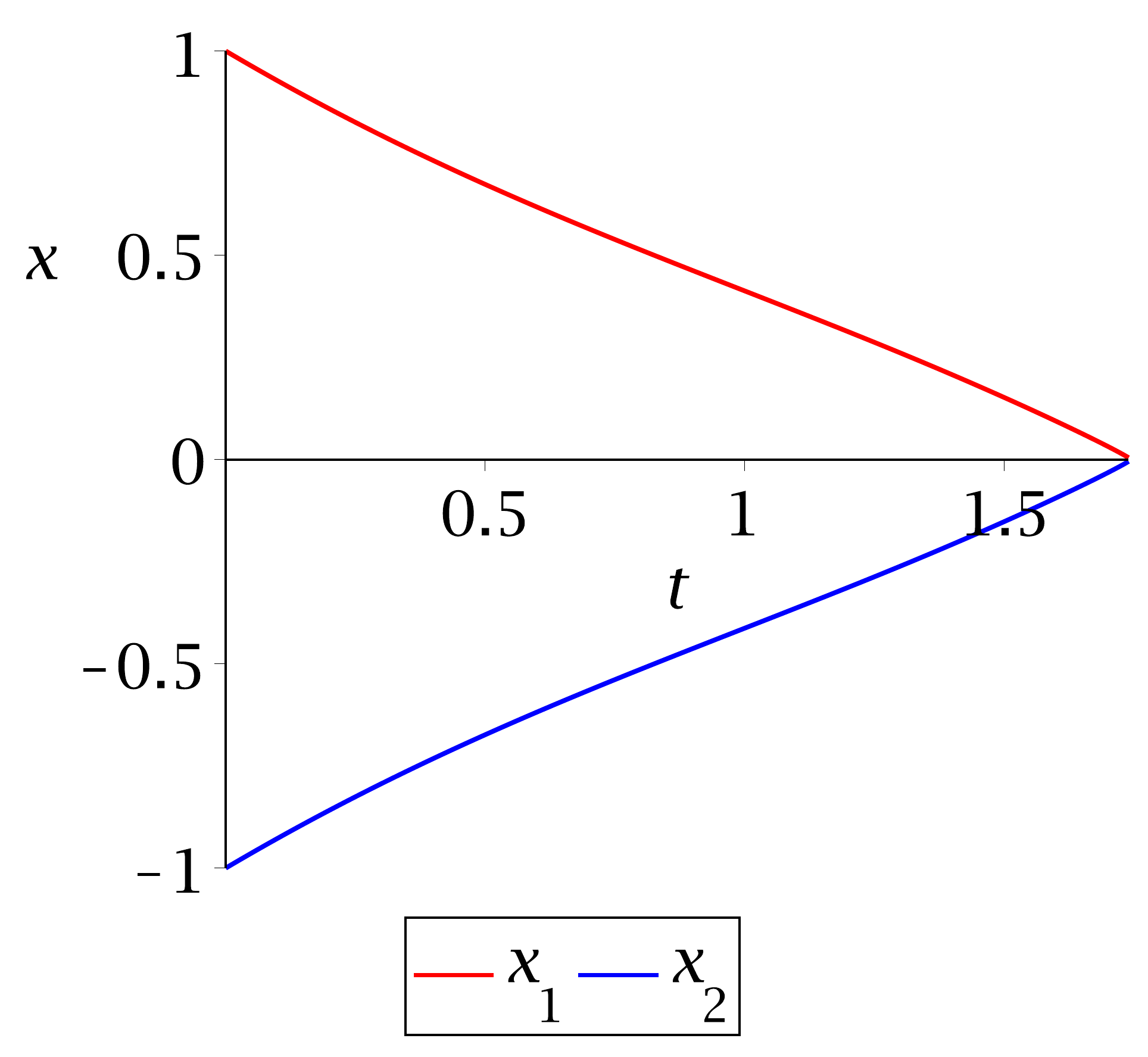}}
\caption{(a) The potential $V(r)$ used to mimic $A\overline{B}$ scattering. (b) Solutions $x_1(t)$ in red and $x_2(t)$ in blue, for $v_0=0$. (c) Solutions $x_1(t)$ and $x_2(t)$ for $v_0=-1.5$.}
\label{ABbar}
\end{figure}

To try and model $A\bar{B}$ scattering, we choose $a<0$ and $g>0$. For example, we fix the parameters to be
\begin{align}
 m=1,\quad g=1,\quad a=-0.1, \quad b=5,\quad c=3,\quad d=1,\quad f=\tfrac{1}{2} .
 \end{align}
The corresponding potential $V(r)$ is displayed in Fig.~\ref{ABbar}(a). It has one maximum, located near $r=1$. There are two different types of behaviour. Firstly, the points can repel and separate to infinity, as shown in Fig.~\ref{ABbar}(b) for initial velocity $v_0=0$. They can also hit the singularity at $r=0$, at which point we cannot evaluate the solution any further. We take this to correspond to the annihilation behaviour of the $A\overline{B}$ kinks, and an example of this type of solution, for initial velocity $v_0=-1.5$ is given in Fig.~\ref{ABbar}(c). Note that there is one behaviour type that we cannot recapture with this model. The $A\overline{B}$ kinks can reflect off each other and then separate to infinity, but we do not see this in the point particle model. 

\begin{figure}\centering
\subfigure[\,]{\includegraphics[height=4cm]{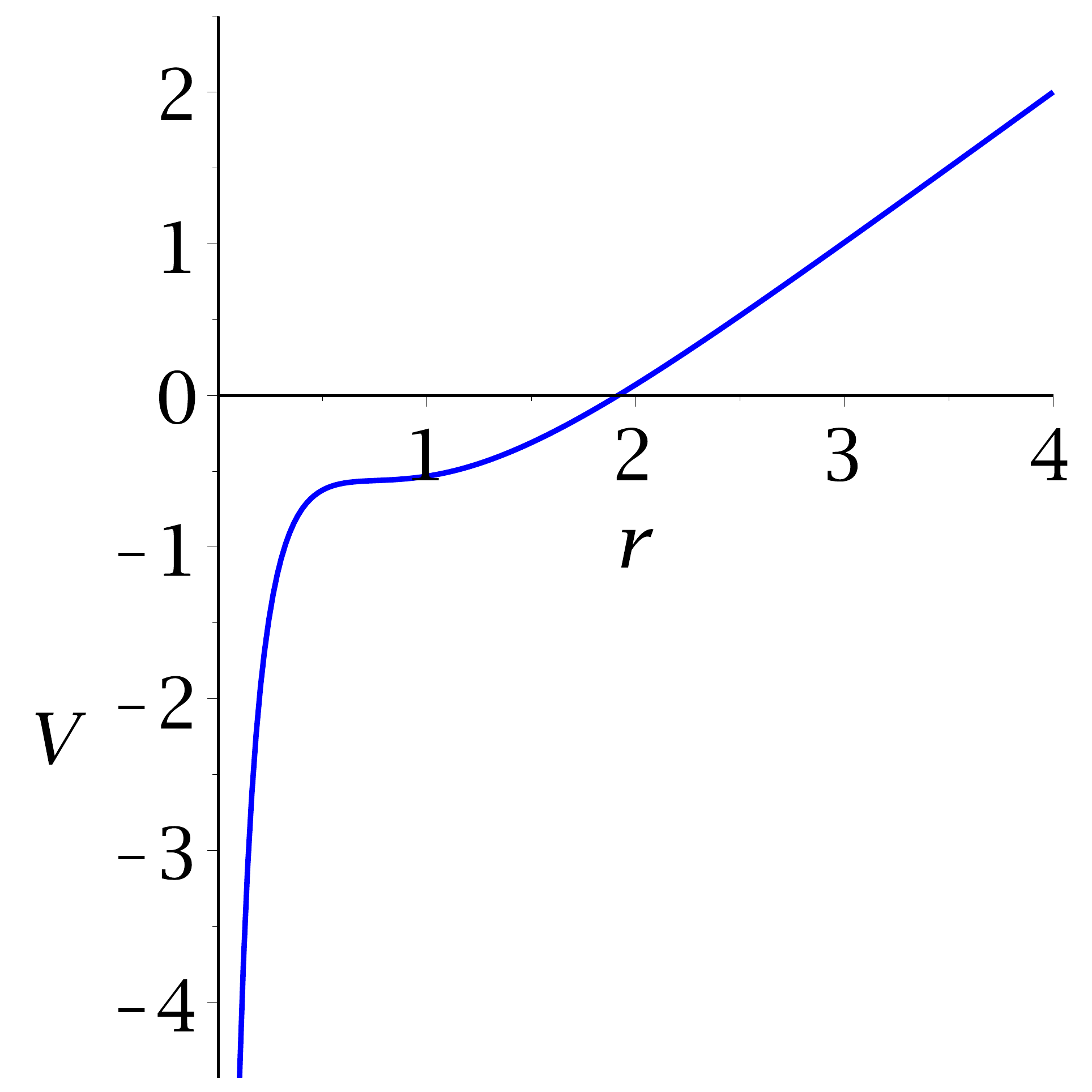}}\qquad
\subfigure[\,]{\includegraphics[height=4cm]{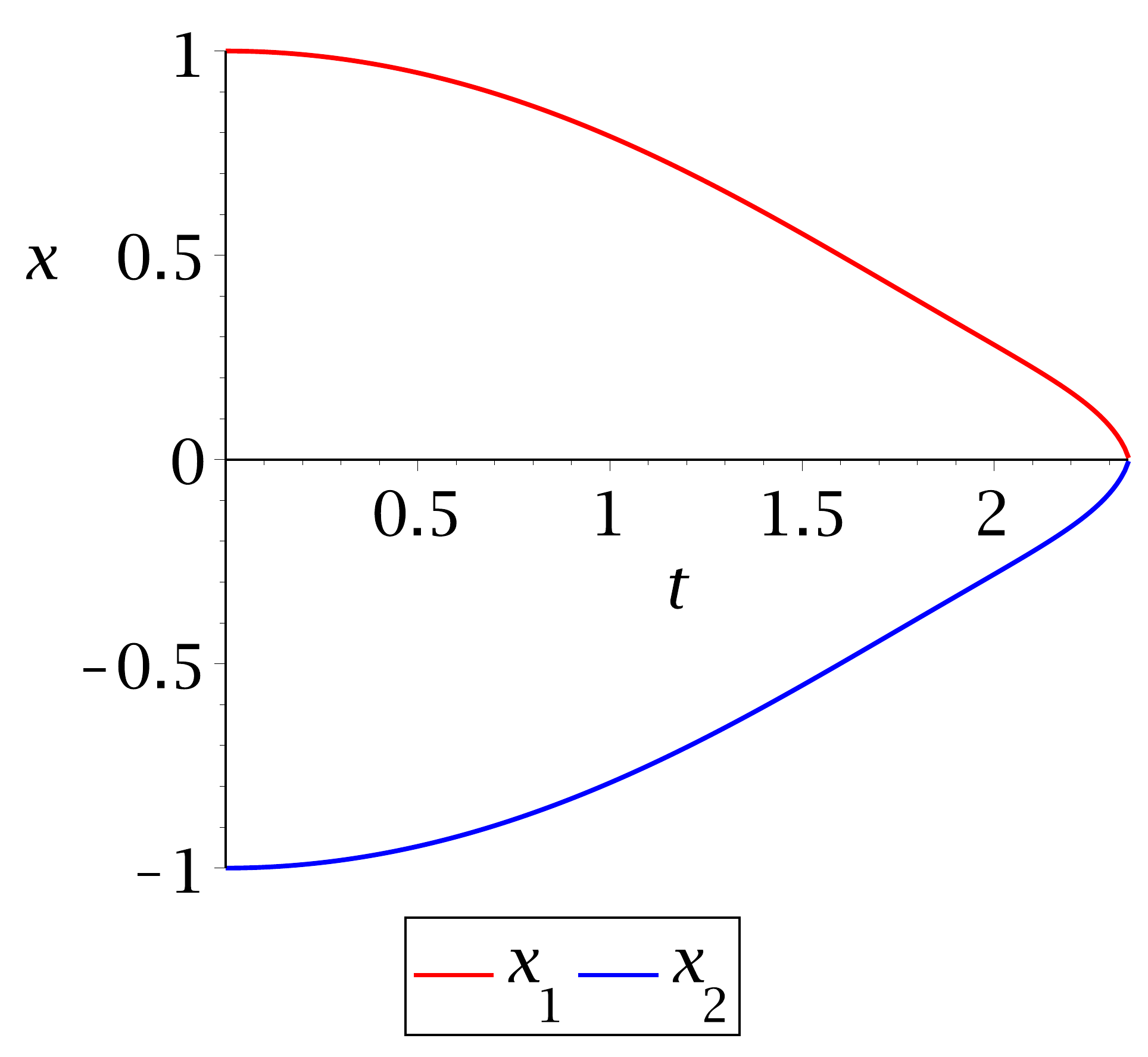}}
\caption{(a) The potential $V(r)$ used to mimic $B\overline{A}$ scattering. (b) Solutions $x_1(t)$ in red and $x_2(t)$ in blue, for $v_0=0$.}
\label{BAbar}
\end{figure}

Finally, when we take $a<0$ and $g<0$, we mimic the $B\overline{A}$ scattering. We choose the parameters
\begin{align}
 m=1,\quad g=-1,\quad a=-4, \quad b=5,\quad c=3,\quad d=1,\quad f=\tfrac{1}{2} .
\end{align}
Fig.~\ref{BAbar}(a) displays the corresponding potential $V(r)$. Note that if the parameters are chosen differently, an extra minimum may appear in the potential. We deliberately avoid this as it would allow for bound state solutions, and this is not an outcome for the $B\overline{A}$ configuration. In the point particle approximation with these parameter values, the only outcome is that the solutions hit the singularity at $r=0$, beyond which we cannot evaluate them any further. As an example, Fig.~\ref{BAbar}(b) presents the solutions for initial velocity $v_0=0$. The $B\overline{A}$ kinks do all eventually annihilate, however for certain choices of initial velocity, they can reflect off each other one or two times before annihilating. We are unable to see this behaviour in the toy model.

To summarise, for the $AB$ and $BA$ configurations, we are able to successfully replicate all of the kink trajectories in this toy model. However, for the $A\overline{B}$ and $B\overline{A}$ configurations, we are unable to see the reflection behaviour in the point particle approximation, and can only replicate the repelling and annihilating behaviours. In a point particle model, we are also unable to capture phenomena such as the oscillon, and the bounces in the shepherd field. Possible adjustments to the toy model could include implementing a different friction term. For example, using a friction term such as $ - \frac{d\dot{r}e^{-cr}}{r^2}$, we can force solutions in the $A\overline{B}$ and $B\overline{A}$ cases to grind to a halt before hitting the singularity.

\bibliography{Kink_ref}

\end{document}